\documentclass[12pt]{iopart}
\usepackage{iopams}
\usepackage{graphicx}
\usepackage{amsmath,amssymb}
\usepackage{dcolumn}
\usepackage{bm}
\usepackage{epsfig}
\usepackage{longtable}
\usepackage{subfigure}
\usepackage{multirow}
\usepackage{enumerate}
\usepackage{url}
\usepackage{afterpage}

\newcommand{\bfr}{ {\bf r}} 
\newcommand{\bfrp}{ {\bf r'}} 
\newcommand{\tp}{ {t^\prime}} 
\newcommand{\bfrpp}{ {\bf r^{\prime\prime}}} 
\newcommand{\bfR}{ {\bf R}} 
\newcommand{\sigmap}{\sigma^\prime} 
\newcommand{\omegap}{\omega^\prime} 

\newcommand\opd{d}
\newcommand\im{i}
\newcommand\ep[1]{e^{#1}}

\newcommand\abs[1]{\lvert#1\rvert}
\DeclareMathOperator\erfc{erfc}

\allowdisplaybreaks

\begin{document}
\bibliographystyle{iopart-num}

\title[]{Resolution-of-identity approach to Hartree-Fock, hybrid density functionals, RPA, MP2, and \textit{GW} with numeric atom-centered orbital basis functions}
\date{\today}

\author{Xinguo Ren\footnote{Corresponding author: xinguo@fhi-berlin.mpg.de}, Patrick Rinke, 
        Volker Blum, J{\"u}rgen Wieferink,
        Alexandre Tkatchenko, Andrea Sanfilippo, 
        Karsten Reuter\footnote{Present address: Lehrstuhl f\"ur Theoretische Chemie, 
        Technische Universit\"at M\"unchen, Lichtenbergstr.~4, D-85747 Garching, Germany}
        and Matthias Scheffler}
	\address{Fritz-Haber-Institut der Max-Planck-Gesellschaft, Faradayweg~4-6, D-14195 Berlin, Germany}

\begin{abstract}
Efficient implementations of electronic structure methods are essential for 
first-principles modeling of molecules and solids.
We here present a particularly efficient common framework
for methods beyond semilocal density-functional theory, including Hartree-Fock (HF), 
hybrid density functionals, random-phase approximation (RPA), second-order M{\o}ller-Plesset
perturbation theory (MP2), and the $GW$ method. This computational framework allows us to use 
compact and accurate numeric atom-centered orbitals (popular in many implementations of 
semilocal density-functional theory) as basis functions.  The essence of our framework is
to employ the ``resolution of identity (RI)'' 
technique to facilitate the treatment of both the two-electron Coulomb repulsion integrals (required in all these approaches) as well as the linear density-response function (required for RPA and $GW$). This is possible because
these quantities can be expressed in terms of products of single-particle basis functions, 
which can in turn be expanded in a set of auxiliary basis functions (ABFs). The construction 
of ABFs lies at the heart of the RI technique, and here we propose a simple prescription for
constructing the ABFs which can be applied regardless of whether 
the underlying radial functions have a specific analytical shape (e.g., Gaussian) or are numerically 
tabulated.  We demonstrate the accuracy of our RI implementation for Gaussian and NAO basis functions, 
as well as the convergence behavior of our NAO basis sets for the above-mentioned methods. 
Benchmark results are presented for the ionization energies of 50 selected atoms and molecules from 
the G2 ion test set as obtained with $GW$ and MP2 self-energy methods, and the G2-I atomization 
energies as well as the S22 molecular interaction energies as obtained with the RPA method.
\end{abstract}

\submitto{\NJP}


\pagestyle{plain}
\pagenumbering{arabic}

\maketitle
\tableofcontents

\section{\label{sec:introduction}Introduction}
        Accurate quantum-mechanical predictions of the properties of molecules
and materials (solids, surfaces, nano-structures, etc.) from first
principles play  an essential role in chemistry and condensed-matter
research today. Of particular importance are computational approximations to the 
many-body Schr\"odinger or Dirac equations that are tractable and yet retain
quantitatively reliable atomic-scale information about the system --- if not for all 
possible materials and properties, then at least for a relevant subset. 

Density-functional theory (DFT) \cite{Hohenberg/Kohn:1964,Kohn/Sham:1965} is one 
such successful avenue. It maps the interacting many-body problem onto an effective single-particle
one where the many-body complexity is hidden in the unknown exchange-correlation (XC)
term, which has to be approximated in practice.
Existing approximations of the XC term roughly fall into a  hierarchial 
scheme \cite{Perdew/Schmidt:2001}. Its local-density (LDA) \cite{Kohn/Sham:1965} and
generalized gradient approximations
(GGAs) \cite{Langreth/Mehr:1983,Becke:1988a,Lee/Yang/Parr:1992,Perdew/Burke/Ernzerhof:1996} 
are now well recognized 
workhorses with a broad application range in computational molecular
and materials science. However, several qualitative failures are well
known: To name but a few, certain adsorbate
geometries \cite{Feibelman:2001}, $f$-electron
systems \cite{McMahan/etal:1998,Johansson/etal:1995,Savrasov/etal:2001,Jiang/etal:2009,Petit/etal:2010}, 
or van der Waals interactions \cite{London:1937,Lifshitz:1956,Gunnarsson/Lundqvist:1976,Kristyan/Pulay:1994,Dobson:1994,Tkatchenko/etal:2011} 
are not described correctly at this level of theory. Thus, there is much ongoing work to extend the
reach of density-functional theory, e.g.,
meta-GGAs \cite{Tao/etal:2003,M06L,Perdew09}, formalisms to include van
der Waals interactions \cite{Wu/Yang:2002,Grimme:2006b,Tkatchenko/Scheffler:2009,Dion/etal:2004,Vydrov/Vanvoorhis:2009,Tkatchenko/etal:2010}, hybrid functionals \cite{Becke:1993,Perdew/Ernzerhof/Burke:1996,Heyd/Scuseria/Ernzerhof:2003,Zhao/Truhlar:2006a,M06,Grimme:2006}, or approaches based on the random-phase approximation (RPA) 
 \cite{Bohm/Pines:1953,Gell-Mann/Brueckner:1957,Langreth/Perdew:1975,Furche:2001,Harl/Kresse:2009,Ren/etal:2011,Paier/etal:2011,Eshuis/Bates/Furche:2011} that deal with the non-local correlations in a more systematical and non-empirical way.

Another avenue are the approaches of
quantum chemistry, that start with Hartree-Fock theory \cite{Slater:1960,Szabo/Ostlund:1989}. These
approaches offer a systematically convergable hierarchy of methods by
construction, said to reach ``gold standard'' accuracy for many
molecular systems at the level of coupled-cluster theory \cite{Koester/Kummel:1960,Cizek:1966,Bartlett/Musial:2007} 
[often, taken to include singles, doubles, and perturbative triples, CCSD(T) \cite{Raghavachari/etal:1989}]. CCSD(T) theory is significantly more accurate than DFT-LDA/GGA for many molecular systems but also significantly more
costly (formally scaling as ${\cal O}(N^7)$ with system size). It has its 
own shortcomings as well. 
For instance, systematic, material-specific
failures can occur in cases where the underlying
Hartree-Fock solution itself is not a good reference to start with (for example,
many open-shell systems), and a multireference extension of the approach
 \cite{Shavitt/Bartlett:2009} becomes necessary.

A third avenue for electronic structure calculations is the quantum Monte Carlo (QMC) method, 
in particular the diffusion QMC method \cite{Foulkes/etal:2001,Needs/etal:2010}. This
is a stochastic approach that deals with the many-body wavefunction directly. The diffusion QMC method
can often deliver high accuracies, and provide data and insights for problems
which are difficult for other approaches.  Its widespread use, however, is also impeded by the rather 
high computational costs. Moreover, the fixed-node approximation and the underlying
pseudopotential approximation are known issues which limit the practical accuracy of QMC.  
Regarding the computational cost, the QMC methods,
the algorithm of which is intrinsically parallel, are in a better position to benefit from the
development of petaflop supercomputers  \cite{Gillan/etal:2011}.

With the successes, but also the failures or shortcomings of the aforementioned
avenues, much attention is currently devoted to the
construction of further, \emph{systematic} and \emph{generally
applicable} methods or theoretical frameworks that 
can offer better accuracies than conventional DFT, but have lower
numerical costs and are free of the limitations of CCSD(T) and QMC. 
Among the various possible pathways, many-body perturbation theory (MBPT) based on an
efficiently attainable and trustful electronic reference state offers
such an avenue.
In particular,
approaches based on the RPA, which bridge the DFT and MBPT worlds
 \cite{Bohm/Pines:1953,Gell-Mann/Brueckner:1957,Langreth/Perdew:1977,Gunnarsson/Lundqvist:1976},
have recently enjoyed considerable attention for ground-state total-energy calculations.
For electron addition and removal energies, a self-energy based approach
that is consistent with the RPA total-energy treatment is Hedin's $GW$ approximation \cite{Hedin:1965}. 
$GW$ is especially popular in the solid state community \cite{Hybertsen/Louie:1986,Godby/Schlueter/Sham:1986,Schilfgaarde/Kotani/Faleev:2006,Shishkin/Marsman/Kresse:2007} and has become the method of choice for the calculation of quasiparticle band structures as measured in direct and inverse photoemission \cite{Aulbur/Jonsson/Wilkins:2000,Rinke/etal:2005,Rinke/pssb}.

Although RPA and $GW$ are receiving much attention in the
community today, the systematic investigation of diagrammatic
perturbation theory from first principles for real materials is only just
beginning. Its full promise lies in the fact that it is intermediate in
cost between DFT and coupled-cluster theories, and applicable in practice
to molecular and condensed materials alike -- including open-shell systems and
metals. 

Besides the more generally applicable RPA and $GW$ approaches, another correlation
method that is widely used in computational chemistry is 
second-order M{\o}ller-Plesset theory (MP2) \cite{Moller/Plesset:1934,Szabo/Ostlund:1989},
which belongs to the category of Hartree-Fock based quantum chemistry approaches mentioned 
above. MP2 does not reach the CCSD(T) accuracy, but its more favorable
computational scaling makes it applicable to larger system sizes. 
In analogy to the $GW$ self-energy, a MP2 self-energy approach \cite{Suhai:1983,Szabo/Ostlund:1989}
that is compatible with the MP2 total energy is possible as well.
As will be demonstrated more clearly in the next section, RPA, $GW$, and 
MP2 (both total and self-energy) are related approaches both 
diagramatically and numerically.
The development of numerical frameworks that enables
the implementation of all these approaches on an equal footing with high
numerical efficiency and accuracy is thus highly desirable.

In the present work, we describe the underpinnings of such a unified
numerical framework that is promising to boost the efficiency for 
all the above-mentioned methods, by allowing for their implementations 
with compact and efficient NAO basis sets.
Our specific 
implementation is based on the ``Fritz Haber Institute \emph{ab
initio} molecular simulations'' (FHI-aims) \cite{Blum/etal:2009}
program package. While we make reference to FHI-aims basis sets throughout much
of this work, the numerical foundation presented here is general:
applicable to any other type of atom-centered basis set. We
note that many production-quality implementations of hybrid
functionals, Hartree-Fock, and MBPT are based on analytical basis
functions such as Gaussian-type orbitals (GTOs) or
plane waves, and typically rely on pseudopotential-type approaches. In contrast, we
here aim for an all-electron, full-potential treatment with 
NAO basis sets that does not sacrifice accuracy compared to the alternatives. 

For DFT-LDA/GGA, NAOs are well established and can be found in several
implementations \cite{Delley:1990,Koepernik/Eschrig:1999,Soler/etal:2002,Ozaki/etal:2008,Blum/etal:2009,Kenney/Horsfield:2009,Chen/Guo/He:2010}. This is however not the case for HF and the MBPT approaches we are 
going to address in this paper.  Specifically we will present in the following
\begin{enumerate} [i)]
\item an atom-centered \emph{resolution of identity} (RI) framework
analogous to what is pursued in quantum chemical
methods \cite{Boys/Shavitt:1959,Whitten:1973,Dunlap/Connolly/Sabin:1979,Mittmire/Sabin/Trickey:1982,Vahtras/Almlof/Feyereisen:1993,Feyereisen/Fitzgerald/Komornicki:1993,Weigend/Haser/Patzelt/Ahlrichs:1998,Weigend:2002}.
This framework allows us to reduce all four-center two-electron Coulomb integrals
to precomputed three- and two-center integrals. Our scheme differs from
the quantum chemistry approach \cite{Weigend/Haser/Patzelt/Ahlrichs:1998,Weigend:2002}
in the auxiliary basis set construction, which is essential for
retaining the flexibility to work with any atom-centered 
basis function shape, rather than being restricted to analytical
shapes only. 
\item an assessment of the accuracy of the NAO basis sets used
for normal LDA/GGA calculations in FHI-aims, and
intended to be transferable regardless of the specific underlying
materials or functionals, for Hartree-Fock, MP2, hybrid functionals,
RPA, and $GW$. 
\end{enumerate}
Reference to relevant work by other groups in 
electronic-structure theory is made throughout this work.

The present paper demonstrates our approach for molecular systems
(non-periodic), and makes extensive use of established GTO basis sets
for comparison and reference purposes. In addition, we provide benchmark $GW$
\emph{vertical} (geometry of the ionized molecule not relaxed) ionization energies 
(IEs) for a subset of the G2 ion test set \cite{Curtiss/etal:1998}, and 
benchmark binding energies (RPA) for the G2-I and the S22 molecular test
set \cite{Jurecka/etal:2006}.
We restrict ourselves to algorithms that have standard scaling with system size
[${\cal O} (N^4)$ for HF, ${\cal O} (N^5)$ for MP2, etc]. Regarding total
energy \emph{differences} we restrict ourselves to a
discussion of counterpoise-corrected \cite{Boys/Bernardi:1970} results when
assessing the accuracy of MBPT methods that utilize the full (also
unoccupied) spectrum. 

In the following, we first recapitulate the HF, MP2, RPA, and $GW$ 
methods and highlight the structural similarity and difference
of the three correlation methods (section~\ref{sec:th}). We then introduce the basics of RI
and the RI formulation of the above methods (section~\ref{sec:RI_and_ABF}). 
Our own RI prescription and its accuracy is the subject of 
section~\ref{sec:our_RI}.  Section~\ref{sec:res} 
demonstrates the overall accuracy of our approach with NAO basis sets
for a variety of test systems for HF, MP2, hybrid density functionals, RPA, and $GW$.
The benchmark results for a subset of the G2 and S22 molecular test sets using
our approaches are presented in section~\ref{sec:benchmark}. Finally we conclude our paper in section~\ref{sec:conc}.

\section{\label{sec:th}Theoretical framework: HF, hybrid density functionals, MP2, RPA, and \textit{GW}}
	\subsection{Many-electron Hamiltonian and many-body perturbation theory}

Hartree-Fock (HF) theory, hybrid density functionals, and MBPT (MP2, RPA, and $GW$)
are all approximate ways to solve the interacting many-electron Hamiltonian  
  \begin{equation}
     \hat{H} = \sum_{i=1}^{N_e}\left[-\frac{1}{2} \nabla^2_i +
       v_i^\text{ext} \right] + \sum_{i<j}^{N_e} v^\text{ee}_{ij} \, ,
    \label{Eq:full_MB_hamiltonian}
  \end{equation}
where $N_e$ is the number of electrons that interact via the Coulomb interaction
$v^\text{ee}_{ij} \equiv 1/|\bfr_i-\bfr_j|$, and 
$v_i^\text{ext} \equiv v_\text{ext} (\bfr_i)$ is a local, multiplicative external
potential, usually due to the nuclei. 
Hartree atomic units are used 
throughout this paper. The numerical cost for an exact
solution of the Hamiltonian~(\ref{Eq:full_MB_hamiltonian}) scales
exponentially with system size (number of electrons). The 
systems for which such a solution is possible are thus heavily restricted
in size. In general, accurate approximations are needed. The most common
approximations first resort to the solution of a mean-field,
non-interacting  Hamiltonian $\hat{H}^0$ that yields an approximate
ground-state wave function $|\Phi_0 \rangle$:  
 \begin{align}    \label{Eq:nonint_manybody_Eq}
     \hat{H}^{0}  = \sum_{i=1}^{N_e} \hat{h}^{0}_i & =  \sum_{i=1}^{N_e} \left[ -\frac{1}{2} \nabla^2_i + 
     v_i^\text{ext}  + v^\text{MF}_i \right] \\ \nonumber 
    \hat{H}^{0} |\Phi_0 \rangle &  =  E_0^{(0)}|\Phi_0 \rangle \, .
 \end{align}
The ``(0)'' in $E_0^{(0)}$ implies the fact that this is the ground
state energy of the mean-field Hamiltonian. 
A suitable $\hat{H}^{0}$ should be solvable
with relative ease. $|\Phi_0 \rangle$ is a single Slater determinant 
formed from the lowest $N_e$ single-particle spin-orbitals determined by
 \begin{equation}
   \hat{h}^0|\psi_{n\sigma}\rangle = \epsilon_{n\sigma}|\psi_{n\sigma}\rangle \, 
   \label{Eq:SP_eigen_EQ}
 \end{equation}
where $\sigma$ denotes the spin index and $\hat{h}^0$ is the effective single-particle 
Hamiltonian noted in the bracket of (\ref{Eq:nonint_manybody_Eq}). The form
of (\ref{Eq:nonint_manybody_Eq}-\ref{Eq:SP_eigen_EQ}) is, of
course, precisely that of Kohn-Sham(KS) DFT (with a local mean-field
potential $v_i^\text{MF}$), or of HF theory and hybrid
functionals (with a non-local mean-field potential $v_i^\text{MF}$).

The purpose of starting with (\ref{Eq:full_MB_hamiltonian}-\ref{Eq:SP_eigen_EQ}) is 
to establish our notation for the following sections and to distinguish between
\begin{itemize}
\item $\hat{H}$ (the many-electron Hamiltonian), 
\item the mean-field Hamiltonian $\hat{H}^{0}$,
the solutions of which are many-electron wave functions given by
single Slater determinants,
and define an excited-state spectrum of their own
(obviously not the same as that of $\hat{H}$),
\item the effective single-particle Hamiltonian $\hat{h}^0$, which
generates the single-particle orbitals $\psi_{n\sigma}$ and orbital energies
$\epsilon_{n\sigma}$. 
\end{itemize}
In MBPT, one starts from $\hat{h}^0$ and the associated eigenenergies and eigenfunctions to systematically
approximate the properties of $\hat{H}$, e.g., its true ground-state
energy $E_0$  in MP2 or RPA or its singple-particle excitations in $GW$. The interacting
many-electron Hamiltonian $\hat{H}$ is partitioned into a mean-field
Hamiltonian  $\hat{H}^{0}$ as given by (\ref{Eq:nonint_manybody_Eq}) and
an interaction term $\hat{H}'$,  
 \begin{eqnarray}
     \hat{H} &= & \hat{H}^0 + \hat{H}' \nonumber \\
     \hat{H}' &=& \sum_{i<j}^{N_e}\frac{1}{|\bfr_i-\bfr_j|} - \sum_{i=1}^{N_e} v^\text{MF}_i\, . 
  \label{Eq:non_int_hamiltonian} 
 \end{eqnarray}
In the remainder of this section we collect the basic formulae
that define the mean-field Hamiltonians, perturbation theory for
ground state properties (MP2, RPA), and perturbation theory for
excited states (electron addition and removal energies, through either $GW$ or
MP2 self-energies). From a numerical point of view, the underpinning
of all these methods is the same: an efficient, accurate basis set
prescription, and an efficient expansion of the two-electron Coulomb
operator, which is the primary focus of this paper.

\subsection{Mean-field Hamiltonians of HF or DFT}

In HF theory the ground-state wave function of the Hamiltonian in
 (\ref{Eq:full_MB_hamiltonian}) is approximated by a single Slater
determinant $|\Phi_0\rangle$ and $E_0^{(0)}$ is obtained by a variational optimization,
leading to 
   \begin{equation}
    \langle \bfr|\hat{f}|\psi_n\rangle= \left[-\frac{1}{2} \nabla^2 + v_\text{ext}(\bfr) + v^\text{h}(\bfr) \right] \psi_{n\sigma}(\bfr)+   
        \int d\bfrp \Sigma^\text{x}_{\sigma}(\bfr, \bfrp) \psi_{n\sigma}
        (\bfrp) = \epsilon_{n\sigma} \psi_{n\sigma}(\bfr) \, .
     \label{Eq:HF_eq_realspace}
   \end{equation}
$\hat{f}$ here denotes the HF single-particle Hamiltonian, 
and $v^\text{h}(\bfr)$ is the Hartree potential, 
  \begin{equation}
    v^\text{h}(\bfr) = \int \frac{n(\bfrp)}{|\bfr-\bfrp|} d\bfrp \, 
    \label{Eq:hartree_potential}
  \end{equation}
with the electron density
  \begin{equation}
     n(\bfr) = \sum_{n\sigma}^\text{occ} |\psi_{n\sigma}(\bfr)|^2 \, ,
    \label{Eq:charge_density}
  \end{equation}
and $\Sigma^\text{x}_{\sigma}$ is the non-local, exact-exchange potential 
  \begin{equation}
       \Sigma^\text{x}_{\sigma}(\bfr, \bfrp)  = - \sum_n^\text{occ} 
       \frac{\psi_{n\sigma} (\bfr)\psi_{n\sigma}^{\ast}(\bfrp)}{|\bfr-\bfrp|} \, .
    \label{Eq:exact_exchange}
  \end{equation}
Equations (\ref{Eq:HF_eq_realspace})-(\ref{Eq:exact_exchange}) form a set of
non-linear equations that have to  be solved
iteratively. $v^h(\bfr)$ and $\Sigma^\text{x}_{\sigma}(\bfr,\bfrp)$
together yield the HF potential $v^\text{HF}$, a special case of the
mean-field potential $v_i^\text{MF}$ in
(\ref{Eq:nonint_manybody_Eq}) and (\ref{Eq:non_int_hamiltonian}). The HF  wavefunction
$|\Phi_0\rangle$ is given by the Slater determinant formed by the $N_e$
spin-orbitals $\psi_{n\sigma}$ with lowest energies $\epsilon_{n\sigma}$. 

At self-consistency, the HF total energy is
  \begin{equation}
     \label{Eq:HF_total_energy}
     E_\text{HF} = \langle \Phi_0 |\hat{H}^{0} + \hat{H}'| \Phi_0 \rangle 
                = \sum_{n\sigma}^\text{occ} \epsilon_{n\sigma} - E^\text{h} - E^\text{x}
     \nonumber
  \end{equation}
where the Hartree energy $E^\text{h}$ and exact-exchange energy $E^\text{x}$ are given
respectively by
  \begin{eqnarray}
    E^\text{h} &=& \frac{1}{2} \int d\bfr \:  n(\bfr) v_\text{h}(\bfr)  \nonumber    
 \\
    E^\text{x} &=& \frac{1}{2} \sum_{n\sigma}^\text{occ} \iint d\bfr d\bfrp \psi_{n\sigma}^{\ast}(\bfr)
               \Sigma^\text{x}_{\sigma}(\bfr,\bfrp) \psi_{n\sigma}(\bfrp) \nonumber \\  
            &=&  -\frac{1}{2} \sum_{mn\sigma}^\text{occ} \iint d\bfr
               d\bfrp \frac{\psi_{n\sigma}^{\ast}(\bfr)\psi_{m\sigma}(\bfr)
                                  \psi_{m\sigma}^{\ast}(\bfrp)\psi_{n\sigma}(\bfrp)}{|\bfr-\bfrp|}
               \, .   \label{Eq:HF_exchange_energy}
  \end{eqnarray} 
In general, the single-particle spin-orbitals $\psi_{n\sigma}(\bfr)$ are expanded in terms of a
set of basis functions $\{\varphi_i(\bfr)\}$  
  \begin{equation}
    \psi_{n\sigma}(\bfr) = \sum_{i} c^{i}_{n\sigma} \varphi_i(\bfr) \, .
    \label{Eq:AO_to_MO}
  \end{equation}
where $c^{i}_{n\sigma}$ are the expansion coefficients. In terms
of these basis functions, the HF effective potential can be expressed
in a matrix form 
  \begin{eqnarray}
   V^\text{HF}_{ij,\sigma} &=& \iint d\bfr d\bfrp \varphi_i(\bfr) 
   \left[ v^\text{h}(\bfr)\delta(\bfr-\bfrp)+\Sigma^\text{x}_\sigma(\bfr,\bfrp)\right] \varphi_j(\bfrp) \nonumber \\
   &=&   v^\text{h}_{ij,\sigma}+\Sigma^\text{x}_{ij,\sigma},
  \end{eqnarray}
where in particular the exact-exchange matrix is given by
  \begin{equation}
   \Sigma^\text{x}_{ij,\sigma}
   = \sum_{k l} (ik|lj)D_{kl,\sigma}\, .
  \label{Eq:exchange_matr}
  \end{equation}
In (\ref{Eq:exchange_matr}) $D_{kl,\sigma}$ is the density matrix
\begin{equation}
  \label{Eq:dens_mat}
    D_{kl,\sigma} = \sum_{n}^\text{occ}
    c_{n\sigma}^{k}c_{n\sigma}^{l\ast} \, ,
\end{equation}
and $(ij|kl)$ is the short-hand notation of quantum chemistry for 4-center 2-electron integrals
  \begin{equation}
    (ij|kl) = \iint \frac{\varphi_i(\bfr)\varphi_j(\bfr)\varphi_k(\bfrp)\varphi_l(\bfrp)}
              {|\bfr-\bfrp|} d\bfr d\bfrp\, .
    \label{Eq:4center_integral_first}
  \end{equation}
Very similar equations arise in KS-DFT with a
local $v_i^\text{MF}$, or in the generalized KS scheme \cite{GDFT:1993} with a fraction of
$\Sigma^\text{x}_{\sigma}(\bfr,\bfrp)$ in the potential. In
principle, the exact KS-DFT would yield the exact many-electron
ground-state energy $E_0$ and ground-state density $n_0(\bfr)$. In practice, the
XC energy functional and potential have to be approximated. 
The effective single-particle orbitals from either
HF or from approximate KS-DFT are convenient starting points for MBPT.

\subsection{Perturbation theory for the
  many-electron ground-state energy: MP2}

Assuming that $\hat{H}'$ is benign and can be treated as a perturbation
on top of $\hat{H}^0$,
the ground-state energy for the interacting system can be obtained
using Rayleigh-Schr{\"o}dinger perturbation theory (RSPT), or more precisely 
Brueckner-Goldstone perturbation theory \cite{Brueckner:1955,Goldstone:1957}. 
According to the Goldstone theorem \cite{Goldstone:1957}, in a diagrammatic
expansion of the ground-state total energy, only the ``linked" diagrams
need to be taken into account. And this guarantees the
the size-extensivity of the theory, i.e., the total energy scales 
correctly with the system size.
M{\o}ller-Plesset (MP) perturbation theory is a special case of RSPT \cite{Moller/Plesset:1934}, where the reference Hamiltonian $\hat{H}^0$ 
the HF Hamiltonian $\hat{H}_\text{HF}=\sum_i \hat{f}_i$. Terminating
the expansion at second order gives the MP2 theory, 
in which the (second-order) 
correlation energy is given by
 \begin{equation}
    E^{(2)}_0 = \sum_{k > 0} \frac{|\langle \Phi_k|\hat{H}'|\Phi_0 \rangle|^2}{E^\text{(0)}_{0} - E^\text{(0)}_{k}}\, .
   \label{Eq:mp2_c_energy_general}
 \end{equation}
 Here $\Phi_k$ are the Slater determinants representing the excited states of $H^0=H^\text{HF}$, and $E^\text{(0)}_k$ 
are the corresponding excited-state energies. $H'$ is given by (\ref{Eq:non_int_hamiltonian}) with
 $v^\text{MF}=v^\text{HF}$. Among all possible excited-state configurations $\Phi_k$, only double
excitations contributes in (\ref{Eq:mp2_c_energy_general}). This
is because singly-excited $\Phi_k$ do not
couple to the ground-state $\Phi_0$ (Brillouin's theorem \cite{Szabo/Ostlund:1989} for the HF
reference), whereas even higher-excited configerations (triples, quadruples, etc.) do not contribute due to the two-particle nature of the operator $H'$. As such, equation (\ref{Eq:mp2_c_energy_general})
can be expressed in terms of single-particle spin-orbitals,
 \begin{equation}
  E^{(2)}_0  = 
  \frac{1}{2} \sum_{mn}^\text{occ}\sum_{ab}^\text{unocc}\sum_{\sigma,\sigmap}
     (ma,\sigma|nb,\sigmap)    \left[ \frac{ (am,\sigma|bn,\sigmap)-
   (bm,\sigma|an,\sigmap)\delta_{\sigma\sigmap}}
   {\epsilon_{m\sigma} + \epsilon_{n\sigmap} -\epsilon_{a\sigma}-\epsilon_{b\sigmap}} \right] \hspace{1.5cm}
  \label{Eq:mp2_c_energy}
 \end{equation}
where $(ma,\sigma|nb,\sigmap)$ are 2-electron Coulomb repulsion integrals for molecular orbitals 
 \begin{equation}
  (ma,\sigma|nb,\sigmap)=\iint d\bfr d\bfrp \frac{\psi_{m\sigma}^{*}(\bfr)\psi_{a\sigma}(\bfr)
            \psi_{n\sigmap}^{*}(\bfrp)\psi_{b\sigmap}(\bfrp)}{|\bfr-\bfrp|} \, .
   \label{Eq:MO_4orb_integral}
 \end{equation}
The two terms in (\ref{Eq:mp2_c_energy})
correspond to the 2nd-order Coulomb (direct) and 2nd-order exchange energy, respectively.  

\subsection{Perturbation theory for the
  many-electron ground-state energy: RPA}

MP2 corresponds to the 2nd-order term in a perturbation theory where the perturbation expansion is 
essentially based on  the bare Coulomb operator (with HF effective potential subtracted). 
As such, the MP2 correlation energy diverges for the homogeneous electron gas and 
metals with zero direct energy gap. To overcome this problem in 
the framework of MBPT it is essential to sum up the diverging terms in the perturbation series to infinite order. One such example, which has gained considerable popularity recently \cite{Furche:2001,Fuchs/Gonze:2002,Furche/Voorhis:2005,Scuseria/Henderson/Sorensen:2008,Janesko/Henderson/Scuseria:2009,Toulouse/etal:2009,Paier/etal:2010,Marini/Gonzalez/Rubio:2006,Harl/Kresse:2008,Harl/Kresse:2009,Lu/Li/Rocca/Galli:2009,Dobson/Wang:1999,Rohlfing/Bredow:2008,Ren/Rinke/Scheffler:2009,Schimka/etal:2010,Zhu/etal:2010}, is the RPA \cite{Bohm/Pines:1953,Gell-Mann/Brueckner:1957,Harris/Griffin:1975,Langreth/Perdew:1977,Gunnarsson/Lundqvist:1976}, that 
in the context of MBPT corresponds to 
an infinite summation of ``ring" diagrams. The choice of the non-interacting
reference Hamiltonian $\hat{H}^0$ can be, e.g., HF or DFT with any desired XC functional.

Apart from the diagrammatic representation, RPA can also be formulated in other ways, e.g., 
as the simplest time-dependent Hartree approximation in the context of the adiabatic-connection
fluctuation-dissipation theorem \cite{Harris/Griffin:1975,Langreth/Perdew:1977,Gunnarsson/Lundqvist:1976},
or as a subclass of terms in coupled-cluster theory with double
excitations \cite{Scuseria/Henderson/Sorensen:2008}. In the context of DFT \cite{Langreth/Perdew:1977}, 
RPA calculations can be performed self-consistently by means of the optimized effective potential 
approach \cite{Niquet/Fuchs/Gonze:2003_1,Hellgren/Barth:2007,Kuemmel/Kronik:2008,Hellgren/Barth:2010}.

In close-packed notation, the RPA correlation energy (cRPA) can be expressed in terms of the
RPA dielectric function $\varepsilon$, or alternatively the
non-interacting density response function $\chi^0$ on the imaginary
frequency axis 
 \begin{eqnarray}
  E_\text{c}^\text{RPA} &=& \frac{1}{2\pi}\int_0^\infty d\omega \text{Tr} 
    \left[\text{ln} \left(\varepsilon(i\omega)\right) + \left(1-\varepsilon(i\omega)\right) \right] \nonumber \\
    &=& \frac{1}{2\pi}\int_0^\infty d\omega \text{Tr} 
    \left[\text{ln}\left(1-\chi^0 (i\omega)v\right) + \chi^0 (i\omega)v\right] \nonumber \\
   &=&  - \frac{1}{2\pi}\int_0^\infty d\omega \sum_{n=2}^{\infty}\frac{1}{n}\text{Tr} \left[(\chi^0 (i\omega)v)^n \right]\, .
  \label{Eq:cRPA}
 \end{eqnarray}
The real-space (Adler-Wiser \cite{Adler:1962,Wiser:1963}) representation of $\chi^0$ reads
 \begin{align}
   \chi^0(\bfr,\bfrp,i\omega)&=\langle \bfr |\chi^0(i\omega) |\bfrp \rangle \nonumber \\
   & = 
      \sum_{\sigma}\sum_m^\text{occ}\sum_a^\text{unocc} 
    \frac{\psi_{m\sigma}^{\ast}(\bfr)\psi_{a\sigma}(\bfr)\psi_{a\sigma}^{\ast}(\bfrp)
     \psi_{m\sigma}(\bfrp)} {i\omega - \epsilon_{a\sigma} + \epsilon_{m\sigma} } + 
       \text{c.c.}\, , 
   \label{Eq:chi^0_realspace_imag}
 \end{align}
where c.c. denotes ``complex conjugate", and $\psi_n(\bfr)$ and
$\epsilon_n$ are single-particle orbitals and orbital energies as
implied by (\ref{Eq:SP_eigen_EQ}). 
The RPA dielectric function
$\varepsilon$ is linked to $\chi^0$ through
  \begin{equation}
      \varepsilon(\bfr, \bfrp, i\omega) = \delta(\bfr-\bfrp) - \int d\bfrpp  
            v(\bfr,\bfrpp)\chi^\text{0}(\bfrpp,\bfrp,i\omega) \, .
      \label{Eq:dielec_func_realspace}
  \end{equation}
Using (\ref{Eq:MO_4orb_integral}) and (\ref{Eq:chi^0_realspace_imag}),
the lowest-order term in (\ref{Eq:cRPA}) can be expressed as
 \begin{equation}
  \text{Tr}\left[\chi^0(i\omega)v\right]=\iint d\bfr d\bfrp \chi^0(\bfr,\bfrp,i\omega) v(\bfrp,\bfr)
  = \sum_\sigma \sum_m^\text{occ}\sum_a^\text{unocc} \frac{(ma,\sigma|am,\sigma)}
           {i\omega-\epsilon_{a\sigma}+\epsilon_{m\sigma}} + \text{c.c.}\, .
 \end{equation}
Higher-order terms in (\ref{Eq:cRPA}) follow analogously.
There is thus a straightforward path to compute $\chi^0$, and hence
the RPA correlation energy, once the selected non-interacting
reference state is solved. 
In practice the RPA correlation energy is
always combined with the exact-exchange energy (EX) in
(\ref{Eq:HF_exchange_energy}), henceforth denoted (EX+cRPA), but evaluated with the same
single-particle orbitals as used in $E_\text{c}^\text{RPA}$. The
choice of input orbitals will in the following be marked by
(EX+cRPA)@MF, where MF specifies the mean-field approach used to
compute the single-particle orbitals. The application of EX+cRPA to various
realistic 
systems, as well as the development of schemes that go beyond simple EX+cRPA, is an active field \cite{Furche:2001,Fuchs/Gonze:2002,Furche/Voorhis:2005,Jiang/Engel:2007,Furche:2008,Scuseria/Henderson/Sorensen:2008,Janesko/Henderson/Scuseria:2009,Marini/Gonzalez/Rubio:2006,Gonzalez/Fernandez/Rubio:2007,Harl/Kresse:2008,Harl/Kresse:2009,Toulouse/etal:2009,Lu/Li/Rocca/Galli:2009,Dobson/Wang:1999,Rohlfing/Bredow:2008,Ren/Rinke/Scheffler:2009,Schimka/etal:2010,Zhu/etal:2010,Ismail-Beigi:2010,Ren/etal:2011}. %
For instance, when the so-called single-excitation and second-order screened exchange
contributions \cite{Grueneis/etal:2009,Paier/etal:2010} are added to EX+cRPA, the resulting accuracy 
is impressive \cite{Ren/etal:2011,Paier/etal:2011}.

\subsection{Perturbation theory for electron addition or removal energies:
  \textit{GW} or MP2}

Ground-state energies aside, one is often interested in the properties
of electronically excited states. Part of this information is in principle
accessible by taking the difference of the total energies of $N$-electron
system and $N\pm 1$-electron systems using approaches discussed above.
In practice, this approach mainly works for computing core-level excitations and/or
the first ionization energy and electron affinity for (small) finite systems.
In contrast, Green function techniques are more convenient and powerful for dealing
with electronic excitations in general. The basic theory of 
Green functions is well documented in textbooks \cite{Abrikosov/Gorkov/Dzyaloshinski:1975,Fetter/Walecka:1971}. Here we
collect the contextual equations, based on which
practical approximations can be introduced.

The single-particle Green function of an interacting many-electron system is defined as 
 \begin{equation}
   G(\bfr,t;\bfrp,\tp)  =  -i\langle N |\hat{T}
   \hat{\psi}(\bfr,t)\hat{\psi}^{\dagger}(\bfrp,\tp) |N \rangle 
 \end{equation}
where $|N\rangle=|\Psi_0(\bfr_1,\dots,\bfr_N)\rangle$ denotes the \emph{interacting} ground-state wave
function of an $N$-electron system (solution to
(\ref{Eq:full_MB_hamiltonian})). $\hat{\psi}(\bfr,t)$ 
and $\hat{\psi}^{\dagger}(\bfrp,\tp)$ are field operators in the
Heisenberg picture that annihilate and create an electron at
space-time point ($\bfr,t$) and ($\bfrp,\tp$), respectively. $\hat{T}$ is the
time-ordering operator. The Green function
$G(\bfr,t;\bfrp,\tp)$ measures the probability amplitude of a hole 
created at ($\bfr$, $t$) propagating to ($\bfrp,\tp$) for $t<\tp$, or an electron 
added at ($\bfrp,\tp$) propagating to ($\bfr$, $t$) for $t>t'$. The poles of
its Fourier transform, $G(\bfr,\bfrp,\omega)$, correspond to 
the single-particle excitation energies as measured for example in
direct and inverse photoemission experiments.  

For Hamiltonians with a time-independent external potential, as considered in this paper, the Green function depends only on the difference between $t$ and $t'$, $ G(\bfr,t;\bfrp,\tp)= G(\bfr,\bfrp;t-\tp)$.
Its Fourier transform gives the frequency-dependent Green function $G(\bfr,\bfrp;\omega)$ that
satisfies the Dyson equation,
 \begin{equation}
  \left[\omega +\frac{1}{2}\nabla^2 -v_\text{ext}(\bfr) - v^\text{h}(\bfr) \right] 
 G(\bfr,\bfrp;\omega)  
  - \int d \bfrpp \Sigma(\bfr,\bfrpp,\omega)G(\bfrpp,\bfrp;\omega) = \delta(\bfr-\bfrp)\, .
 \end{equation}
Here $v^\text{h}(\bfr)$ is the electrostatic Hartree potential defined in (\ref{Eq:hartree_potential}) and $\Sigma(\bfr,\bfrpp,\omega)$ is the dynamical,
  non-local, complex self-energy that contains all the many-body XC effects. 

$G(\bfr,\bfrp;\omega)$ and $\Sigma(\bfr,\bfrp;\omega)$ for many-body interacting systems
can in principle be obtained using diagramatic Feynman-Dyson perturbation theory. The perturbation series is
built on a non-interacting Green function $G^0(\bfr,\bfrp,\omega)$
that corresponds to the  \emph{non-interacting} single-particle  Hamiltonian
$\hat{h}^0$. With single-particle orbitals $\psi_{n\sigma}(\bfr)$
and orbital energies $\epsilon_{n\sigma}$ of $\hat{h}^0$, one has  
  \begin{equation}
   G^\text{0}_{\sigma}(\bfr,\bfrp,\omega)=\sum_n
  \frac{\psi_{n\sigma}(\bfr)\psi^\ast_{n\sigma}(\bfrp)}{\omega-\epsilon_{n\sigma}-i\eta~\text{sgn}(\epsilon_\text{F}-\epsilon_{n\sigma})
  } \, ,
   \label{Eq:non_int_GF}
  \end{equation}
where $\epsilon_\text{F}$ is the Fermi energy, and $\eta$ a positive infinitesimal. 
For a given $G^0$, corrections to the single-particle excitation energies
can be computed from approximate perturbative expansions of
$\Sigma(\bfr,\bfrp;\omega)$. Examples are the $GW$ method and the
2nd-order approximations discussed below.   

In the $GW$ approximation proposed by Hedin \cite{Hedin:1965}, the self-energy assumes the form
  \begin{equation}
     \label{Eq:Sigma_GW}
     \Sigma^{GW}_\sigma(\bfr,\bfrp,\omega) = \frac{i}{2\pi}\int d \omega 
           G_\sigma(\bfr, \bfrp, \omega+\omegap)
      W(\bfr, \bfrp, \omegap)e^{i\omega\eta}\, .
  \end{equation}
Here $W$ is the screened Coulomb potential at the RPA level
  \begin{equation}
      W(\bfr, \bfrp, \omega) = \int d\bfrpp \varepsilon^{-1}(\bfr, \bfrpp, \omega) 
           v(\bfrpp,\bfr) \, ,
      \label{Eq:screened_coulomb_realspace}
  \end{equation}
with the dynamical dielectric function $\varepsilon$ as 
defined in (\ref{Eq:chi^0_realspace_imag}) and (\ref{Eq:dielec_func_realspace}), but on the real frequency axis with $i\omega$ replaced by $\omega+i\eta$ ($\eta \rightarrow 0^+$).  

In practice, one-shot perturbative $GW$ calculations (often referred to
as  $G^0W^0$) based on a fixed, non-interacting reference state
(DFT with popular functionals or HF) are often performed. With $\chi^0$ computed from the non-interacting Green function in (\ref{Eq:non_int_GF}), the ensuing $W$ can be expanded in powers of $\chi^0v$
 \begin{equation}
    W=v+v\chi^0 v + v\chi^0 v\chi^0 v + \cdots \, .
    \label{Eq:symbolic_W}
 \end{equation}
The $GW$ approximation can thus be regarded as an
infinite series in the bare Coulomb potential $v$, 
or alternatively, as is obvious from (\ref{Eq:Sigma_GW}),
as first-order perturbation in terms of the screened Coulomb potential $W$. 

Once the $G^0W^0$ self-energy is obtained from (\ref{Eq:Sigma_GW}), the corrections to
the single-particle orbital energies are given by
  \begin{equation}
   \label{Eq:qpe}
   \epsilon_{n\sigma}^{G^0W^0} = \epsilon_{n\sigma} + \langle \psi_{n\sigma}| \Sigma_\sigma^{G^0W^0}
   \left(\epsilon_{n\sigma}^{G^0W^0}\right) - v^\text{xc}|
   \psi_{n\sigma}\rangle \, ,
  \end{equation}
where the XC part $v^\text{xc}$ of
the reference mean-field potential has to be subtracted. \ref{Eq:qpe} approximates the quasiparticle wavefunctions with the single-particle orbitals of the reference state. This is often justified, but may break down in certain cases \cite{White/Godby/Rieger/Needs:1997,Rohlfing/Wang/Krueger/Pollmann:2003,Fratesi/Brivio/Rinke/Godby,Pulci/Reining/Onida/DelSole/Bechstedt:2001,ClusterImStates:2004,Gatti/Bruneval/Olevano/Reining:2007}. 

Rewriting the diagonal elements of the $G^0W^0$ self-energy on the
imaginary frequency axis in terms of four-center Coulomb integrals, we obtain
 \begin{align}
  \Sigma_{n\sigma}^{G^0W^0}(i\omega) &= 
           \iint d\bfr d\bfrp \psi_{n\sigma}^\ast(\bfr)\Sigma_\sigma^{G^0W^0} (\bfr, \bfrp, i\omega)
             \psi_{n\sigma}(\bfrp) \nonumber \\
  &= -\frac{1}{2\pi} \sum_m  \int_{-\infty}^{\infty} d\omegap G^0_{m\sigma}(i\omega+i\omegap)
           (nm,\sigma|W^0(i\omegap)|mn,\sigma) 
  \label{Eq:G0W0_imag_freq}
 \end{align}
where $G^0_{m\sigma}(i\omega)=1/(i\omega +\epsilon_\text{F}-\epsilon_{m\sigma})$, and
 \begin{equation}
  (nm,\sigma|W^0(i\omega)|mn,\sigma)=  
  \iint d\bfr d\bfrp \psi_{n\sigma}^\ast(\bfr) \psi_{m\sigma}(\bfr)
         W(\bfr,\bfrp,i\omega) \psi_{m\sigma}^\ast(\bfrp) \psi_{n\sigma}(\bfrp)\, .
     \label{Eq:4index_screened_coulomb}
 \end{equation}
This expression can be analytically continued to the real-frequency
axis using Pad{\'e} approximation or a two-pole model \cite{Rojas/Godby/Needs:1995}.

The $G^\text{0}W^\text{0}$ approach, as a highly popular choice for quasiparticle excitation calculations, 
in particular for solids \cite{Aulbur/Jonsson/Wilkins:2000,Onida/Reining/Rubio:2002,Rinke/etal:2005,Rinke/pssb,Giantomassi/etal:2011}, is akin to the RPA approach for computing the ground-state correlation energy.
Similarly, one can also introduce a 2nd-order perturbation theory for the self-energy \cite{Szabo/Ostlund:1989}
consistent with the MP2 correlation energy. 
Based on the HF noninteracting Green function [equation (\ref{Eq:non_int_GF}) with HF orbitals], the 2nd-order self-energy can be expressed as
 \begin{eqnarray}
   \Sigma^{(2)}(1,2) &=& -i\int d3 d4 G^\text{HF}(1,2)G^\text{HF}(3,4)G^\text{HF}(4,3)v(1,3)v(2,4) + \nonumber \\
  &~& i\int d3 d4 G^\text{HF}(1,4)G^\text{HF}(4,3)G^\text{HF}(3,2)v(1,3)v(2,4)\, .
          \label{Eq:2nd_self-energy_symbo}
 \end{eqnarray}
Here for notational simplicity we have used $1=(\bfr_1,t_1)$, and $v(1,2)=v(\bfr_1-\bfr_2)\delta(t_1,t_2)$. 
After integrating out the internal (spatial and time) coordinates in
(\ref{Eq:2nd_self-energy_symbo}), the final
expression for $\Sigma^{(2)}$ (in frequency domain) within the HF molecular orbital basis is given by 
   \begin{align}
      \Sigma^\text{(2)}_{mn,\sigma} (\omega) 
         & = \sum_{l}^\text{occ}\sum_a^\text{unocc} \sum_{p,\sigmap} (mp,\sigma|la,\sigmap)(pn,\sigma|al,\sigmap) \times  \nonumber \\
      &~~  \left[ \frac{\theta(\epsilon_\text{F}-\epsilon_p)} 
                          {\omega+\epsilon_{a\sigmap}-\epsilon_{l\sigmap} - \epsilon_{p\sigma} -i\eta} 
                 +  \frac{\theta(\epsilon_p-\epsilon_\text{F})}
                         {\omega + \epsilon_{l\sigmap} - \epsilon_{a\sigmap} - \epsilon_{p\sigma} + i\eta} 
                      \right] \nonumber \\
     &~ - \sum_{l}^\text{occ}\sum_a^\text{unocc} \sum_{p} (mp,\sigma|la,\sigma)(pl,\sigma|an,\sigma) \times
           \nonumber \\
     &~~ \left[ \frac{\theta(\epsilon_\text{F}-\epsilon_p)} 
                          {\omega+\epsilon_{a\sigma}-\epsilon_{l\sigma} - \epsilon_{p\sigma} -i\eta}  
                 +  \frac{\theta(\epsilon_p-\epsilon_\text{F})}
                         {\omega + \epsilon_{l\sigma} - \epsilon_{a\sigma} - \epsilon_{p\sigma} + i\eta} 
                      \right]\, ,
          \label{Eq:2nd_self-energy}
    \end{align}
where $\theta(x)$ is the Heaviside step function and $\eta\rightarrow 0^+$. 
Again, the two lines
in (\ref{Eq:2nd_self-energy}) correspond to correlations arising from the 2nd-order direct
and 2nd-order exchange interaction, respectively. With this self-energy single-particle
excitation energies --  here denoted MP2 quasiparticle energies -- can
be obtained by adding a correction to the HF orbital energies 
 \begin{equation}
   \epsilon^\text{MP2}_{n\sigma} = \epsilon_{n\sigma}^\text{HF} + \Sigma^\text{(2)}_{nn,\sigma}(\epsilon^\text{MP2}_{n\sigma})\, .
   \label{Eq:MP2_SP_energy}
 \end{equation}
Similar to the MP2 correlation energy, the quality of the MP2 self-energy as given by (\ref{Eq:2nd_self-energy})
will deteriorate as the single-particle energy gap shrinks.

By means of the Dyson equation $G=G^0+G^0\Sigma G$ self-consistent Green functions and self-energies could be obtained. The advantage is that the self-consistent Green function will then be independent of
$G^0$ and satisfies particle number, momentum and energy
conservation \cite{Baym/Kadanoff:1961,Baym:1962}. For an 
in-depth discussion, we refer to \cite{Dahlen/Leeuwen/Barth:2005,Dahlen/Leeuwen/Barth:2006} and references therein.

\section{\label{sec:RI_and_ABF}Resolution of identity for HF, MP2, RPA, and \textit{GW}}

\subsection{Background}

In this section we present the basic resolution of 
identity (RI) formalism: the auxiliary basis sets, different
variants of RI, and the working equations for HF, hybrid density functionals,
MP2, RPA, and $GW$. Similar accounts have been given
in the literature on one or another of the above
methods, and we encourage readers to consult them \cite{Feyereisen/Fitzgerald/Komornicki:1993,Weigend/Haser/Patzelt/Ahlrichs:1998,
Weigend:2002,Aryasetiawan/Gunnarsson:1994a,Friedrich/etal:2009,Umari/Stenuit/Baroni:2009,
Eshuis/Yarkony/Furche:2010}. Our aim here is to lay out a complete
description of all necessary specifics \emph{in one
consistent notation} that will allow us to present our own
developments (next sections) in a self-contained way.

The common ingredient of all techniques introduced in section~\ref{sec:th}
(Hartree-Fock, hybrid functionals, MP2, RPA, $GW$) are the 
four-orbital Coulomb integrals  
  \begin{equation}
     (mn,\sigma|ab,\sigmap) = \sum_{ijkl} (ij|kl) 
     c^{i\ast}_{m\sigma} c^{j}_{n\sigma} c^{k\ast}_{a\sigmap} c^{l}_{b\sigmap},
    \label{Eq:AO2MO_integral}
  \end{equation}
where $(ij|kl)$ are the two-electron integrals in a basis set representation as
defined in (\ref{Eq:4center_integral_first}), and $c^{i}_{m\sigma}$
are the eigen-coefficients for the molecular orbitals.
Computing (and possibly storing) these 4-center 2-electron integrals
can be a major bottleneck for approaches beyond LDA and GGAs.

For analytical GTOs algorithms have been developed to handle
$(mn,\sigma|ab,\sigmap)$ integrals efficiently and on-the-fly
\cite{Head-Gordon/Pople:1988,Adams/Adamson/Gill:1997}. 
More general NAOs, however, are not amenable to such
algorithms.  In the context of HF, we note that RI is not the
only technique available to deal with the four-center integrals: 
Making use of the translational properties of spherical harmonics,
Talman and others \cite{Talman84-4center,Talman:2003,Talman:2007,Toyoda/Ozaki:2009} have
developed techniques based on multipole expansions of basis
functions. Multi-center NAO integrals can then be treated partially analytically.
Alternatively, efficient Poisson solvers \cite{Shang/Li/Yang:2010}
have recently been used to enable direct NAO HF calculations
through four-center integrals for simple
systems \cite{Talman:2003,Shang/Li/Yang:2010}.  Finally, a yet
different numerical route (based on expanding orbital products
directly) has been adopted along the line of time-dependent DFT in the
linear-response framework \cite{Delley:2010}. RI is, however,
most successful at reducing the computational load compared to
direct four-center integral based methods,
most prominently for MP2 in quantum chemistry \cite{Weigend/Haser/Patzelt/Ahlrichs:1998}, 
which is why we pursue this route here. 


\subsection{\label{sec:auxbasis}Auxiliary basis}

The resolution of identity (RI) or
synonymously density
fitting technique \cite{Boys/Shavitt:1959,Whitten:1973,Dunlap/Connolly/Sabin:1979,Mittmire/Sabin/Trickey:1982,Vahtras/Almlof/Feyereisen:1993,Feyereisen/Fitzgerald/Komornicki:1993,Weigend/Haser/Patzelt/Ahlrichs:1998,Weigend:2002}  
amounts to representing pair products of atomic basis
functions $\varphi_i(\bfr)\varphi_j(\bfr)$ in terms of 
auxiliary basis functions (ABFs),
 \begin{equation}
   \rho_{ij}(\bfr) \equiv \varphi_i({\bfr})\varphi_j({\bfr}) \approx \tilde{\rho}_{ij}(\bfr)
    \equiv \sum_{\mu} C_{ij}^{\mu} P_{\mu}({\bfr}).
   \label{Eq:RI_expansion}
 \end{equation}
$\mu=1,2,\ldots,N_{\text{aux}}$ labels the auxiliary basis 
functions $\{P_{\mu}\}$, $C_{ij}^{\mu}$ are the expansion coefficents,   
and $\rho_{ij}(\bfr)$ and $\tilde{\rho}_{ij}(\bfr)$ here denote
pair products of basis functions and their approximate expansion in
ABFs. The evaluation of the 4-center integrals in
(\ref{Eq:4center_integral_first}) then reduces to  
  \begin{eqnarray}
    (ij|k l) & \approx & \sum_{\mu\nu} C_{ij}^{\mu} (\mu|\nu)
     C_{k l}^{\nu}, 
    \label{Eq:RI_expansion_general} \\
    (\mu|\nu)& = & V_{\mu\nu}=\int \frac{P_\mu(\bfr) P_\nu(\bfrp)}{|\bfr-\bfrp|}
     d\bfr d\bfrp \, .
    \label{Eq:coulomb_matrix}
  \end{eqnarray}
To determine the expansion coefficients $ C_{ij}^{\mu}$,
three-center integrals involving the ABFs and the pair products of the NAOs 
are required. Thus the expensive (both in time and memory, if there is a
need to pre-compute numerical matrix elements) 4-center integrals
reduce to the much cheaper 3-center and 2-center ones in RI. The key reason
for the success of RI lies in the fact that the set of all 
possible pair products $\{\varphi_i(\bfr)\cdot\varphi_j(\bfr)\}$, as a set of
basis functions in three-dimensional function space, is heavily
linearly dependent. Their number scales quadratically with system
size, while a non-redundant basis set that expands the same
three-dimensional space should scale linearly with system size. For
example, the non-interacting response function $\chi^0$ in
(\ref{Eq:chi^0_realspace_imag}), as well as the screened Coulomb
interaction $W$ in (\ref{Eq:screened_coulomb_realspace}), is
written in terms of orbital pair products, and hence can be
represented in terms of the ABFs. As will be shown below, this
naturally leads to a RI implementation for RPA and $GW$.

Next we will present RI formulations for all pertinent methods in this paper 
before presenting our specific choice for the ABFs subsequently.

\subsection{\label{sec:2ecoul}Metric and variational principle in RI}
   For a given set of ABFs $\{P_\mu(\bfr)\}$, the way to determine the expansion coefficients 
$C_{ij}^{\mu}$ is not unique. Different variational procedures give rise to different versions of RI and 
different working equations for computing the $C_{ij}^{\mu}$ 
\cite{Boys/Shavitt:1959,Whitten:1973,Dunlap/Connolly/Sabin:1979,Mittmire/Sabin/Trickey:1982,Vahtras/Almlof/Feyereisen:1993,Feyereisen/Fitzgerald/Komornicki:1993,Weigend/Haser/Patzelt/Ahlrichs:1998,Weigend:2002}. 

The expansion error of basis products in terms of the ABFs
  [equation (\ref{Eq:RI_expansion})] is
  \begin{equation}
   \delta \rho_{ij}(\bfr)= \tilde{\rho}_{ij}(\bfr)-\rho_{ij}(\bfr)
    = \sum_{\mu} C_{ij}^{\mu} P_{\mu}(\bfr) -\varphi_i({\bfr})\varphi_j({\bfr}) \, .
   \label{Eq:RI_expansion_error}
  \end{equation}
One choice for the construction of the expansion coefficients $C_{ij}^{\mu}$ is
to minimize the residual $\delta \rho_{ij}(\bfr)$. A simple
least-square fit amounts to minimizing the norm of the residual,
  $\int |\delta \rho_{ij}(\bfr)|^2 d\bfr$ and  yields
  \begin{equation}
    C_{ij}^{\mu}  =  \sum_\nu \langle ij|\nu\rangle S_{\nu\mu}^{-1}\, , 
    \label{Eq:SVS_coefficient} 
  \end{equation}
    where 
  \begin{equation}
   \langle ij|\mu\rangle  =  \int \varphi_i(\bfr)\varphi_j(\bfr)P_\mu(\bfr) d\bfr\, , 
   \label{Eq:3center_overlap}
  \end{equation}
    and
  \begin{equation}
   S_{\mu\nu}  =  \int P_\mu(\bfr) P_\nu(\bfr) d\bfr \, . 
   \label{Eq:ABF_overlap}
 \end{equation}
Combining (\ref{Eq:RI_expansion_general}) and
(\ref{Eq:SVS_coefficient}), one arrives at
the following approximation to the four-center Coulomb integrals,
 \begin{equation}
   (ij|kl)  = \sum_{\mu\nu\mu^\prime\nu^\prime}
     \langle ij|\mu\rangle S_{\mu\nu}^{-1} V_{\nu\nu^\prime} 
      S_{\nu^\prime\mu^\prime}^{-1} \langle k l |\mu^\prime \rangle \, .
   \label{Eq:RI_SVS}
 \end{equation}
In the literature equation (\ref{Eq:RI_SVS}) is therefore referred to as the
``SVS" version \cite{Vahtras/Almlof/Feyereisen:1993} of RI (``RI-SVS"
   in the following) because of the appearance of the inverse $S$ before and after
  the $V$ matrix in (\ref{Eq:RI_SVS}). 

A better criterion for obtaining $C_{ij}^{\mu}$ is to directly minimize the RI error
of the 4-center integrals themselves,  
 \begin{equation}
  \delta I_{ij,k l} =  
  (\tilde{\rho}_{ij}|\tilde{\rho}_{k l})-(\rho_{ij}|\rho_{k l}) \, .
 \end{equation}
As shown by Whitten \cite{Whitten:1973} $\delta I_{ij,k l}$  has an upper bound:
 \begin{equation}
  \delta I_{ij,kl} <  (\delta \rho_{ij}|\delta \rho_{ij} )^{1/2}     
   (\delta \rho_{k l} | \delta \rho_{k l})^{1/2}  
    + (\tilde{\rho}_{k l}|\delta \rho_{k l})^{1/2}
   (\delta \rho_{ij}|\delta \rho_{ij} )^{1/2} 
    +  (\tilde{\rho}_{ij}|\delta \rho_{ij})^{1/2} 
   (\delta \rho_{k l} | \delta \rho_{k l})^{1/2} \, .
 \end{equation}
The minimization of $\delta I_{ij,k l}$ can thus be achieved by 
independently minimizing $\delta U_{ij}=(\delta \rho_{ij}|\delta \rho_{ij} )$
and $\delta U_{k l}=(\delta \rho_{k l} | \delta \rho_{k l})$--
the self-repulsion of the basis pair density residuals. Minimizing $\delta U_{ij}$
with respect to $C_{ij}^{\mu}$ leads to \cite{Whitten:1973,Dunlap/Connolly/Sabin:1979,Mittmire/Sabin/Trickey:1982,Vahtras/Almlof/Feyereisen:1993}
  \begin{equation}
      C_{ij}^{\mu} = \sum_{\nu}(ij|\nu)V_{\nu\mu}^{-1} \, ,
      \label{Eq:V_coefficient}
  \end{equation}
where 
  \begin{equation}
     (ij|\nu)=\iint \frac{\phi_i(\bfr)\phi_j(\bfr)P_\nu(\bfrp)}{|\bfr-\bfrp|} d\bfr d\bfrp \, ,
  \end{equation}
and the Coulomb matrix $V$ is defined in (\ref{Eq:coulomb_matrix}).
Combining (\ref{Eq:RI_expansion_general}) and (\ref{Eq:V_coefficient})
one obtains the following decomposition of the 4-center ERIs
   \begin{equation}
     (ij|k l) \approx \sum_{\mu\nu}
     (ij|\mu)V^{-1}_{\mu\nu}(\nu|k l)\, .
     \label{Eq:RI_V}
   \end{equation}
Equation (\ref{Eq:RI_V}) is based on the global Coulomb metric and corresponds to the
``RI-V" method \cite{Vahtras/Almlof/Feyereisen:1993} mentioned earlier. 
``RI-V" has long been known to be superior to ``RI-SVS" in
the quantum chemistry
community \cite{Whitten:1973,Dunlap/Connolly/Sabin:1979,Mittmire/Sabin/Trickey:1982,Vahtras/Almlof/Feyereisen:1993}. 
This can be most easily seen by inspecting the error introduced in the 
self-Coulomb-repulsion of the NAO pairs,
  \begin{eqnarray}
  \delta I_{ij,ij} = (\tilde{\rho}_{ij}|\tilde{\rho}_{ij})- (\rho_{ij}|\rho_{ij})
       = 2(\delta\rho_{ij}|\tilde{\rho}_{ij})-(\delta\rho_{ij}|\delta{\rho}_{ij}) \, .
     \label{Eq:RI_coul_error}
 \end{eqnarray}
In the RI-V approximation, the first term in the above equation vanishes, 
and the non-zero contribution comes only from the second order of
$\delta\rho_{ij}$. This can be readily verified as follows,
  \begin{eqnarray} 
   (\delta\rho_{ij}|\tilde{\rho}_{ij}) & = & (\tilde{\rho}_{ij}|\tilde{\rho}_{ij}) -
             (\rho_{ij}|\tilde{\rho}_{ij}) 
       = \sum_{\mu\nu} C_{ij}^{\mu}V_{\mu\nu}C_{ij}^{\nu} -\sum_{\nu} (ij|\nu)C_{ij}^{\nu} \nonumber \\
       &=& \sum_{\nu} (ij|\nu)C_{ij}^{\nu} - \sum_{\nu} (ij|\nu)C_{ij}^{\nu} = 0 \nonumber 
  \end{eqnarray}
where (\ref{Eq:RI_expansion}) and particularly (\ref{Eq:V_coefficient}) have been used.
In contrast, in RI-SVS the term linear in $\delta \rho$ is non-zero and represents the dominating contribution 
to the total error.  

Our preferred flavor of RI is therefore RI-V, based on the Coulomb
metric \cite{Whitten:1973,Dunlap/Connolly/Sabin:1979,Mittmire/Sabin/Trickey:1982,Vahtras/Almlof/Feyereisen:1993,Feyereisen/Fitzgerald/Komornicki:1993,Weigend/Haser/Patzelt/Ahlrichs:1998,Weigend:2002},
on which all working equations for HF and other approaches presented further down
in this section  are based. Before proceeding we reiterate that RI-V continues to be
 the \emph{de facto} standard in quantum chemical 
calculations, due to its well-established accuracy and
reliability \cite{Weigend/Haser/Patzelt/Ahlrichs:1998,Weigend:2002}. 

That said, the long-range nature of the Coulomb interaction does
present a bottleneck for implementations that scale better than the
textbook standard (e.g., better than $O(N^4)$ for Hartree-Fock). In
order to avoid delocalizing each localized two-center basis function
product entirely across the system through $C_{ij}^{\mu}$, more
localized approaches would be desirable. Research into better-scaling
RI expansions that retain at least most of the accuracy of the Coulomb
metric is thus an active field, for example by Cholesky decomposition
techniques or an explicitly local treatment of the expansion of each
product (see, e.g., \cite{Werner/Manby/Knowles:2003,Jung/Shao/Head-Gordon:2006, 
Sodt/etal:2006,Sodt/Head-Gordon:2008,Aquilante/etal:2009,Neese/etal:2009,Foerster:2008,Foerster/Koval:2009}
for details). In fact, a promising Coulomb-metric based, yet localized, variant
of RI has been implemented in FHI-aims. In this approach products of
orbital basis functions are only expanded into auxiliary basis
functions centered at the two atoms at which the orbital
basis functions are centered, but the appropriate RI sub-matrices are
still treated by the Coulomb metric \cite{Billingsley/Bloor:1971}. As expected,
the error cancellation in this approach is not as good as that in full
RI-V, but---for Hartree-Fock and hybrid functionals---certainly more than an
order of magnitude better than in RI-SVS, creating a competitive
alternative for cases where RI-V is prohibitive. More details would go beyond the scope of this paper and will be presented in a forthcoming publication \cite{Wieferink/etal:inpreparation}.


\subsection{HF and hybrid functionals}

The key quantity for HF and hybrid functionals is the exact-exchange
matrix -- the representation  of the non-local exact-exchange potential
[equation (\ref{Eq:exact_exchange})] in terms of basis functions as given
in \ref{Eq:exchange_matr}). Its RI-V expansion follows by inserting
(\ref{Eq:RI_V}) into (\ref{Eq:exchange_matr}):
\begin{equation}
\Sigma^\text{x}_{ij,\sigma} = \sum_{k l}\sum_{\mu\nu}
(ik|\mu)V_{\mu\nu}^{-1}(\nu|jl)
D_{kl} = \sum_{\mu} \sum_{k l} M_{ik}^{\mu}M_{jl}^{\mu}D_{k l}
\label{Eq:exchange_RI_expan}
\end{equation}
where 
\begin{equation}
M_{ik}^{\mu} = \sum_{\nu} (ik|\nu)V_{\nu\mu}^{-1/2}
 =\sum_{\nu} C_{ik}^{\nu} V_{\nu\mu}^{1/2} \, .
\label{Eq:ovlp_trans}
\end{equation}
$\Sigma^\text{x}_{ij,\sigma}$ must thus be recomputed for each
iteration within the self-consistent field (scf) loop. The required
floating point operations scale as $N_\text{b}^3\cdot
{N_\text{aux}}$. The transformation matrix $M_{ik}^{\mu}$,
on the other hand, is constructed only once (prior to the scf loop), requiring
$N_\text{b}^2\cdot N_\text{aux}^2$ operations.

The numerical efficiency can be further improved by inserting the expression for the density matrix 
(\ref{Eq:dens_mat}) into (\ref{Eq:exchange_RI_expan}):
\begin{equation}
\Sigma^\text{x}_{ij,\sigma}=\sum_{n}^\text{occ}\sum_{\mu}\left(\sum_{k}M_{ik}^{\mu}
c_{n \sigma}^k\right)\left(\sum_{l}
M_{jl}^{\mu}c_{n \sigma}^l\right) = \sum_{n}^\text{occ}\sum_{\mu}B_{in\sigma}^{\mu}B_{jn\sigma }^{\mu} \, .
\label{Eq:exchange_RI_expan_2}
\end{equation}
The formal scaling in (\ref{Eq:exchange_RI_expan_2}) is now
$N_{\text{ occ}}\cdot {N_\text{b}}^2 \cdot N_\text{aux}$, with $N_\text{occ}$
being the number of occupied orbitals, and thus improved by a factor 
$N_\text{b}/N_\text{occ}$ (typically 5 to 10).  Once the exact-exchange matrix is obtained, the exact-exchange energy
follows through
\begin{equation}
  E_\text{x}^\text{HF} = -\frac{1}{2} \sum_{ij,\sigma} \Sigma^\text{x}_{ij,\sigma} D_{ij,\sigma} \, .
\end{equation}

For a variety of physical problems, combinding HF exchange with
semi-local exchange and correlation of the GGA type gives much better results than
with pure HF or pure GGAs \cite{Becke:1993}. Various flavors of these 
so-called hybrid functionals exist in the literature. The simplest
one-parameter functionals are of the following form
 \begin{equation}
 E^\text{hyb}_\text{xc}=E^\text{GGA}_\text{xc} + \alpha (E^\text{HF}_x -
  E^\text{GGA}_\text{x}) \, .
 \end{equation}
In the PBE0 hybrid functional \cite{Perdew/Ernzerhof/Burke:1996}, 
the GGA  is taken to be PBE, and the mixing parameter $\alpha$ is set
to $1/4$. Naturally, the computational cost of hybrid functionals is dominated by the 
HF exchange. Once HF exchange is implemented, it is straightforward to also perform 
hybrid functional calculations.

\subsection{MP2 (total-energy correction and self-energy)}

To compute the MP2 correlation energy in (\ref{Eq:mp2_c_energy})
and the MP2 self-energy in (\ref{Eq:2nd_self-energy}) using the RI technique,
the MO-based 4-orbital 2-electron Coulomb integrals are decomposed as follows
\begin{equation}
 (ma,\sigma|nb,\sigma^\prime)=\sum_{\mu}O^{\mu}_{ma,\sigma}O^{\mu}_{nb,\sigma^\prime} \, .
\label{Eq:mo_eri_expan}
\end{equation}
The 3-orbital integrals can be evaluated by
\begin{equation}
 O^{\mu}_{ma,\sigma}=\sum_{ij}M^{\mu}_{ij}c_{m\sigma}^{i\ast}c_{a\sigma}^j \, .
\label{Eq:MO_3index_trans}
\end{equation}
following (\ref{Eq:AO2MO_integral}), (\ref{Eq:RI_V}), and (\ref{Eq:ovlp_trans}).
Plugging (\ref{Eq:mo_eri_expan}) into (\ref{Eq:mp2_c_energy}), one obtains the RI-V version
of the MP2 correlation energy
\begin{eqnarray}
  E^{(2)}_0 & = &
  \frac{1}{2} \sum_{mn}^\text{occ}\sum_{ab}^\text{unocc}\sum_{\sigma,\sigmap}
     \left(\sum_{\mu}O^{\mu}_{ma,\sigma}O^{\mu}_{nb,\sigma^\prime}\right)  \times \nonumber  \\
  &&   \left[ \frac{ \left(\sum_{\mu}O^{\mu}_{am,\sigma}O^{\mu}_{bn,\sigma^\prime}\right)-
    \left(\sum_{\mu}O^{\mu}_{bm,\sigma}O^{\mu}_{an,\sigma}\right)\delta_{\sigma\sigmap}}
   {\epsilon_{m\sigma} + \epsilon_{n\sigmap} -\epsilon_{a\sigma}-\epsilon_{b\sigmap}} \right] 
   \, . \nonumber \\
  \label{Eq:RI_mp2_c_energy}
 \end{eqnarray}
In practice, one first transforms the atomic orbital-based integrals $M^{\mu}_{ij}$ to MO-based ones
$O^{\mu}_{ma,\sigma}$. The transformation scales formally as
  $N_\textnormal{occ}\cdot N_\textnormal{b}^2\cdot N_\text{aux}$ and the summation in (\ref{Eq:RI_mp2_c_energy}) as  $N_\text{occ}^2\cdot (N_\text{b}-N_\text{occ})^2 \cdot  N_\text{aux}$. Like in Hartree-Fock, the scaling exponent  ${\cal O}(N^5)$ is therefore not reduced by RI-V. However, the \emph{prefactor} in  RI-MP2 is  one to two orders of
  magnitude smaller than in full MP2.  

The computation of the MP2 self-energy at each frequency point proceeds analogously to that of the correlation energy.
In our implementaton, we first calculate the MP2 self-energy on the imaginary frequency axis,
 $\Sigma^\text{(2)}_{mn,i\sigma}$, and then 
continue analytically to the real frequency axis using either a ``two-pole'' model \cite{Rojas/Godby/Needs:1995},
or the Pad{\'e} approximation. Both approaches have been implemented in FHI-aims and can used to cross-check each 
other to guarantee the reliability of the final results.

\subsection{RPA and \textit{GW} }

To derive the working equations for the RPA correlation energy and the
$GW$ self-energy in the RI approximation, it is illuminating to first consider the RI-decomposition of
Tr$\left[\chi^0(i\omega)v\right]$. Combining 
(\ref{Eq:chi^0_realspace_imag}) and (\ref{Eq:mo_eri_expan}) we obtain
\begin{equation}
  \text{Tr}\left[\chi^0(i\omega)v\right]=
   \sum_{\mu} \sum_\sigma \sum_m^\text{occ}\sum_a^\text{unocc} \frac{O^\mu_{ma,\sigma}O^\mu_{am,\sigma}}
         {i\omega-\epsilon_{a\sigma}+\epsilon_{m\sigma}} + \text{c.c.} \, ,
\end{equation}
where $O^\mu_{ma,\sigma}$ is given by (\ref{Eq:MO_3index_trans}). 
Next we introduce an auxiliary quantity $\Pi(i\omega)$:
 \begin{equation}
    \Pi(i\omega)_{\mu\nu} = \sum_\sigma \sum_m^\text{occ}\sum_a^\text{unocc} \frac{O^\mu_{ma,\sigma}O^\nu_{am,\sigma}}
         {i\omega-\epsilon_{a\sigma}+\epsilon_{m\sigma}} + \text{c.c.} \, ,
   \label{Eq:Pi_matrix}
 \end{equation}
which allows us to write
 \begin{equation}
  \text{Tr}\left[\chi^0(i\omega)v\right]=\text{Tr}\left[v^{1/2}\chi^0(i\omega)v^{1/2}\right] = 
  \text{Tr}\left[\Pi(i\omega)\right] \, .
 \end{equation}
Thus the matrix $\Pi$ can be regarded as the matrix representation of the composite quantity 
$v^{1/2}\chi^0(i\omega)v^{1/2}$ using the ABFs. It is then easy to see that the RPA correlation
energy (\ref{Eq:cRPA}) can be computed as
 \begin{eqnarray}
   E_\text{c}^\text{RPA} &= &
  \frac{1}{2\pi}\int_0^\infty d\omega \text{Tr} 
    \left[\text{ln}\left(1-\Pi(i\omega)\right) + \Pi(i\omega)\right] \nonumber \\
    &=& \frac{1}{2\pi} \int_0^\infty d\omega \left\{ \ln\left[\text{det}(1-\Pi(i\omega))\right] 
          + \text{Tr}\left[ \Pi(i\omega) \right] \right\} \, . \nonumber \\
  \label{Eq:E_c_rpa_ri}
 \end{eqnarray}
 using the general property Tr$\left[ \ln
 (A)\right]=\ln\left[\text{det}(A)\right]$ for any matrix $A$. This is very convenient
 since all matrix operations in \ref{Eq:E_c_rpa_ri} occur within the compressed
 space of ABFs and the computational effort is therefore significantly reduced.

In practice, we first construct the auxiliary quantity $\Pi(i\omega)$  [equation (\ref{Eq:Pi_matrix})] on a suitable  imaginary frequency grid where we
 use a modified Gauss-Legendre grid (see appendix~\ref{sec:appC} for
 further details) with typically 20-40 frequency points. For
 fixed frequency grid size, the number of required operations is proportional to
 $N_\text{occ}\cdot N_\text{unocc} \cdot N_\text{aux}^2$
 ($N_\text{unocc}$ is the number of unoccupied orbitals using the full spectrum of our Hamiltonian matrix). 
 The next step is to compute the determinant of the matrix 
 $1 -\Pi(i\omega)$ as well as the trace of $\Pi(i\omega)$.
 What remains is a simple integration over the imaginary
 frequency axis. Thus our RI-RPA implementation is dominated by
 the step in (\ref{Eq:Pi_matrix}) that has a formal ${\cal
 O}(N^4)$ scaling. An ${\cal O}(N^4)$-scaling  algorithm of RI-RPA was
 recently derived from a different
 perspective \cite{Eshuis/Yarkony/Furche:2010}, based on the plasmonic formulation 
 of RPA correlation energy \cite{Furche:2008} and a transformation analogous to the 
 Casimir-Polder integral \cite{Casimir/Polder:1948}.
An even better scaling can be achieved  by taking advantage of the sparsity of the
 matrices involved \cite{Foerster/Koval:2009}.
  
Finally we come to the RI-V formalism for $GW$. To make (\ref{Eq:4index_screened_coulomb}) tractable, we expand the
screened Coulomb interaction $W(i\omega)$  in terms of the ABFs. Using
(\ref{Eq:symbolic_W}) and  
$\Pi(i\omega)=v^{1/2} \chi_0(i\omega) v^{1/2}$, we obtain
 \begin{equation}
   W_{\mu\nu}(i\omega) = \iint d\bfr d\bfrp P_\mu^\ast(\bfr)W(\bfr,\bfrp,i\omega)P_\nu(\bfr) \nonumber \\
          = \sum_{\mu^\prime \nu^\prime} V^{1/2}_{\mu\mu^\prime} [1-\Pi(i\omega)]^{-1}_{\mu^\prime \nu^\prime}
          V^{1/2}_{\nu^\prime\nu} \, .
 \end{equation}
To apply the RI-decomposition to (\ref{Eq:4index_screened_coulomb}), we expand 
  \begin{equation}
    \psi_{n\sigma}^\ast(\bfr) \psi_{m\sigma}(\bfr)=\sum_{ij}\sum_{\mu} P_{\mu}(\bfr)
            C_{ij}^{\mu}c_{n\sigma}^{i\ast} c_{m\sigma}^j \, ,
   \label{Eq:psi_pair_expan}
  \end{equation}
where (\ref{Eq:AO_to_MO}) and (\ref{Eq:RI_expansion}) are used. Combing (\ref{Eq:4index_screened_coulomb}),
(\ref{Eq:ovlp_trans}), (\ref{Eq:MO_3index_trans}), and \ref{Eq:psi_pair_expan}) gives
 \begin{equation}
  (nm,\sigma|W^0(i\omega)|mn,\sigma)=  
  \sum_\sigma \sum_{\mu\nu} O_{nm,\sigma}^\mu \left[1-\Pi(i\omega)\right]^{-1}_{\mu\nu} O_{mn,\sigma}^\nu \, .
  \label{Eq:W_RI}
 \end{equation}
Inserting (\ref{Eq:W_RI}) into (\ref{Eq:G0W0_imag_freq}), one finally arrives at the RI-version of 
the $G^0W^0$ self-energy 
 \begin{eqnarray}
  \Sigma_{n\sigma}^{G^0W^0}(i\omega) &= & -\frac{1}{2\pi} \sum_m  \int_{-\infty}^{\infty} d\omegap
          \frac{1}{i\omega+i\omegap+\epsilon_\text{F}-\epsilon_{m\sigma}} \times \nonumber \\
  &&  \sum_{\mu\nu}  O_{nm,\sigma}^\mu \left[1-\Pi(i\omega)\right]^{-1}_{\mu\nu} O_{mn,\sigma}^\nu \, .
  \label{Eq:G0W0_matrix_elem}
 \end{eqnarray}
As stated above for the MP2 self-energy, the expression is analytically continued to the real-frequency
axis before the quasiparticle energies are computed by means of (\ref{Eq:qpe}).

 \section{\label{sec:our_RI}Atom-centered auxiliary basis for all-electron NAO calculations}
   \subsection{\label{sec:basis}Orbital basis set definitions}
   For the practical implementation, all the aforementioned objects (wave
functions, effective single-particle orbitals, Green function,
response function, screened Coulomb interaction etc.) are expanded 
either in a single-particle basis set or an auxiliary basis set. We first 
summarize the nomenclature used for our NAO basis sets \cite{Blum/etal:2009}
before defining a suitable auxiliary basis prescription for RI in the
next subsections. 

NAO basis sets
$\{\varphi_{i}(\bfr)\}$ to expand the single-particle spin orbitals
$\psi_{n\sigma}(\bfr)$ [equation (\ref{Eq:AO_to_MO})] are of the general form  
  \begin{equation}
    \varphi_i(\bfr) = \frac{u_{s(a)
    l \kappa}(r)}{r}Y_{lm}(\hat{\bfr}_a) .
    \label{Eq:atomic_basis_function}
  \end{equation}
$u_{s(a) l \kappa}$ is a radial function centered at atom $a$,
and $Y_{lm}(\hat{\bfr}_a)$ is a spherical harmonic. The index
$s(a)$ denotes the element species $s$ for an atom $a$, and $\kappa$
enumerates the different radial functions for a given  
species $s$ and an angular momentum $l$. The unit vector
$\hat{\bfr}_a=(\bfr-\bfR_a)/|\bfr-\bfR_a|$ refers to the
position $\bfR_a$ of atom $a$. The basis index $i$ thus combines
$a$, $\kappa$, $l$, and $m$.  

For numerical convenience, and without losing generality, we use
real-valued basis functions, meaning that the $Y_{lm}(\Omega)$ denote
the real (for $m=0,\cdots,l$) and the imaginary part (for
$m=-l,\cdots,-1$) of complex spherical harmonics. For NAOs, $u_{s(a)
l \kappa}(r)$ need not adhere to 
any particular analytic shape, but are tabulated functions (in
practice, tabulated on a dense logarithmic grid and evaluated
in between by cubic splines). Of course, Gaussian, Slater-type, or even
muffin-tin orbital basis sets are all special cases of the generic
shape (\ref{Eq:atomic_basis_function}). All algorithms in this
paper could be used for them. In fact, we employ the Dunning GTO 
basis sets (see \cite{Wilson/vanMourik/Dunning:1996,Schuchardt/etal:2007} and references therein) for
comparison throughout this work.   

Our own implementation, FHI-aims, \cite{Blum/etal:2009} provides hierarchical sets of
all-electron NAO basis functions. The hierarchy starts from
the \textit{minimal} basis composed of the radial functions for all
core and valence electrons of the free atoms. Additional groups of 
basis functions, which we call \emph{tiers} (quality levels) can be added for
increasing accuracy (for brevity, the
notation is \textit{minimal}, \textit{tier} 1, \textit{tier} 2,
etc.). Each higher level includes  
the lower level. In practice, this hierarchy
defines a recipe for systematic, variational convergence down to
meV/atom accuracy for total energies in LDA and GGA calculations. The
\textit{minimal} basis (atomic core and valence radial functions) is
different for different functionals, but one could as well use,
e.g., LDA \emph{minimal} basis functions for calculations using
other functionals in cases their minimal basis functions are not available. 
We discuss this possibility for HF below. 

To give one specific example, consider the nitrogen atom (this case and more are spelled out in
detail in Table 1 of Ref. \cite{Blum/etal:2009}). There are 5 \textit{minimal} basis functions, of 1$s$, 2$s$, and
2$p$ orbital character, respectively. In a shorthand notation, we
denote the number of radial functions for given angular momenta
as (2$s$1$p$) (two $s$-type radial functions, one $p$-type
radial function). At the \textit{tier} 1 basis level, one $s$, $p$,
and $d$ function  is added to give a total of 14 basis functions 
($3s2p1d$). There are 39 basis functions ($4s3p2d1f1g$) in
\textit{tier} 2, 55 ($5s4p3d2f1g$) in \textit{tier} 3, and 80
($6s5p4d3f2g$) in \textit{tier} 4.

\subsection{\label{sec:auxil_basis}Construction of the auxiliary basis }
   In the past, different communities have adopted different strategies
for building auxiliary basis sets.
For the GTO-based RI-MP2 method \cite{Weigend/Haser/Patzelt/Ahlrichs:1998}, 
which is widely used in the quantum chemistry community,
a variational procedure has been used to generate optimal gaussian-type 
atom-centered ABF sets. In the condensed matter community a so-called ``product basis" has been employed in the context of all-electron $GW$ implementations based on the linearized muffin-tin orbital (LMTO) and/or augmented plane-wave (LAPW) method \cite{Andersen:1975} to represent the response function and the Coulomb potential within the muffin-tin 
spheres \cite{Aryasetiawan/Gunnarsson:1994a,Gomez-Abal/etal:2010,Friedrich/etal:2009}.
Finally, it is even possible to generate ABFs only \emph{implicitly},
by identifying the ``dominant directions" in the orbital product
space through singular value decomposition (SVD) \cite{Foerster:2008,Foerster/Koval:2009,Foerster/etal:2011}. 
As will be illustrated below, our procedure to construct the ABFs combines features 
from both communities. Formally it is similarly to the ``product basis" construction in the $GW$ community, 
but instead of the simple overlap metric the Coulomb metric is used to remove the
linear dependence of the ``products" of the single-particle basis functions, and to build all
the matrices required for the electronic structure schemes in this paper.

Our procedure employs numeric atom-centered ABFs whereby the infrastructure that is
already available to treat the NAO orbital basis sets can be utilized in many respects. 
Specifically the ABFs are chosen as
    \begin{equation}
       P_{\mu}(\bfr) = \frac{\xi_{s(a) l \kappa }(r)}{r}Y_{lm}(\hat{\bfr}_a)   
       \label{Eq:auxiliary_function}
    \end{equation}
just like for the one-particle NAO basis functions in
(\ref{Eq:atomic_basis_function}), but of course with different radial functions. 
To distinguish the auxiliary basis functions from the NAO basis functions 
we denote the radial functions of the ABFs as  $\xi_{s(a) l \kappa }$. 

The auxiliary basis should primarily expand products of basis functions centered
on the same atom \emph{exactly}, but at the same time be
sufficiently flexible to expand all other two-center basis function products
with a negligible error. 
In contrast to the ABFs used in the GTO-based RI-MP2 method 
\cite{Weigend/Haser/Patzelt/Ahlrichs:1998},
in our case the construction of auxiliary basis functions follows
from the definition of the orbital basis set. At each level of NAO
basis, one can generate a corresponding ABF set, denoted as
\textit{aux\_min}, \textit{aux\_tier} 1, \textit{aux\_tier} 2, etc.
We achieve this
objective as follows:
\begin{enumerate}
  \item
     For each atomic species (element) $s$, and for each $l$ below a 
     limit $l^\text{max}_s$, we form all possible ``on-site'' pair
     products of atomic radial functions 
     $\{\tilde{\xi}_{s l\kappa }(r)=u_{s k_1l_1}(r)u_{s k_2l_2}(r)\}$. The
     allowed values of $l$ are given by the possible multiples of
     the spherical harmonics associated with the orbital basis
     functions corresponding to $u_{s k_1l_1}$ and $u_{s k_2l_2}$,
     i.e., $|l_1-l_2| \le l \le |l_1+l_2|$.  
  \item 
    Even for relatively small orbital basis sets, the number of resulting
    auxiliary radial functions $\{\tilde{\xi}_{sl \kappa }(r)\}$ is
     large. They are non-orthogonal and heavily linear dependent. We
     can 
    thus use a Gram-Schmidt like procedure (separately for each $s$
    and $l$) to keep only radial function components
     $\xi_{sl\kappa}(r)$ that are not 
    essentially represented by others, with a threshold for the
    remaining norm $\varepsilon^\text{orth}$, below which a given radial
    function can be filtered out. In doing so the Coulomb metric is used
    in the orthogonalization procedure. The result is a much smaller set of
    linearly independent, orthonormalized radial functions
     $\{\xi_{sl\kappa}(r)\}$ that expand the required function space.
  \item The radial functions $\{\xi_{s(a),l\kappa}\}$ are multiplied
    with the spherical harmonics $Y_{lm}(\hat{\bfr}_a)$ as in (\ref{Eq:auxiliary_function}).
  \item The resulting $\{P_\mu(\bfr)\}$ are orthonormal if they
     are centered on the same atom, but not if situated on different
     atoms. Since we use large ABF sets, linear dependencies could
     also arise between different atomic centers, allowing us to
     further reduce the ABF space through 
     SVD of the applied metric ($S$ in the case of RI-SVS,
     $V$ in the case of RI-V). For the molecule-wide SVD we use a
     second threshold $\varepsilon^\text{svd}$, which is not the same as
     the on-site Gram-Schmidt threshold $\varepsilon^\text{orth}$.  
\end{enumerate}
For a given set of NAOs, the number of the corresponding ABFs depends on
the angular momentum limit $l^\text{max}_s$ in step 1 and the
Gram-Schmidt orthonormalization threshold
$\varepsilon^\text{orth}$, and to a small extent on
$\varepsilon^\text{svd}$. For RI-V, as documented in the literature 
\cite{Weigend:2002} and 
demonstrated later in this work (section~\ref{sec:N2_benchmark_GTO}), it is 
sufficient to keep $l^\text{max}_s$ just one higher than the highest angular momentum of
the one-electron NAOs. Usually $\varepsilon^\text{orth}=10^{-2}$ or $10^{-3}$
suffices for calculations of energy differences. Nevertheless 
both $l^\text{max}_s$ and $\varepsilon^\text{orth}$
can be treated as explicit convergence parameters if
needed. In practice, we keep $\varepsilon^\text{svd}$ as small as
possible, typically 10$^{-4}$ or 10$^{-5}$, only large enough to
guarantee the absence of numerical instabilities through an
ill-conditioned auxiliary basis. 
The resulting auxiliary basis size is typically 3-6 times that of the NAO
basis. This is still a considerable size and could be reduced by
introducing optimized ABF sets as is sometimes done for GTOs. On
the other hand, it is the size and quality of our auxiliary basis that
guarantees low expansion errors for RI-V, as we will show in our
benchmark calculations below. We therefore prefer to keep the safety
margins of our ABFs to minimize the  expansion errors, bearing in mind
that the regular orbital basis introduces expansion errors that are
always present.

 \subsection{\label{sec:numerical_integral}Numerical integral evaluation}
   With a prescription to construct ABFs at hand, we need to compute the
overlap integrals $C_{ij}^{\mu}$, defined in
(\ref{Eq:SVS_coefficient}) for RI-SVS or
in (\ref{Eq:V_coefficient}) for RI-V, respectively. We also need
their Coulomb matrix given by (\ref{Eq:coulomb_matrix}) in general, and
additionally the ``normal'' overlap matrix $S_{\mu\nu}$ given by
(\ref{Eq:ABF_overlap}) in RI-SVS. Having efficient algorithms for these
tasks is enormously important, but since many pieces of our eventual
implementation exist in the literature, we here only give a brief
summary and refer to separate appendices for details. 

Since our auxiliary basis set $\{P_\mu\}$ is atom-centered, we
obtain the Coulomb potential $Q_\mu(\bfr)$ of each $P_\mu(\bfr)$ by a
one-dimensional integration for a single multipole
(\ref{sec:Coulomb}). The required three-center integrals 
\begin{equation} 
  (ij|\mu) = \int \phi_i(\bfr)\phi_j(\bfr)Q_\mu(\bfr) d\bfr
 \label{Eq:3center_integral}
\end{equation}
are carried out by standard overlapping atom-centered grids as used in
many quantum-chemical codes for the exchange-correlation matrix
in DFT \cite{Becke:1988,Delley:1990,Blum/etal:2009,Havu/etal:2009}, see
\ref{sec:three-center}. The same strategy works for two-center
integrals 
\begin{equation} 
  (\mu|\nu) = \int P_\mu(\bfr)Q_\nu(\bfr) d\bfr.
 \label{Eq:2center_integral}
\end{equation}
As an alternative, we have also implemented two-center integrals following
the ideas developed by Talman \cite{Talman:2003,Talman:2007}, which are
described in \ref{sec:two-cent-integr} and \ref{sec:logsbt}.
In summary, we thus have accurate matrix elements at hand
that are used for the remainder of this work.

 \subsection{\label{sec:auxil_basis_acc}Accuracy of the auxiliary basis: expansion of a single product}
   In this section we examine the quality of our prescription for generating
the ABFs as described in section~\ref{sec:auxil_basis}. Our procedure guarantees that the
ABFs accurately represent the ``on-site" products of the NAO
basis pairs by construction, but it is not \textit{a priori} clear how 
the ``off-site" pairs are represented. The purpose of this section is
to demonstrate the quality of our ABFs to represent the ``off-site" pairs.

\begin{figure}
 \begin{center}
 \includegraphics[width=0.7\textwidth]{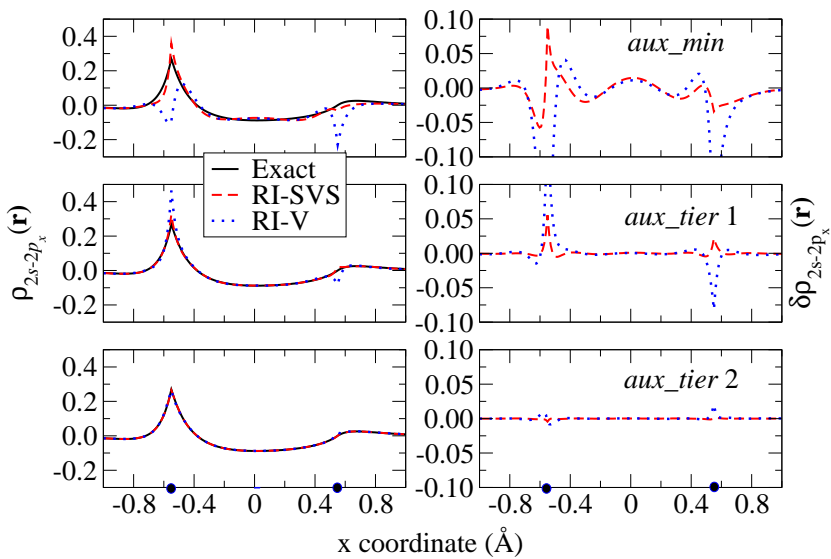}
 \end{center}
 \caption{(Color online) Left panels: the product of the atomic $2s$ and $2p_x$ functions centered 
          respectively on the two atoms in a N$_2$ molecule along the bonding direction, and its 
          approximate behaviours from the RI-SVS and RI-V expansions for three hierarchical levels of ABFs 
          (from \textit{aux\_min} to \textit{aux\_tier} 2). Right panels: The corresponding 
          deviations of the RI-SVS and RI-V expansions from the reference curve. The positions of 
          the atoms are marked by blue dots at the $x$-axis.
         }
 \label{Fig:N2_auxil_basis_accu} 
\end{figure}

In the left panels of figure \ref{Fig:N2_auxil_basis_accu} we plot
$\rho_{2s\text{-}2p_x} (\bfr)$ for a simple N$_2$
molecule at the equilibrium bonding distance ($d=1.1$ \AA) -- the product of
the atomic $2s$ function from the left atom and the atomic $2p_x$
function from the right atom. We compare this directly taken product
to its ABF expansions, both in RI-SVS [equation (\ref{Eq:SVS_coefficient})]
and in RI-V [equation (\ref{Eq:V_coefficient})]. The particular product
$\rho_{2s\text{-}2p_x} (\bfr)$ is part of the \emph{minimal} basis of
free-atom like valence radial functions. As we increase the orbital
basis set (adding \emph{tier} 1, \emph{tier} 2, etc.), the
exact product remains the same, whereas
its ABF expansion will successively improve, since the
auxiliary basis set is implicitly defined through the underlying
orbital basis. Three different levels of ABF sets are shown
(\textit{aux\_min}, \textit{aux\_tier} 1, \textit{aux\_tier} 2 
from the top to bottom panels). The onsite threshold
$\varepsilon^\text{orth}$ is set to 10$^{-2}$, and the global
SVD threshold $\varepsilon^\text{svd}$ is set
to 10$^{-4}$, yielding 28, 133, and 
355 ABFs, respectively.  In the right panels of
figure \ref{Fig:N2_auxil_basis_accu}, the corresponding 
$\delta \rho_{2s\text{-}2p_x} (\bfr)$ -- the deviations of the ABF
expansions from the reference curve -- are plotted. One can clearly see
two trends: First, the quality of the ABF expansion improves
as the number of ABFs increases. Second, at the same level of ABF, 
the absolute deviation of the RI-V expansion
is larger than the RI-SVS expansion. This is an expected behaviour
for the simple pair product: RI-V is designed to minimize the  
error of the Coulomb integral, see section~\ref{sec:2ecoul}. In either
method, the remaining expansion errors are centered around the
nucleus, leading to a relatively small error in overall energies (in
the 3-dimensional integrations, the integration weight $r^2dr$ is
small). 

Table \ref{Tab:RI_error} gives the errors of the Coulomb 
repulsion $\delta I_{ij,ij}$ of the $2s\text{-}2p_x$ NAO basis pair
under the RI approximation for the three ABF basis sets of
figure \ref{Fig:N2_auxil_basis_accu}. Table \ref{Tab:RI_error} also
includes the influence of 
the threshold parameters $\varepsilon^\text{orth}$ and
$\varepsilon^\text{svd}$, separate for RI-SVS (top half) and RI-V (bottom
half). The error diminishes quickly with increasing ABF basis size, 
and is 2-3 orders of magnitude smaller in
RI-V than in RI-SVS. By decreasing $\varepsilon_\text{orth}$, the number
of ABFs at each level increases, improving particularly the accuracies at
the \textit{aux\_min} and \textit{aux\_tier} 1 levels. The global
SVD threshold $\varepsilon^\text{svd}$ comes
into play only for the larger basis sets. In general, both
control parameters have a much bigger effect on RI-SVS than RI-V,
underscoring the desired variational properties of RI-V \cite{Boys/Shavitt:1959,Whitten:1973,Dunlap/Connolly/Sabin:1979,Mittmire/Sabin/Trickey:1982,Vahtras/Almlof/Feyereisen:1993}.

\begin{table}
  \caption{\label{Tab:RI_error}
    The errors $\delta I_{2s\text{-}2p_x,2s\text{-}2p_x}$ (in eV) introduced in the RI-SVS and RI-V 
       for calculating the self-repulsion of NAO pair products
       $(\rho_{2s\text{-}2p_x}|\rho_{2s\text{-}2p_x})$ for N$_2$ 
       ($d=1.1$ \AA) at three levels
       of ABF basis sets. The number of ABFs that survive the
       SVD is also shown.}
 \begin{indented}
 \item[]
  \begin{tabular}{@{}lccc}
   \br \\[-1.0ex]
   ABF sets  & \textit{aux\_min}  & \textit{aux\_tier} 1 & \textit{aux\_tier} 2 \\[0.5ex]
   \mr \\[-0.7ex]
   & \multicolumn{3}{c}{RI-SVS} \\ \cline{2-4} \\[0.2ex]
   & \multicolumn{3}{c}{$\varepsilon^\text{orth}=10^{-2}$,~~$\varepsilon^\text{svd}=10^{-4}$} \\[1.2ex]
  Error  & $-54\times 10^{-2}$   & $ 85\times 10^{-3}$  & $-47\times 10^{-4}$   \\
  \# of ABFs  & 28   &  133  & 355  \\ \cline{2-4} \\[0.2ex]
   & \multicolumn{3}{c}{$\varepsilon^\text{orth}=10^{-2}$,~~$\varepsilon^\text{svd}=10^{-5}$} \\[1.2ex]
  Error  & $-54\times 10^{-2}$   & $84\times 10^{-3}$  & $-24\times 10^{-5}$  \\
  \# of ABFs & 28   &  134  &  363 \\ \cline{2-4} \\[0.2ex]  
  & \multicolumn{3}{c}{$\varepsilon^\text{orth}=10^{-3}$,~~$\varepsilon^\text{svd}=10^{-5}$} \\[1.2ex]
  Error  & $-11 \times 10^{-2}$   & $-23\times 10^{-3}$  & $13 \times 10^{-3}$   \\
  \# of ABFs  & 36  & 151  & 417  \\[-0.2ex]
   \mr\\[-0.7ex]
   & \multicolumn{3}{c}{RI-V} \\ \cline{2-4} \\[0.2ex]
   & \multicolumn{3}{c}{$\varepsilon^\text{orth}=10^{-2}$,~~$\varepsilon^\text{svd}=10^{-4}$} \\[1.2ex]
   Error  &  $-68\times 10^{-3}$  & $-16\times 10^{-5}$  & $-10\times 10^{-7}$  \\
   \# of ABFs & 28 & 133 & 356 \\ \cline{2-4} \\[-0.7ex]
   & \multicolumn{3}{c}{$\varepsilon^\text{orth}=10^{-2}$,~~$\varepsilon^\text{svd}=10^{-5}$} \\[1.2ex]
  Error  & $-68 \times 10^{-3}$   & $-16 \times 10^{-5}$  & $-40 \times 10^{-7}$ \\ 
  \# of ABFs  & 28   & 133  & 359 \\ \cline{2-4} \\[-0.7ex]
  & \multicolumn{3}{c}{$\varepsilon^\text{orth}=10^{-3}$,~~$\varepsilon^\text{svd}=10^{-5}$} \\[1.2ex]
  Error  & $-32\times 10^{-4}$   & $-20 \times 10^{-6}$  & $-20\times 10^{-7}$   \\
  \# of ABFs  & 36  & 151  & 417 \\
   \br
  \end{tabular}
  \end{indented}
\end{table}

  \subsection{\label{sec:N2_benchmark_GTO}Accuracy of the auxiliary basis: energies and thresholds}
     Next we turn to the accuracy of our auxiliary basis 
prescription for actual HF and MP2 (total and binding energy)
calculations.
For this purpose, we employ all-electron GTO basis
sets, when possible, to be able to refer to completely independent and accurate 
implementations from quantum chemistry without invoking the RI approximation (referred to
as ``RI-free" in the following). Our specific choice here is the GTO-based
NWChem \cite{NWChem2010} code package, where ``RI-free" results can be obtained 
using traditional methods of quantum chemistry.
In the following we compare our RI-based HF and MP2 results with their
``RI-free" counterparts produced by NWChem, in order to benchmark the accuracy of the RI 
implementation in FHI-aims.  All results presented in this
section correspond to the cc-pVQZ basis set of
Dunning \textit{et
al.} \cite{Dunning:1989,Wilson/vanMourik/Dunning:1996,Schuchardt/etal:2007} The convergence
behaviour with respect to NAO / GTO single-particle basis set size will be
the topic of next section.

\begin{table}
\caption{\label{tab_N2_auxil_acc}
   The deviation of HF total ($\Delta E_\text{tot}$) and binding energies ($\Delta E_\text{b}$) 
   (in meV) from NWChem reference values for N$_2$ at bond length $d=$1.1 {\AA}. 
  RI-V and RI-SVS calculations are done using the Dunning
  cc-pVQZ basis set. \cite{Dunning:1989,Wilson/vanMourik/Dunning:1996} 
  The reference $E_\text{tot}$ and $E_\text{b}$ values (in eV) from 
  NWChem calculations are shown at the bottom. 
  The binding energies are BSSE-corrected using the counterpoise method.}
\begin{indented}
\item[]
\begin{tabular}{@{\extracolsep{\fill}}ccccc}
\br \\[-2.0ex]
 \multirow{3}{*}{$\varepsilon^\text{svd}$}& \multicolumn{2}{c}{$\Delta E_\text{tot}$} & \multicolumn{2}{c}{$\Delta E_\text{b}$} \\[0.5ex]
 & RI-V & RI-SVS & RI-V & RI-SVS \\[1.0ex] \cline{2-5} \\[-2.0ex]
 & \multicolumn{4}{c}{$\varepsilon^\text{orth}=10^{-2}$} \\[1.0ex]
$10^{-4}$ & -0.11 & 80.45 & -0.07 & -4.16 \\
$10^{-5}$ & -0.11 & 81.15 & -0.04 & -3.22 \\
$10^{-6}$ & -0.16 & 16.95 & -0.04 & -2.20 \\ [1.0ex] \cline{2-5} \\[-2.0ex]
 & \multicolumn{4}{c}{$\varepsilon^\text{orth}=10^{-3}$} \\[1.0ex]
$10^{-4}$ & ~0.14 & 67.82 & -0.10 & -0.18 \\
$10^{-5}$ & -0.16 & 81.28 & -0.03 & -1.87 \\
$10^{-6}$ & -0.16 & 72.42 & -0.03 & -0.49 \\ [1.0ex] \cline{2-5} \\[-2.0ex]
&  \multicolumn{2}{c}{$E_\text{tot}=-2965.78514$}  &  \multicolumn{2}{c}{$E_\text{b}=-4.98236$} \\
\br
\end{tabular}
\end{indented}
\end{table}

We first check the quality of the ABF prescription for light 
(first and second-row) elements. We again choose N$_2$ as a
first illustrating example.
Table \ref{tab_N2_auxil_acc} presents RI-HF total and binding
energy \emph{errors} for the N$_2$ molecule at bond length
$d$=1.1~{\AA}. The reference numbers given at the bottom of the table
are from ``RI-free" NWChem calculations. All other numbers were
obtained with FHI-aims and the ABF prescription of
section~\ref{sec:auxil_basis}. Total and binding energy errors
are given for several different choices of thresholding
parameters $\varepsilon^\text{orth}$ and $\varepsilon^\text{svd}$
(see section~\ref{sec:auxil_basis}).
All binding energy errors are
obtained after a counterpoise correction \cite{Boys/Bernardi:1970} to
remove any possible basis set superposition errors (BSSE) (see also
lowest panel of figure \ref{Fig_N2_HF_BE_cc-pVQZ} below). 

For N$_2$ at equilibrium bonding distance, the salient results can be summarized as follows. 
First, we see that RI-V with our ABF prescription implies  \emph{total energy} 
errors for Hartree-Fock of only $\sim$0.1-0.2 meV, while the corresponding RI-SVS errors are much
larger. Both methods can, however, be adjusted to yield sub-meV
binding energy errors, with those from RI-V being essentially
zero. This is consistent with our observations for the error in the
self-repulsion integrals in section~\ref{sec:auxil_basis_acc}. Since
RI-V performs much better than RI-SVS, we will only report RI-V
results for the remainder of this paper.

The excellent quality of our ABFs is not restricted to the
equilibrium region of N$_2$. In the top panel of
figure \ref{Fig_N2_HF_BE_cc-pVQZ} the restricted HF total energies are
plotted for a range of bonding distances. The RI-V numbers are in very
good agreement with the reference throughout. For greater clarity,
the total-energy deviation of the RI-V result from
the reference is plotted in the middle panel of
figure \ref{Fig_N2_HF_BE_cc-pVQZ}. One can see that the total-energy
errors are in general quite small, but the actual sign and magnitude
of the errors vary as a function of bond length.
The deviation is shown for two different choices of the
ABFs, $(\varepsilon^\text{orth}$=10$^{-2}$, $\varepsilon^\text{svd}$=10$^{-4})$
(standard accuracy) and $(\varepsilon^\text{orth}=10^{-3}$, $\varepsilon^\text{svd}$=10$^{-5})$
(somewhat tighter accuracy). It is evident that the tighter settings
produce a smoother total energy error at the sub-meV level, but,
strikingly, there is no meaningful difference for the
counterpoise corrected \emph{binding energy} of N$_2$
(bottom panel of figure \ref{Fig_N2_HF_BE_cc-pVQZ}). 

\begin{figure}
  \begin{center}
  \includegraphics[width=0.6\textwidth,clip]{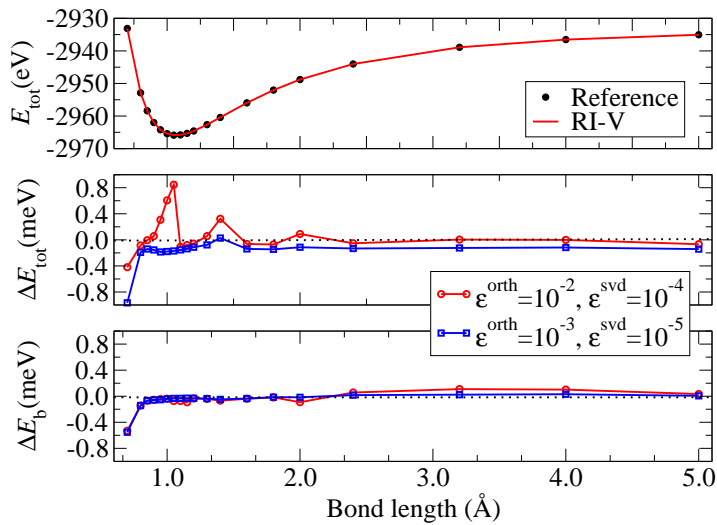}
  \end{center}
 \caption{(color online) Upper panel: RI-V HF total energies as a function of bond length
          for N$_2$, in comparison with NWChem reference values. 
          Middle panel: the deviation of the RI-V HF total energies from the reference values
          for two sets of thresholding parameters. Lower panel: the deviation of
          the BSSE-corrected RI-V HF binding energies from the reference values
          for the same sets of thresholding parameters.  The cc-pVQZ basis is used in
          all the calculations. Note that the dependence of the total energies
          on the thresholding parameters is not visible in the upper panel.}
\label{Fig_N2_HF_BE_cc-pVQZ} 
\end{figure}
\begin{figure}
  \begin{center}
  \includegraphics[width=0.6\textwidth,clip]{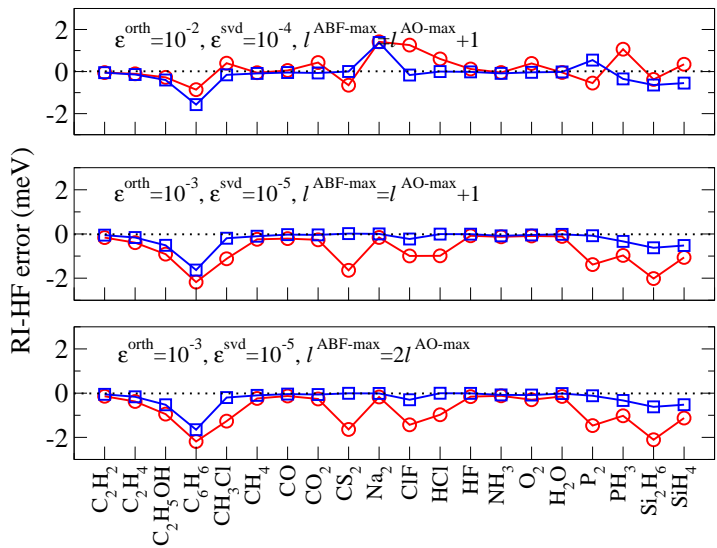}
  \end{center}
 \caption{(color online) Deviations of RI-V HF total energies (red circles) and atomization
          energies (blue squares) from the corresponding reference values for 20 small molecules. 
          The three panels illustrate the dependence of the RI errors on the truncation 
          parameters $\varepsilon^\text{orth}$, $\varepsilon^\text{svd}$, and the highest ABF angular 
          momentum $l^\text{ABF-max}$ ($l^\text{AO-max}$ denotes the highest angular momentum of
          the single-particle atomic orbitals).  Experimental equilibrium geometries and the 
          gaussian cc-pVQZ basis are used. }
\label{Fig_RI-HF_acc_cc-pVQZ} 
\end{figure}
\begin{figure}
  \begin{center}
  \includegraphics[width=0.6\textwidth,clip]{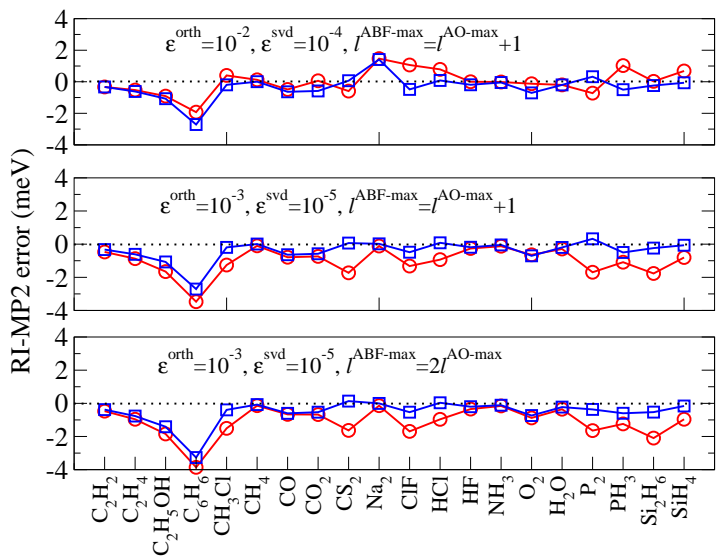}
  \end{center}
 \caption{(color online) Deviations of RI-V MP2 total energies and atomization
          energies from the corresponding reference values for 20 small molecules. 
          Nomenclature and labelling are the same as figure \ref{Fig_RI-HF_acc_cc-pVQZ}.}
\label{Fig_RI-MP2_acc_cc-pVQZ} 
\end{figure}
%
Next we demonstrate the quality of our ABFs for a set of molecules consisting
of first and second-row elements. In figure \ref{Fig_RI-HF_acc_cc-pVQZ} and
\ref{Fig_RI-MP2_acc_cc-pVQZ} the non-relativistic RI-V HF and MP2 total energy errors and 
atomization energy errors for 20
molecules are shown. In all cases the total energy 
error is below 1 $meV/atom$, demonstrating that the $meV$-accuracy in total energy can
routinely be achieved for the RI-V approximation with our ABFs for light
elements. In addition, it is evident that varying the auxiliary basis convergence settings
has essentially no influence on the low overall residual error, which is attributed to
other small numerical differences between two completely different
codes (analytical integrations in NWChem vs. numerical integrations in
FHI-aims, for example). Furthermore, it is also clear that it is sufficient
to choose the highest angular momentum in the ABF construction ($l^\text{ABF-max}$) 
to be just 1 higher than that of the single-particle atomic orbitals ($l^\text{AO-max}$).

\begin{figure}
  \begin{center}
  \includegraphics[width=0.6\textwidth,clip]{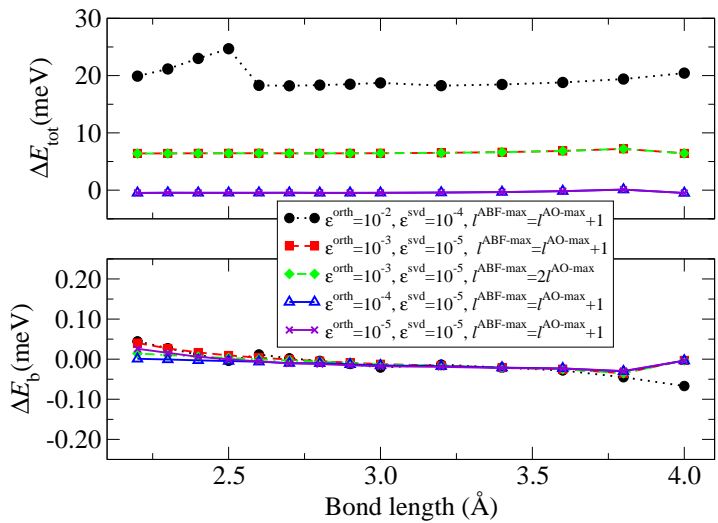}
  \end{center}
 \caption{(color online) Upper panel: Deviations of non-relativistic RI-V HF total energies from the 
          NWChem reference values as a function of bond length for Cu$_2$ for four sets of thresholding parameters.  
          Lower panel: the deviation of
          the BSSE-corrected RI-V HF binding energies from the reference values
          for the same sets of thresholding parameters.  The cc-pVQZ basis is used in
          all the calculations.}
\label{Fig_Cu2_HF_BE_cc-pVQZ} 
\end{figure}

Having established the quality of our ABFs for light elements, we now proceed to check their performance 
for the heavier elements where some noteworthy feature is emerging. In figure \ref{Fig_Cu2_HF_BE_cc-pVQZ}
we plot the errors in the RI-V HF total energies and binding energies for Cu$_2$ as a function of 
the bond length. Again the cc-pVQZ basis and the NWChem reference values are used here.
Using the thresholding parameters $(\varepsilon^\text{orth}$=10$^{-2}$, $\varepsilon^\text{svd}$=10$^{-4})$,
the RI-V HF total energy error can be as large as $\sim 15$ $meV/atom$ for copper, in contrast to
the $< 1$ $meV/atom$ total energy accuracy for light elements. This is because for
Cu, deep core electrons are present and the absolute total-energy scale is 1-2 orders of magnitude 
larger than that of light elements. The residual basis components
eliminated in the on-site Gram-Schmidt orthonormalization 
procedure (see section~\ref{sec:auxil_basis}) thus give bigger contributions to the total energy on
an absolute scale (although not on a relative scale). Indeed by decreasing $\varepsilon^\text{orth}$
the total-energy error gets increasingly smaller, and 1-1.5 $meV/atom$ total-energy accuracy can be achieved at
$\varepsilon^\text{orth}=10^{-4}$, as demonstrated in the upper panel of figure \ref{Fig_Cu2_HF_BE_cc-pVQZ}.
In contrast, similar to the case of light elements, the errors in the BSSE-corrected binding energies are 
significantly below 0.1 $meV/atom$ along a large range of bonding distances, regardless of the choice of thresholding 
parameters. And also increasing the highest angular momentum for ABFs beyond $l^\text{AO-max}+1$ does not
give noticeable improvements.

Finally we look at the quality of our ABFs for even heavier elements
-- the Au dimer. Due to the strongly localized core states in Au, 
all-electron GTO basis sets that are converged to the same level of accuracy as for N and Cu above are,  
to our knowledge, not available for Au. 
Furthermore, relativity is no longer negligible and must at least be treated at a scalar relativistic level. 
However, different flavors of relativistic implementations can differ heavily
in their absolute total energy (even if all chemically relevant energy 
differences are the same). 
An independent reference for all-electron GTO-based HF total energy with the
same relativistic treatments available in FHI-aims is not readily obtainable.
Under such circumstances, we demonstrate here the total energy
convergence with respect to our own set of thresholding parameters.

\begin{figure}
  \begin{center}
  \includegraphics[width=0.6\textwidth,clip]{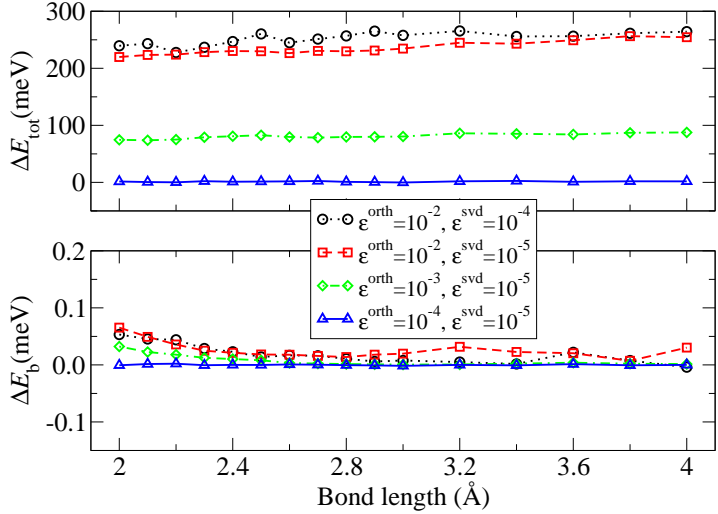}
  \end{center}
 \caption{(color online) Upper panel: Deviations of scaled ZORA RI-V HF total energies from the reference values
          as a function of bond length for Au$_2$ for four sets of thresholding parameters. The reference 
          values here are also obtained with the RI-V HF approach with very tight thresholds 
          $(\varepsilon^\text{orth}=10^{-5}, \varepsilon^\text{svd}=10^{-5})$.
          Lower panel: the deviation of the BSSE-corrected RI-V HF binding energies from the reference values.
          The FHI-aims \textit{tier} 2 basis and $l^\text{ABF-max}=l^\text{AO-max}+1$ are used in all the calculations.}
\label{Fig_Au2_HF_ABF_acc} 
\end{figure}

In figure \ref{Fig_Au2_HF_ABF_acc} we plot the deviations of the RI-V HF
total and binding energies for Au$_2$ obtained with FHI-aims using
NAO \emph{tier} 2 basis with somewhat less tight thresholding parameters from those obtained with a very tight threshold setting $(\varepsilon^\text{orth}=10^{-5}$, $\varepsilon^\text{svd}=10^{-5})$.
The relativistic effect is treated using the scaled zeroth-order regular approximation (ZORA)
\cite{Lenthe/Baerends/Snijders:1994} (see section~\ref{sec:N2_benchmark_GTO}),
but this detail is not really important for the discussion here.
From figure \ref{Fig_Au2_HF_ABF_acc} one can see that with thresholding
parameters that are sufficient for light elements
($\varepsilon^\text{orth}$=10$^{-2}$, $\varepsilon^\text{svd}$=10$^{-4}$),
the error in the total energy is even bigger -- one order of magnitude larger than in
the case of Cu$_2$. However, by going to tighter and tighter on-site
ABF thresholding parameter $\varepsilon^\text{orth}$, the total-energy
error can again to be reduced to the $meV/atom$ level. Similar to the Cu$_2$
case, the accuracy in the BSSE-corrected binding energy is still
extraordinarily good, well below 0.1 meV regardless of the choice of
thresholding parameters. The counterpoise correction can thus be used,
in general, as a simple, readily available convergence accelerator for binding
energies. 

We conclude this section with the following remarks: our procedure for constructing the ABFs gives highly
accurate and reliably results for the RI-V approximation. For light elements, one can readily get $meV/atom$ 
accuracy in total energies and sub-$meV/atom$ accuracy in binding energies, for a wide range of thresholding
parameters. For heavy elements, $meV/atom$ accuracy requires tigher thresholding parameters, particulary for the on-site orthonormalization $\varepsilon^\text{orth}$. However, sub-$meV$ BSSE-corrected binding energy accuracy can always be obtained independent of the choice of thresholding parameters. 
Choosing converged yet efficient thresholding parameters thus obviously depends on the element in question. In fact, 
$\varepsilon^\text{orth}$ can be chosen differently for each element in the same calculation. For RI-V and 
light elements ($Z=1$-10), we employ $\varepsilon^\text{orth}=10^{-2}$ in the following. For heavier elements ($Z>18$), 
we resort to $\varepsilon^\text{orth}=10^{-4}$, which yields negligibly small errors in total energies even for
heavy elements, and $\varepsilon^\text{orth}=10^{-3}$ for elements in between. 
The additional, system-wide SVD threshold $\varepsilon^\text{svd}$ 
is set to $10^{-4}$ or tighter for the remainder of this paper. As shown above, its accuracy implications are 
then negligible as well.

\section{\label{sec:res}NAO basis convergence for HF, hybrid density functionals, MP2, RPA, and \textit{GW}}
   Having established the quality of our ABFs for given \textit{orbital}
basis sets, we next turn to the quality of our actual NAO \emph{orbital}
basis sets for HF, hybrid density functionals, MP2, RPA, and $GW$ calculations. 
Below we will separate the discussions of self-consistent ground-state calculations
(HF and hybrid density functionals) and correlated calculations
(MP2, RPA, and $GW$). As will be demonstrated below, with our basis prescription, 
a convergence of the \emph{absolute} HF total energy to a high accuracy ($mev/atom$) is possible
for light elements. In other cases (heavy elements or correlated methods), the 
total energy convergence is not achieved at the moment, but we show that 
energy \emph{differences}, which are of more physical relevence, can be achieved
to a high quality.

  \subsection{\label{sec:basis_hf}HF and hybrid density functional calculations}
     Here we will demonstrate how well the generic NAO basis set
libraries described in section~\ref{sec:basis} and
in \cite{Blum/etal:2009} perform for HF and hybrid density
functional calculations. As described earlier, our \emph{orbital} basis sets contain 
a \emph{functional-dependent minimal} basis, composed of the core and valence orbitals of
the free atom, and additional \emph{functional-independent} optimized basis sets (\emph{tiers}).
Thus, besides the discussion of the convergence behaviour of the generic optimized
\emph{tiers} basis sets, here we will also address the influence of the choice of
the \emph{minimal} basis which is in practice generated by certain atomic solver . 
For all-electron calculations for 
molecular systems with a given electronic-structure method, it would 
be best if the core basis functions were generated from the atomic solver
using the same method. In this way, the behaviour of the molecular core wavefunctions
in the vicinity of the nuclei would be accurately described at a low price.
This is the case for LDA and most GGA calculations in FHI-aims.
Similarly, for HF molecular calculations,
it would be ideal if the \emph{minimal} basis was generated by the HF atomic solver.
Unfortunately at the moment the HF atomic solver is not yet available in our code,
and in practice we resort to the \emph{minimal} basis generated from other types
of atomic solvers. This is not a fundamental problem, and the only price one has
to pay is that more additional \emph{tiers} basis functions are needed
to achieve a given level of basis convergence.
Nevertheless one should keep in mind that there is a better strategy here and
in principle our basis prescription should work even better than what is reported here.

In the following the NAO basis convergence for HF will be examined, and
along the way the influence of the \emph{minimal} basis 
will be illustrated by comparing those generated by DFT-LDA and 
Krieger-Li-Iafrate (KLI) \cite{Krieger/Li/Iafrate:1992} 
atomic solvers.  The KLI method solves approximately the
exact-exchange optimized-effective-potential (OEP) equation 
\cite{Talman/Shadwick:1976,Kuemmel/Kronik:2008}, by replacing the orbital-dependent denominator 
in the Green function (at zero frequency) appearing in the OEP equation by an orbital-independent
parameter. This in practice reduces the computational efforts 
considerably without losing much accuracy \cite{Engel:2003}. The KLI atomic core
wavefunctions resemble the HF ones much better than the LDA ones do.
As demonstrated below, by moving from the LDA \emph{minimal} basis to the KLI ones
in HF calculations, the abovementioned problem is alleviated to some extent.

\subsubsection{Light elements}
\begin{figure}
    \begin{center}
    \includegraphics[width=0.6\textwidth,clip]{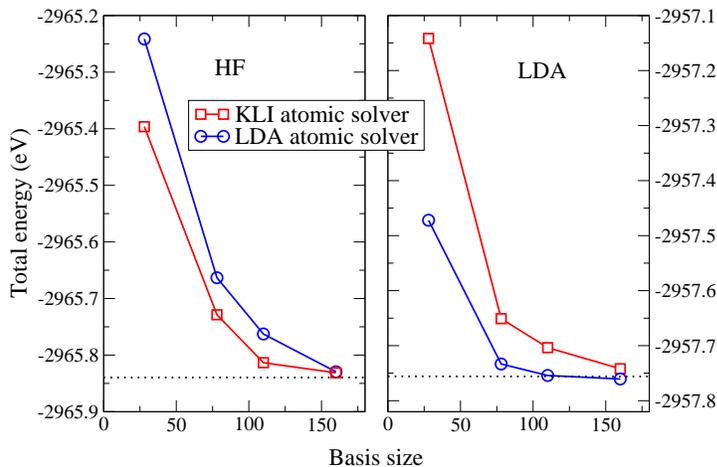}
    \end{center}
    \caption {(color online) NAO basis convergence test: HF and LDA total energies for N$_2$ at $d=1.1${\AA}
              as a function of increasing NAO basis set size (\textit{tier} 1 to \textit{tier} 4). 
              The two convergence curves correspond to starting \textit{minimal} basis sets generated
              by LDA and KLI atomic solvers, respectively.
              The dotted horizontal line marks the HF and LDA total energy computed using NWChem and the 
              cc-pV6Z basis,
              which gives a reliable estimate of the basis-set limit within 2 meV \cite{Halkier/etal:1999}. }
    \label{Fig:N2_LDA-HF_Etot}
\end{figure}

We first check the convergence behaviour of
our NAO basis sets for light elements in HF and hybrid
functional calculations.  In figure \ref{Fig:N2_LDA-HF_Etot} (left panel) 
the HF \emph{total} energy of N$_2$ as a function of increasing basis set size
is plotted, starting with two different sets of \emph{minimal} bases --
generated respectively from LDA and KLI atomic solvers. All other basis 
functions beyond the \textit{minimal} part (the \emph{tiers}) are the same for both
curves. For comparison, the convergence behaviour of the LDA total
energy of N$_2$ is shown on the right panel for the same basis sets.  
The horizontal (dotted) lines indicate independently computed GTO reference values using
NWChem and the Dunning cc-pV6Z basis set, which gives the best estimate for the HF total
energy at the complete basis set limit \cite{Halkier/etal:1999,basis_strategy}. 

First, with both types of \textit{minimal} basis the HF total energy
can be systematically converged to within a few meV of the 
independent GTO reference value. This is reassuring, as we can
thus use our standard NAO basis sets in a transferable manner even between
functionals that are as different as LDA and
HF. Furthermore, it is evident that the KLI-derived 
minimal basis performs better for HF, and similarly the LDA
derived minimal basis performs better for LDA. As mentioned above,
this is because the closer the starting atomic core basis functions 
to the final molecular core orbitals, the faster the overall basis convergence
is. If the true HF \emph{minimal} basis was used, we should expect an even 
faster basis convergence of the HF total energy, similar to the LDA 
total-energy convergence behaviour starting with the LDA \emph{minimal} basis ((blue) circles in the
right panel of figure \ref{Fig:N2_LDA-HF_Etot}). In this case the BSSE in
a diatomic molecular calculation should also be vanishingly small since
the atomic reference is already accurately converged from the outset
with the \emph{minimal} basis.  In practice, the reliance on KLI-derived \textit{minimal} 
basis functions instead leads to some small BSSE-type errors
in energy differences, as shown below.

\begin{figure}
  \begin{center}
  \includegraphics[width=0.6\textwidth,clip]{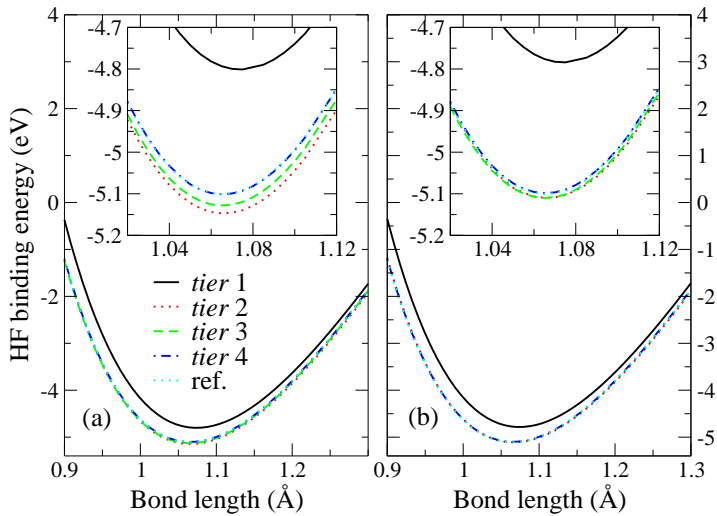}
  \end{center}
  \caption {(color online) HF binding energy of N$_2$ as a function of the bond length 
         for different levels of NAO basis sets (from \textit{tier} 1 to 
       \textit{tier} 4). The reference curve (denoted as ``ref.") is obtained with
        NWChem and a gaussian cc-pV6Z basis. (a): results without BSSE corrections;  (b):
             BSSE corrected results. The insets magnify the equilibrium region.}
    \label{Fig:N2_HF_BE_NAO}
\end{figure}

In figure \ref{Fig:N2_HF_BE_NAO} we present the NAO basis convergence
of the HF binding energy for N$_2$ as a function of bond
distance. Here we start with the
KLI \textit{minimal} basis and then systematically add basis functions
from \textit{tier} 1 to \textit{tier} 4. Results are shown both
without (left) and with (right) a counterpoise correction.     
A substantial improvement of the binding energy is seen between
the \textit{tier} 1 and the  \textit{tier} 2 basis, with only slight
changes beyond  \textit{tier} 2. In the absence of BSSE corrections a
slight overbinding is observed for \textit{tier} 2 and \textit{tier}
3. As noted above, we attribute this to the fact that in our calculations 
we used KLI core basis functions as a practical compromise
and the atomic reference calculation is not sufficiently converged for 
the outset. The basis functions from the neighboring
atom will then still contribute to the atomic total energy and this leads to
non-zero BSSE.  The counterpoise correction will cancel this contribution---which is very
similar for the free atom and for the molecule---almost exactly. The
counterpoise-corrected binding energies for \emph{tier} 2 are in fact
almost the same as for \emph{tier} 4. The latter agrees with results
from a GTO cc-pV6Z basis (NWChem) within 1-2 meV (almost indistinguishable
  in figure \ref{Fig:N2_HF_BE_NAO}).  

\begin{figure}
 \begin{picture}(400,170)(0,0)
  \put(100,0){\includegraphics[width=0.6\linewidth,clip]{graphs/H2O_2_HF-PBE0_NAO_conv.eps}}
  \put(285,115){\includegraphics[width=0.10\linewidth,clip]{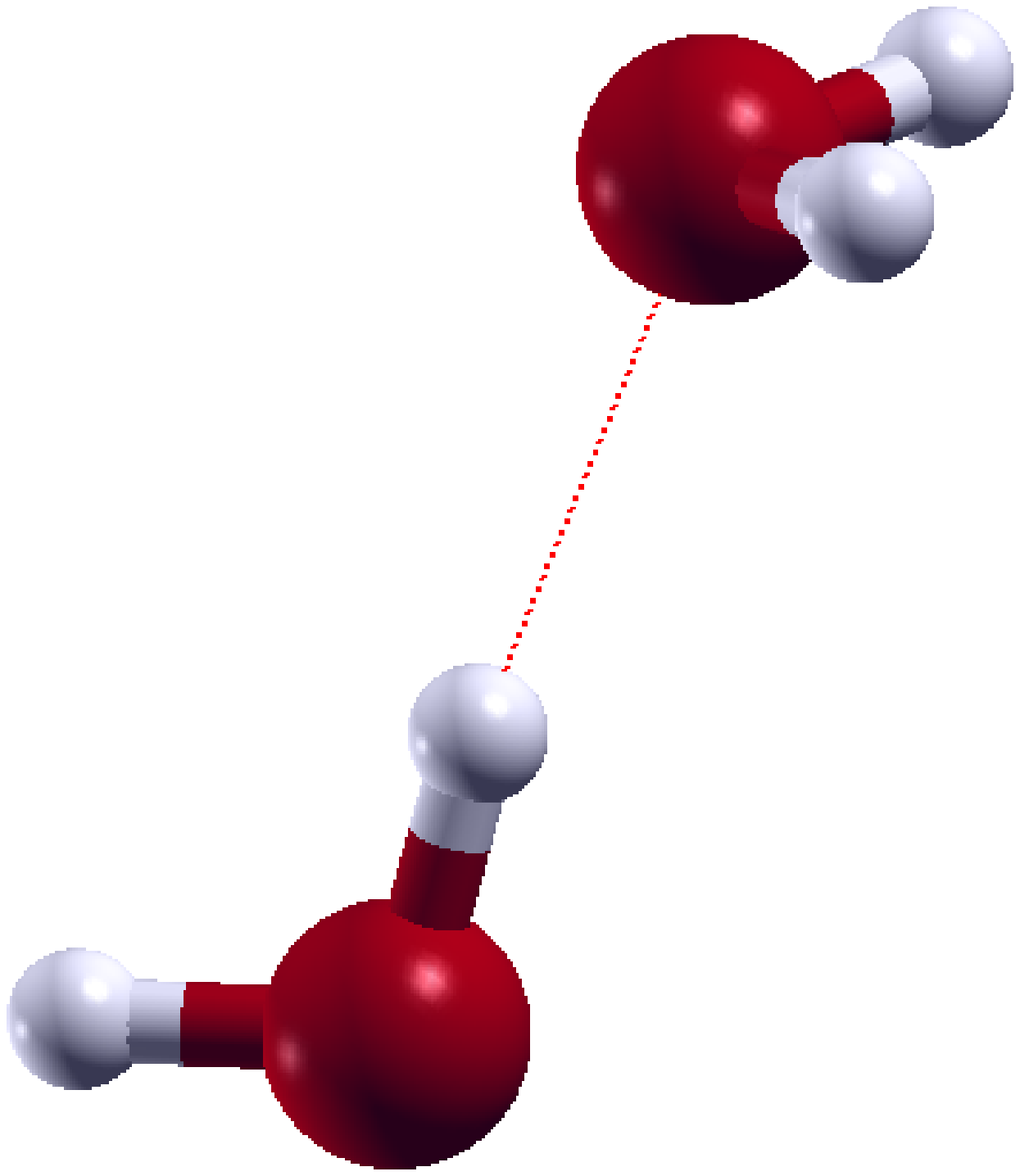}}
 \end{picture}
 \caption {(color online) Convergence of the HF and PBE0 binding energies of the water dimer (at the PBE geometry, 
          pictured in the inset)  as a function of NAO basis size  
         (\textit{tier} 1, 2, 3 for the first three points, and \textit{tier}
       3 for H plus  \textit{tier} 4 for C for the last point). Results 
      both with and without counterpoise BSSE correction are shown. The dotted line
      marks the NWChem/cc-pV6Z value. }
    \label{Fig:H2O_2_HF_BE_NAO}
\end{figure}

Figure \ref{Fig:H2O_2_HF_BE_NAO} demonstrates the same behaviour for a
different test case, the binding energy of the water dimer using HF
(left) and the PBE0 \cite{Perdew/Ernzerhof/Burke:1996,
Perdew/Burke/Ernzerhof:1996} hybrid functional (right). The geometry
of the water dimer has been  
optimized with the PBE functional and a \textit{tier} 2 basis. 
For convenience, the binding energy is computed with reference to H$_2$O
fragments with fixed geometry as in the dimer, not to fully relaxed H$_2$O
monomers. This is sufficient for the purpose of the basis convergence test
here.  Detailed geometrical information for water dimer can be found in
\cite{Blum/etal:2009}.  In figure \ref{Fig:H2O_2_HF_BE_NAO} 
the dotted line again marks the NWChem GTO cc-pV6Z reference results. 
Similar to the case of N$_2$, we observe that the HF binding
energy is fairly well converged at the \textit{tier} 2  level,
particularly after a counterpoise correction, which gives a binding
energy that agrees with the NWChem reference value to within 1-2~meV.
The BSSE arising from insufficient core description is reduced for the PBE0 hybrid
functional, where only a fraction (1/4) of exact-exchange is included.  

In practice, HF calculations at the \textit{tier} 2
level of our NAO basis sets yield accurate results for light elements.
Counterpoise corrections help to cancel residual total-energy errors 
arising from a non-HF \emph{minimal} basis. However, even
without such a correction, the convergence level is already pretty satisfying
(the deviation between the black and the red curve in
figure \ref{Fig:H2O_2_HF_BE_NAO} is well below 10~meV 
for \emph{tier} 2 or higher). 

\subsubsection{Heavy elements}
\begin{figure}
  \label{Fig:Au2_HF_BE}
  \begin{center}
  \includegraphics[width=0.6\linewidth,clip]{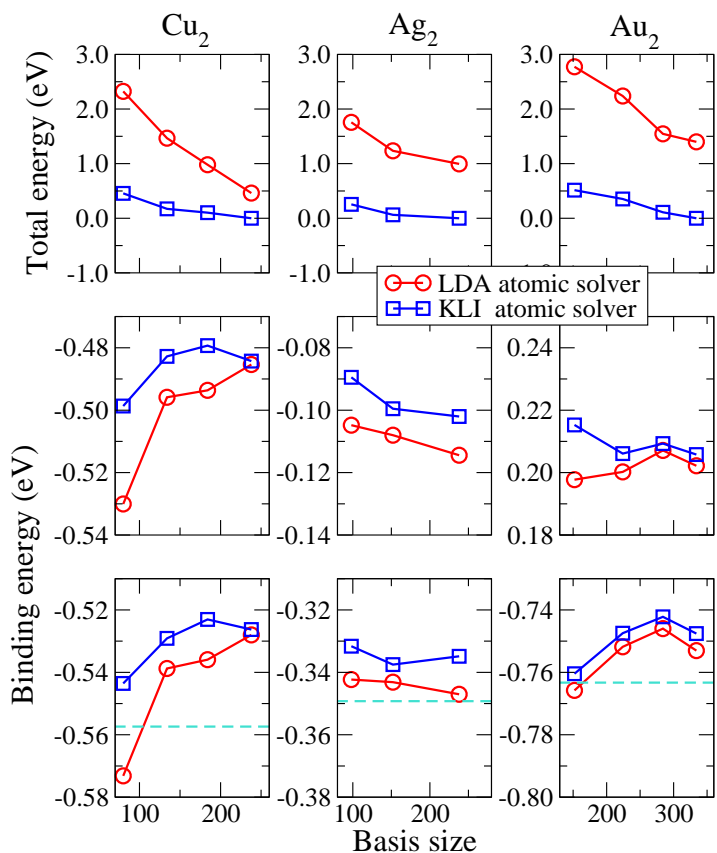}
  \end{center}
  \caption {(color online) 
            Convergence with basis size of the non-relativistic
            (NREL) HF total energies (upper panels), non-relativistic
            binding energies (middle panels) and scalar-relativistic
            (scaled ZORA \cite{Lenthe/Baerends/Snijders:1994,Blum/etal:2009})
            binding energies (bottom panels) of Cu$_2$ (left), Ag$_2$
            (middle), and Au$_2$ (right), at fixed bond length, $d$=2.5 \AA. 
            Similar to figure \ref{Fig:N2_LDA-HF_Etot}, results are shown
            for two sets of \emph{minimal} basis generated using LDA and
            KLI atomic solvers.
            For clarity the NREL total energies are offset by -89197.12 eV, -282872.42 eV, 
            and -972283.08 eV respectively for Cu$_2$, Ag$_2$, Au$_2$, which
            correspond to the actual values of the last data points with KLI \textit{minimal}
            basis. All binding energies are BSSE-corrected. The dashed horizontal lines in the
            bottom panels mark the NWChem reference values using aug-cc-pV5Z-PP basis with ECP. }
    \label{Fig:Cu2_Ag2_Cu2_NAO}
\end{figure}
For heavier elements (Z$>18$), the impact of non-HF core basis
functions on HF \emph{total} energies is larger. However, the
error again largely cancels in energy differences, as will be shown below. In
order to avoid any secondary effects from different
scalar-relativistic approximations to the kinetic energy operator, in
figure \ref{Fig:Cu2_Ag2_Cu2_NAO} (upper panels) we first compare
the convergence of non-relativistic (NREL) HF total energies
with NAO basis size for the coinage metal dimers Cu$_2$, Ag$_2$, and Au$_2$
at fixed binding distance $d$=2.5~{\AA}. (The experimental binding
distances are 2.22~{\AA} \cite{Huber79},
2.53~{\AA} \cite{Simard91,Beutel93} and 2.47~{\AA} \cite{Huber79},
respectively.) Again, we find that KLI-derived minimal basis sets are
noticeably better converged (lower total energies) than LDA-derived
minimal basis sets. In contrast to N$_2$, however, absolute
convergence of the total energy is here achieved in none of
these cases, and the discrepancy increases from Cu (nuclear charge Z$=29$) to
Au (Z$=79$). 

For comparison, the middle and lower panels of
figure \ref{Fig:Cu2_Ag2_Cu2_NAO} show non-relativistic and
scalar-relativistic \emph{binding} energies for all three dimers. The
scalar-relativistic treatment employed is the scaled 
ZORA due to Baerends and
coworkers \cite{Lenthe/Baerends/Snijders:1994} (for details of our own
implementation, see \cite{Blum/etal:2009}). 
In all three cases, the binding energies are converged to
a scale of $\sim$0.02~eV, at least two orders of magnitude better
than total energies. In other words, any residual convergence error
due to the choice of minimal basis (LDA or KLI instead of
HF) cancel out almost exactly. To compare our prescription to
that generally used in the quantum chemstry community where 
effective core potentials (ECP) are used to describe the core electrons
and the relativistic effect, we also marked in figure \ref{Fig:Cu2_Ag2_Cu2_NAO} 
the reference values computed using NWChem and the aug-cc-pV5Z-PP basis 
\cite{Peterson/Puzzarini:2005,Figgen/etal:2005}. The agreement between
our all-electron approach and the GTO-ECP one is pretty decent for
Ag$_2$ and Au$_2$, whereas a larger discrepancy of $\sim$0.03~eV is
observed for Cu$_2$. For this particular case we suspect the remaining disagreement 
is an issue with respect to the atomic reference energy for the Cu atom between both codes.
More work will be done to fully unravel the point.

  \subsection{\label{sec:basis_mp2-rpa-gw}MP2, RPA, and \textit{GW}
       calculations}
       
\begin{figure}
 \begin{center}
 \includegraphics[width=0.6\textwidth,clip]{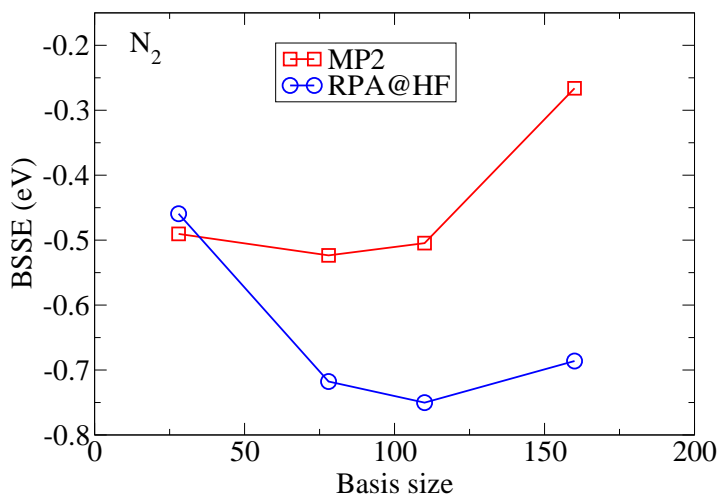}
 \end{center}
 \caption{(color online) BSSEs in MP2 and RPA@HF binding-energy calculations
    for N$_2$ ($d=1.1$ \AA) as a function of
    the NAO basis set size.  The four points corresponds to NAO  \textit{tier} 1 to 
    \textit{tier} 4 basis sets, respectively.} 
  \label{N2_MP2-RPA_BSSE}
\end{figure}

  In the implementation described here, MP2,
RPA, and $GW$ methods require the explicit inclusion of unoccupied
single-particle states. As a consequence,
noticeably larger basis sets are needed to obtain converged results
in these calculations 
\cite{Schwartz:1962,Hill:1985,Kutzelnigg/Morgan:1992,Halkier/etal:1998,Bruneval/Gonze:2008,Berger/Reining/Sottile:2010,Shih/etal:2010,Furche:2001,Furche/Voorhis:2005,Paier/etal:2010}.
Much experience has been gained in the quantum chemistry community to construct
Gaussian basis sets for correlated calculations \cite{Dunning:1989,Weigend/Ahlrichs:2005},
but for NAOs this is not case.
In this section, we show how our standard NAO basis sets perform for
MP2, RPA, and $GW$ calculations, for both light and heavy elements. 
For clarity, we separate the discussions for the convergence of binding energies (in the case of MP2 and RPA) and 
quasiparticle excitations  (in the case of $GW$ and
MP2 self-energy calculations). In contrast to the
cases of HF and hybrid density functionals, 
BSSE corrections for RPA and/or MP2 are essential to obtain reliable binding energies.
This results directly from the larger basis sets required to converge
the MP2 or RPA total energy \cite{Schwartz:1962,Hill:1985,Kutzelnigg/Morgan:1992,Halkier/etal:1998,Furche:2001,Furche/Voorhis:2005,Paier/etal:2010}, yielding larger BSSE for finite basis
set size. 
With our standard NAO basis sets, the actual BSSEs in MP2 and RPA (based on 
the HF reference, denoted as RPA@HF in the following) calculations are plotted
in figure \ref{N2_MP2-RPA_BSSE} for the example of N$_2$. 
The size of BSSEs in these cases is huge and does not
diminish even for the pretty large \textit{tier} 4 basis. It is thus mandatory to 
correct these errors in MP2 and RPA calculations to get reliable binding energies.
As one primary interest in this work is the
applicability of standard NAO basis sets for MP2 and RPA,
all binding energies presented are therefore counterpoise-corrected. 
In all HF reference calculations in this session, the KLI \textit{minimal} basis 
is used.
For RPA, $GW$, and MP2 self-energy calculations, we use 40 imaginary frequency points on a 
modified Gauss-Legendre grid (\ref{sec:appC}), which ensures a high accuracy
for the systems studied here. 

\subsubsection{Binding energies}

\begin{figure}
 \begin{center}
 \includegraphics[width=0.6\textwidth,clip]{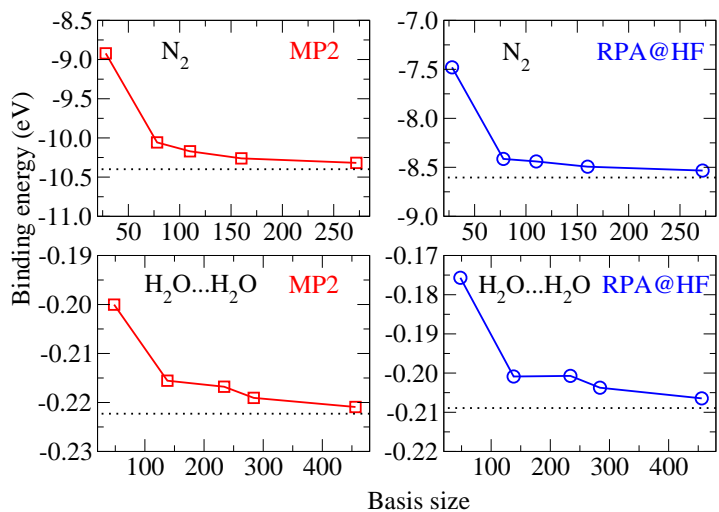}
 \end{center}
 \caption{(color online) Convergence of BSSE-corrected MP2 and RPA@HF binding energies 
    for N$_2$ ($d=1.1$ \AA) and the water dimer (PBE geometry) as a function of
    the NAO basis set size.  The first four points corresponds to NAO  \textit{tier} 1 to 
    \textit{tier} 4 basis sets and the last point corresponds to 
    the composite ``\textit{tier} 4 + a5Z-d" basis. The dotted horizontal line
    marks the aug-cc-pV6Z results. }
  \label{Fig:N2_MP2-RPA_NAO_conv}
\end{figure}

As illustrating examples for light elements, in figure \ref{Fig:N2_MP2-RPA_NAO_conv} 
the BSSE-corrected MP2 and RPA binding energies for N$_2$ and the
water dimer are shown as a function of the NAO basis set size. The dotted line marks reference results computed with 
FHI-aims and the Dunning aug-cc-pV6Z basis.
In the case of MP2, the FHI-aims aug-cc-pV6Z values agree with that of
NWChem to within 0.1 meV. Unfortunately, a similar independent reference
is not available for RPA, but excellent agreement is also seen 
with smaller basis sets, for which reference RPA data are available for N$_2$
\cite{Furche:2001}. Upon increasing the basis size, the biggest
improvement occurs when going from \textit{tier} 1 to \textit{tier}
2, with further, smaller improvements from \textit{tier} 2
to \textit{tier} 4.
For the strongly bonded N$_2$ the MP2 binding energy at \textit{tier} 4 level deviates from
the aug-cc-pV6Z result by $\sim$ 120 meV, or $\sim1\%$ of the
total binding energy. For the hydrogen-bonded water dimer, the
corresponding values are $\sim$ 3 meV and $\sim1.5\%$
respectively. The convergence quality of RPA results with respect to
the NAO basis set size is similar.

Going beyond our FHI-aims standard NAO basis sets, further improvements arise by
adding (\emph{ad hoc}, as a test only) the diffuse functions 
from a GTO aug-cc-pV5Z basis set, denoted  ``a5Z-d'' in the
following. The results computed using this composite ``\textit{tier} 4 + a5Z-d" basis
are shown by the last point in figure \ref{Fig:N2_MP2-RPA_NAO_conv}.  
The deviation between the \textit{tier} 4 and 
aug-cc-pV6Z results  is then reduced by more than a factor of two. 
For the water dimer, for example, ``\textit{tier} 4 + a5Z-d" gives -220.9 meV and -206.5 meV 
for the MP2 and RPA@HF binding energies, comparable to the quality of the
cc-pV6Z basis which yields -221.1 meV and -206.9 meV, respectively. Both then agree with 
the aug-cc-pV6Z results (-222.3 meV for MP2 and -208.9 meV for RPA@HF) to within $\sim 2$ meV.

\begin{figure}
 \begin{center}
 \includegraphics[width=0.6\textwidth,clip]{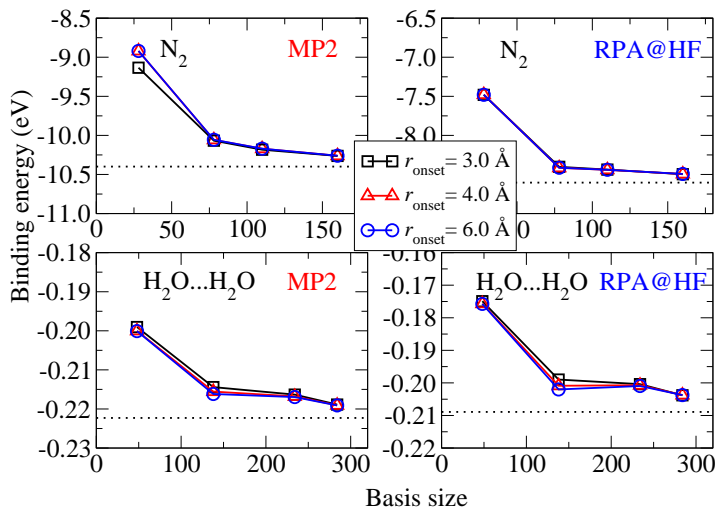}
 \end{center}
 \caption{(color online) Convergence of the BSSE-corrected MP2 and RPA@HF binding energies 
    for N$_2$ ($d=1.1$ \AA) and the water dimer (PBE geometry) as a function of
    the NAO basis set size.   The four points correspond to the \textit{tier} 
    1 to \textit{tier} 4 basis. The NAO basis functions are generated for three
    onset radii (3 \AA, 4 \AA, and 6 \AA) for the confining potential.
    The dotted horizontal line marks the aug-cc-pV6Z results. }
  \label{Fig:N2_MP2-RPA_NAO_conv_cutpot}
\end{figure}

In this context, it is interesting to check if the cut-off radii of
our NAO functions have any influence on the convergence behaviour demonstrated above.  
As described in \cite{Blum/etal:2009}, 
the NAO basis functions are strictly localized in a finite spatial area around
the nuclei, and the extent of this area is controlled by a confining potential.
For the default settings used in the above calculations, this potential sets in at a distance of 4 {\AA} from 
the nucleus and reaches infinity at 6 {\AA}. The question is what would happen if we reduce or increase 
the onset radii of this confining potential?
The answer to this question is illustrated in figure \ref{Fig:N2_MP2-RPA_NAO_conv_cutpot} where 
basis convergence behaviour for N$_2$ and the water dimer are shown for three different 
onset distances of the confining potential.
From figure \ref{Fig:N2_MP2-RPA_NAO_conv_cutpot} one can see that increasing the onset radius of the confining 
potential (i.e., enlarging the extent of the NAO basis functions) from the default value
(4 \AA) has little effect on the convergence behaviour for N$_2$ or (H$_2$O)$_2$. 
Upon reducing it, noticeable changes of the results only occur for \textit{tier} 1
or \textit{tier} 2 in certain cases, but the overall effect is very small and does not change
the general convergence behaviour described above. This finding holds in general for
covalent and hydrogen bonds.  In practice, the onset
radius may always be invoked as an explicit convergence
parameter---for instance, much more weakly bonded (dispersion bonded)
systems benefit from slightly larger radii (5 {\AA} - 6 {\AA}) in our
experience. Further details on this can be found in \cite{Rossi:PhDthesis}. 

\begin{figure}
 \begin{center}
 \includegraphics[width=0.6\textwidth,clip]{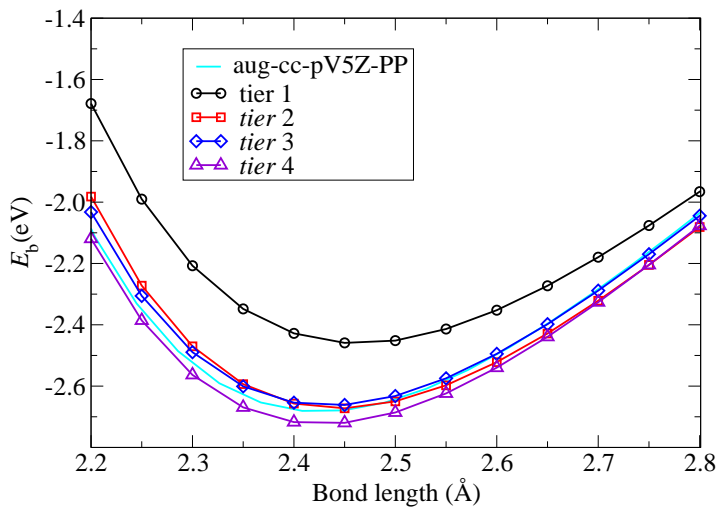}
 \end{center}
 \caption{(color online) Convergence of the BSSE-corrected MP2 binding energy curve
    for Au$_2$ with respect to the \textit{optimized} NAO basis set
 size (\textit{tier} 1 to \textit{tier} 4). Results from an independent, Gaussian-type
 calculation  (aug-cc-pV5Z-PP basis set with ECP, NWChem code) are
 included for comparison.
 }
  \label{Fig:Au2_MP2_NAO_conv}
\end{figure}

We next illustrate the NAO basis convergence for heavy elements, using
Au$_2$ as an example. In figure \ref{Fig:Au2_MP2_NAO_conv} the MP2
binding energy for the Au$_2$ dimer as a function of the bond length
is plotted for different NAO basis sets. Relativity is again treated at the
scaled ZORA level \cite{Lenthe/Baerends/Snijders:1994,Blum/etal:2009}.
The binding curves shown here demonstrate that the same qualitative
convergence behaviour as for our light-element test cases carries
over. 
In essence, significant improvements are gained from \textit{tier} 1 to
\textit{tier} 2, and basis sets between \textit{tier} 2
and \textit{tier} 4 yield essentially converged results. For comparison, we show a
completely independent (NWChem calculations) curve with Gaussian
``aug-cc-pV5Z-PP" basis sets \cite{Peterson/Puzzarini:2005,Figgen/etal:2005}. 
The resulting binding energy curve yields rather close agreement with our
all-electron, NAO basis set results.
\subsubsection{Quasiparticle energies}

\begin{figure}
 \begin{center}
 \includegraphics[width=0.6\textwidth,clip]{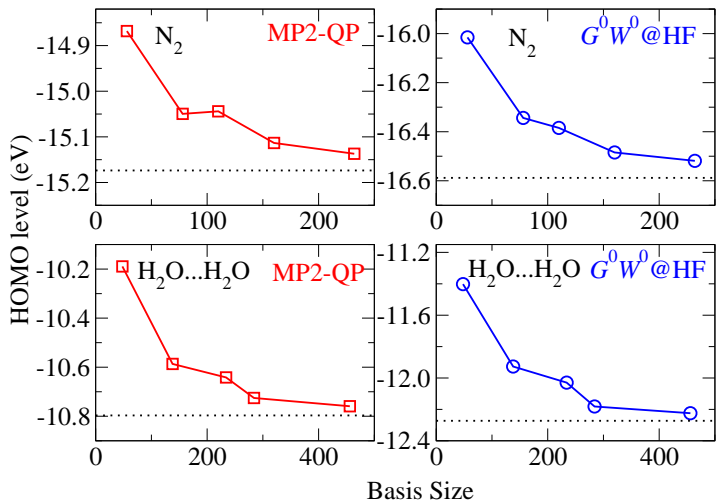}
 \end{center}
 \caption{(color online) Convergence of the quasiparticle HOMO level of N$_2$ and the water 
   dimer obtained with the MP2-QP and $G^0W^0$@HF self-energies versus basis size. 
   The last data point
   corresponds to the composite ``\textit{tier} 4 + a5Z-d" basis. The 
   aug-cc-pV6Z result is marked by the dotted horizontal line. }
  \label{Fig:N2_MP2-GW_HOMO_NAO_conv}
\end{figure}

After the discussion of binding energies, we next examine how the $GW$ and MP2 
quasiparticle energy levels converge with our NAO basis sets. The $G^0W^0@$HF and MP2 quasiparticle 
HOMO levels for N$_2$ and the water dimer are plotted in figure
\ref{Fig:N2_MP2-GW_HOMO_NAO_conv} as a function of basis set size. MP2 is denoted here as ``MP2-QP" to
emphasize that the MP2 self-energy
(\ref{Eq:2nd_self-energy}) is used, rather than a MP2 total-energy difference. 
We again take the results of an ``aug-cc-pV6Z" GTO calculation as a reference. The first four data
points in each sub-plot correspond to the NAO basis sets. The last
point represents the composite ```\textit{tier} 4 + a5Z-d" basis as
described above. 
Once again, the biggest improvement occurs when going
from \textit{tier} 1 to \textit{tier} 2. However, the deviation of the
HOMO levels from the reference values at \textit{tier} 2
level is still considerable, $\sim$ 0.2-0.3 eV for $G^0W^0$ and $\sim$
0.1-0.2 eV for MP2-QP. These errors are brought down to $\sim$0.1 eV
for $G^0W^0$ and $\sim$0.06 eV for MP2-QP by going to a
pure \textit{tier} 4 NAO basis set. The remaining error is then
further reduced by a factor of two by including the diffuse ``a5Z-d"
part of a 5Z GTO basis set. Accounting for the possible
underconvergence of the aug-cc-pV6Z itself (compared to the CBS limit),
we expect an overall $\sim$0.1~eV under-convergence of the composite
```\textit{tier} 4 + a5Z-d" results given here. This is still an
acceptable accuracy, considering the generally known challenge of  
converging the correlation contribution involving virtual states 
using local orbital basis functions \cite{Schwartz:1962,Hill:1985,Kutzelnigg/Morgan:1992,Halkier/etal:1998,Furche:2001,Furche/Voorhis:2005,Paier/etal:2010}.  Therefore
```\textit{tier} 4 + a5Z-d" basis sets were used in the benchmark
calculations presented in the section~\ref{sec:benchmark}. 

\subsubsection{$G^0W^0$ calculations for the benzene molecule}
$G^0W^0$ calculations for molecules have been reported in a number of
publications in recent years using various numerical frameworks and
computer code packages. In particular in the solid-state community, 
calculations have been based on plane wave methods together with
the supercell approach and pseudopotential approximations. Their advantage is a
systematically convergable basis set (plane waves) especially for the
unoccupied spectrum. However, the two approximations mentioned above
(pseudopotential and supercell) can be drastic \cite{GomezAbal08,GW_space-time_method_surf:2007,Freysoldt/etal:2008,Onida/etal:1995,Spataru/etal:2004,Rozzi/etal:2006,Ismail-Beigi/2006}. In particular, the Coulomb
operator in vacuum is not screened in a standard plane wave
approach. As a result, different images of the system interact with
one another across supercell boundaries \cite{Freysoldt/etal:2008,Onida/etal:1995,Spataru/etal:2004,Rozzi/etal:2006,Ismail-Beigi/2006}. In addition, the slow
converence of $G^0W^0$ results with the plane-wave cutoff of the
unoccupied spectrum is notorious \cite{Shih/etal:2010,Friedrich/etal:2011,Samsonidze/etal:2011}.

Table~\ref{Tab:G0W0_benzene} reports literature $G^0W^0$ results for the benzene
molecule\cite{Tiago/Chelikowsky:2006,Umari/Stenuit/Baroni:2009,Umari/Stenuit/Baroni:2010,Samsonidze/etal:2011,Foerster/etal:2011,KeSanhuang:2011}, a particularly often studied case, in comparison to our own
results. We focus on $G^0W^0$@LDA and $G^0W^0$@HF for the HOMO and LUMO levels. 
\begin{table}
\caption{\label{Tab:G0W0_benzene}$G^0W^0$ HOMO and LUMO values for the benzene (C$_6$H$_6$) molecule obtained by FHI-aims and several other numerical approaches as reported in literature. ``A.E" and ``P.P." in the second column
         refer to ``all electron" and ``pseudopotential" respectively.  }
\begin{indented}
\item[]
\begin{tabular}{@{\extracolsep{\fill}}lclcc}
\br \\[-2.0ex]
$G^0W^0$-type & P.P./A.E. & basis type & HOMO (eV) & LUMO (eV) \\
\br
\multirow{9}{*}{$G^0W^0$@LDA} & \multirow{3}{*}{A.E.} & NAO ``\textit{tier} 3"$^a$ & -8.99 & 1.06 \\
                              & & NAO ``\textit{tier} 4"$^a$ &  -9.06 & 0.96 \\
                              & & NAO ``\textit{tier} 4 + a5Z-d"$^a$ & -9.05 & 0.94 \\ [1.0ex]
                               \cline{2-5} \\[-2.0ex]
                              & \multirow{4}{*}{P.P.} & plane waves + NAOs$^b$ & -9.03 & 1.54 \\
                              & & plane waves + extrapol.$^c$ & -9.10 & $\slash$ \\
                              & & plane waves + Lancos$^d$ & -9.40 & $\slash$ \\
                              & & NAO ``TZDP"$^e$  &  -8.78 & 1.24 \\
                              & & real-space grid$^f$  & -9.88 & 0.47 \\ [1.0ex]
\cline{1-5}\\[-2.0ex]
\multirow{2}{*}{$G^0W^0$@HF} & \multirow{2}{*}{A.E.} &  NAO ``\textit{tier} 4"$^a$ & -9.64 & 1.51  \\
                             &                       & Gaussian ``cc-pVTZ"$^g$  &-9.28& $\slash$ \\[1.0ex]
\cline{1-5}\\[-2.0ex]
Experiments                   &                        &                  & -9.24$^h$ & 1.12$^i$ \\[1.0ex]                                        
\br
\end{tabular}
\begin{tabular}{ll}
 $^a$this work & \\
 $^b$Ref. \cite{Samsonidze/etal:2011} : & plane wave basis augmented with  \textsc{siesta}-type localized atomic orbitals \\
 $^c$Ref. \cite{Umari/Stenuit/Baroni:2009} : & plane wave basis extrapolated to infinite energy cutoff \\
 $^d$Ref. \cite{Umari/Stenuit/Baroni:2010} : & plane wave basis plus Lanczos trick to remove the empty states\\
 $^e$Ref. \cite{Foerster/etal:2011} & $^f$Ref. \cite{Tiago/Chelikowsky:2006}  \hspace{5mm} $^g$Ref. \cite{KeSanhuang:2011}  \\
 $^h$Ref. \cite{Nemeth/etal:1993} : &  (negative) ionization energy (IE). The \textit{vertical} IE only differs from \\
                                    &  this value by 0.01 eV according to the NIST database \cite{nist_database}.  \\
 $^i$Ref. \cite{Burrow/etal:1987} : &  (negative) \textit{vertical} electron affinity (IE) 
\end{tabular}
\end{indented}
\end{table}
Based on these results, it is clear that there is a significant degree
of scatter, even between results that are ostensibly converged using
the same fundamental approximations. Our own results for NAO
``\textit{tier} 3'', ``\textit{tier} 4'', and ``\textit{tier} 4 +
a5Z-d'' basis sets suggest internal convergence at the
``\textit{tier} 4'' level: $-$9.06~eV for the HOMO, and 0.94~eV for
the LUMO in $G^0W^0$@LDA, compared to $-$9.64~eV and 1.51~eV in
$G^0W^0$@HF. The LUMO values are unbound and can be interpreted
as experimental resonance energies (here taken from the tabulated
negative vertical electron affinity). 
In either case (HOMO or LUMO, $G^0W^0$@LDA or $G^0W^0$@HF) the results are not far 
from the experimental values.

The same cannot be said for the comparison between the different numerical implementations. 
For instance, the HOMO values in
$G^0W^0$@LDA range from $-$8.78~eV (small numeric basis set and
pseudopotentials) to $-$9.88~eV on a real-space grid. Even within the
plane wave based approaches, the values range from $-$9.03~eV to
$-$9.40~eV. Similar discrepancies arise for $G^0W^0$@HF, and for the
LUMO values. 
Within the scatter evidenced by
Table~\ref{Tab:G0W0_benzene}, we believe that our own results possess
some merit, as the system (i) is isolated (no supercell), (ii) is
treated fully all-electron, and (iii) any residual convergence issue
of the unoccupied spectrum should be exposed by the diffuse Gaussian
basis functions (``a5Z-d''). 
\section{\label{sec:benchmark}Basis set converged benchmark data
    for the G2 and S22 molecular test sets} 

  Our final section utilizes the preceding methodologies and techniques to 
provide accurate all-electron results for molecular test sets close to basis set
convergence. We cover \textit{vertical} ionization energies and binding energies at
different levels of theory, for well-defined, published molecular
geometries. 

\subsection{Vertical ionization energies from HF, MP2, and $G^0W^0$ methods}  

The $G^\text{0}W^\text{0}$ method has been used to calculate
the single-particle properties for various small- and medium-size molecules
with considerable success
\cite{Rohlfing/Louie:1998,Reining/Pulci/Palummo/Onida:2000,Ishii/etal:2001,
Tiago/Chelikowsky:2006,Hahn/Schmidt/Bechstedt:2005,Stan/Dahlen/Leeuwen:2006,Pavlyukh/Hubner:2007,Tiago/etal:2008,Palummo/etal:2009,Umari/Stenuit/Baroni:2009,Rostgaard/Jacobsen/Thygesen:2010,Blase/Attaccalite/Olevano:2011,KeSanhuang:2011,Foerster/etal:2011}.
The dependence of $G^\text{0}W^\text{0}$ on the starting point has also been
looked at in the past \cite{Rinke/etal:2005,Rinke/pssb,Stan/Dahlen/Leeuwen:2006,Blase/Attaccalite/Olevano:2011}.
Here, using a selected collection of atoms and molecules from the G2 ion test set for
ionization energies \cite{Curtiss/etal:1998}, we aim to establish
the overall performance of $G^\text{0}W^\text{0}$ for computing IEs close to the basis set limit, and
systematically examine its dependence on the starting point. The selection of molecules is
based on the availability of experimental geometries and experimental \textit{vertial} IEs. 
With \textit{vertical} IEs we denote ionization energies at fixed geomerty, i.e. no structural relaxation after excitation. This quantity is directly comparable to the $GW$ and MP2-QP and quasiparticle energies.  We will also assess the MP2-QP approach for determining IEs since
this method was used in quantum chemistry in past decades, but
direct comparisons in the literature are scarce \cite{Szabo/Ostlund:1989}.  
Finally, results for IEs determined by MP2 total energy differences
(denoted here simply as ``MP2" in contrast to ``MP2-QP") between the
neutral atoms/molecules and the corresponding positively charged ions are also presented.
We note that IEs given by MP2-QP essentially correspond to the MP2 total energy difference
if the HF orbitals of the neutral systems were used in the ionic calculation (for a discussion,
see, e.g., \cite{Szabo/Ostlund:1989}). All calculations were carried out
at the experimental geometries. As mentioned above, the composite
``\textit{tier} 4 + a5Z-d" basis set is 
used for most of the elements except for a few cases (e.g., rare gas
atoms) where the \textit{tier} 4 NAO basis is not available. 
In these cases the full aug-cc-pV5Z functions are added to the NAO
\textit{tier} 2 basis.  On average we expect the chosen basis setup to guarantee
a convergence of the HOMO quasiparticle levels to $\sim$0.1 eV or better for the calculations presented in the following.

\begin{figure}
 \begin{center}
 \includegraphics[width=0.6\textwidth,clip]{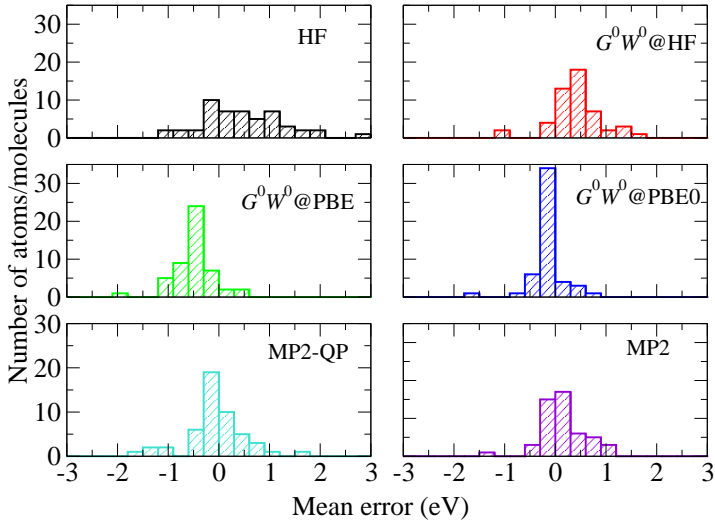}
 \end{center}
 \caption{Histograms of the error distribution for IEs for a set of 50 atoms
  and molecules calculated with HF, $G^\text{0}W^\text{0}$@HF, $G^\text{0}W^\text{0}$@PBE,
   $G^\text{0}W^\text{0}$@PBE0,  MP2-QP, and MP2.}
 \label{Fig:G2_error_histogram}
\end{figure}

In figure \ref{Fig:G2_error_histogram} we present histograms of the error distributions given by HF, 
MP2-QP, $G^\text{0}W^\text{0}$@HF and $G^\text{0}W^\text{0}$@PBE0 for this database. 
The actual IE values are presented in \ref{sec:appB}. 
Compared to HF, one can see that the deviations from the experimental values
are much smaller in $G^\text{0}W^\text{0}$ and MP2-QP. On average HF tends to overestimate IEs. This 
trend is corrected by MP2 and $G^\text{0}W^\text{0}$. Using the same HF
reference, the magnitude of the correction is smaller in $G^\text{0}W^\text{0}$ than in MP2-QP, due
to the renormalization effect  coming from the screened Coulomb
interaction in the $GW$ self-energy. 
Concerning the starting-point dependence of the $G^\text{0}W^\text{0}$
method, $G^\text{0}W^\text{0}$ based on HF  
gives too large IEs on average, whereas $G^\text{0}W^\text{0}$
based on PBE does the
opposite. Consequently, $G^\text{0}W^\text{0}$ based on the hybrid density
functional PBE0 appears to be the best compromise, although a slight
underestimation of the IEs is still visible. Furthermore, comparing MP2-QP
with MP2 shows that the two approaches yields very similar results,
implying that for the light elements reported here orbital relaxation effects are 
not significant \cite{MP2-QP_problem}.
Table~\ref{Tab:G2_error_statistics} summarizes the error statistics. For the
subset of atoms and molecules MP2-QP gives the smallest mean 
error (ME) and $G^0W^0$@PBE0 the smallest mean absolute error
(MAE). 
\begin{table}
  \caption{\label{Tab:G2_error_statistics}Mean error (ME), mean absolute error (MAE), and mean absolute
       percentage error (MAPE) for the ionization energies in the  G2-I subset  
       computed with HF, MP2-QP, MP2, and $G^0W^0$ based on HF, PBE, and PBE0.}
  \begin{indented} 
  \item[]
  \begin{tabular}{lccc}
   \br \\[-1.5ex]
          & ME (eV) & MAE (eV) & MAPE \\       
    \mr \\[-1.5ex]
    HF & 0.48 & 0.70 & 5.8\% \\[0.2ex]
    MP2-QP & -0.04 & 0.40 & 3.3\% \\[0.2ex]
    MP2 & 0.15 & 0.31 & 2.7\% \\[0.2ex]
    $G^\text{0}W^\text{0}$@HF & 0.41 & 0.52 & 4.6\% \\[0.2ex]
    $G^\text{0}W^\text{0}$@PBE & -0.49 & 0.53 & 4.6\% \\[0.2ex]
    $G^\text{0}W^\text{0}$@PBE0 & -0.15 & 0.25 & 2.2\% \\[0.2ex]
   \br
 \end{tabular}
 \end{indented}
\end{table}

\subsection{Benchmark MP2 and RPA results for the G2-I atomization energies}
The 55 atomization energies from the original G2-I set (Table~III of \cite{Curtiss/etal:1991})
serve as a good benchmark database for ground-state total-energy methods for covalently bonded systems.
In Table~\ref{Tab:G2_AE} we present our MP2 and RPA atomization energies for the G2-I set
as computed using FHI-aims and the ``\textit{tier} 4 + a5Z-d" basis set. 
The geometries used in the calculations are those determined by Curtiss {\it et al.} \cite{Curtiss/etal:1991},
i.e., all-electron MP2 optimization with a Gaussian 6-31G* basis set.
The reference data, taken from \cite{Feller/Peterson:1999}, is
derived from experiment and corrected for zero-point vibrations.

Our MP2 and RPA results are in good agreement with those reported by Feller and Peterson
\cite{Feller/Peterson:1998} and Paier \textit{et al.} \cite{Paier/etal:2010}, respectively, both
obtained with GTO basis sets and extrapolated to the complete basis set limit. Table~\ref{Tab:G2_AE}
demonstrates that MP2 performs best for these small covalently bonded molecules. The RPA approach, on the other hand, which has a broader applicability (e.g. bond breaking and/or metals where MP2 fails), exhibits 
significant underbinding.
For RPA we also observe that KS-PBE provides a much better reference than HF. Recently
computational approaches to overcome the underbinding behaviour of the standard RPA scheme have been proposed. These comprise contributions from single excitations \cite{Ren/etal:2011} and second-order screened exchange \cite{Grueneis/etal:2009,Paier/etal:2010}.
 \begin{table}[tp]
  \caption{\label{Tab:G2_AE}
    G2-I atomization energies computed with MP2, RPA@HF, and RPA@PBE. All numbers have been calculated 
   with FHI-aims and ``\textit{tier} 4 + a5Z-d" basis sets. The experimental reference results
    are taken from \cite{Feller/Peterson:1999}. } 
  \begin{indented}
  \item[]
  \begin{tabular}{@{}lcccc}
    \br
     Molecule & MP2 & RPA@HF & RPA@PBE & Exp. \\
    \mr
       BeH &     ~54.3 &     ~55.6 &     ~49.9 &     ~49.8 \\ 
      C$_2$H$_2$ &    412.9 &    374.1 &    377.3 &    403.4 \\ 
      C$_2$H$_4$ &    564.3 &    527.4 &    533.1 &    562.4 \\ 
      C$_2$H$_6$ &    709.9 &    671.4 &    678.7 &    712.7 \\ 
        CH &     \ 80.0 &     \ 75.7 &     \ 80.5 &     \ 83.9 \\ 
     CH2($^1A_1$) &    176.6 &    168.1 &    173.3 &    180.9 \\ 
     CH2($^3B_3$) &    189.3 &    181.0 &    178.2 &    190.2 \\ 
       CH$_3$ &    304.6 &    290.5 &    292.4 &    307.9 \\ 
     CH$_3$Cl &    397.1 &    370.1 &    368.0 &    394.5 \\ 
     CH$_3$OH &    515.7 &    475.8 &    486.2 &    512.8 \\ 
     CH$_3$SH &    466.5 &    432.2 &    446.2 &    473.7 \\ 
       CH$_4$ &    416.1 &    395.8 &    402.4 &    420.1 \\ 
        CN &    165.8 &    138.6 &    170.1 &    181.0 \\ 
        CO &    270.8 &    234.7 &    241.9 &    259.3 \\ 
       CO$_2$ &    412.1 &    350.9 &    358.9 &    389.1 \\ 
        CS &    178.3 &    147.9 &    154.7 &    171.2 \\ 
       Cl$_2$ &    \ 64.5 &    \ 50.1 &    \ 45.8 &    \ 58.0 \\ 
       ClF &    \ 67.2 &    \ 47.2 &    \ 47.4 &    \ 61.5 \\ 
       ClO &    \ 61.3 &    \ 43.3 &    \ 56.7 &    \ 64.6 \\ 
        F$_2$ &    \ 42.6 &    \ 16.0 &    \ 28.4 &    \ 38.2 \\ 
      H2CO &    381.4 &    343.1 &    351.6 &    373.7 \\ 
       H2O &    235.6 &    212.1 &    220.8 &    232.7 \\ 
      H2O2 &    274.4 &    231.7 &    252.2 &    268.7 \\ 
       H2S &    179.8 &    167.6 &    170.0 &    182.5 \\ 
       HCN &    322.3 &    281.4 &    295.6 &    311.7 \\ 
       HCO &    287.2 &    251.3 &    260.3 &    278.5 \\ 
       HCl &    107.0 &     \ 98.0 &     \ 92.7 &    106.4 \\ 
        HF &    145.2 &    129.4 &    130.3 &    141.1 \\ 
      HOCl &    168.5 &    138.3 &    148.8 &    164.6 \\ 
       Li$_2$ &    \ 18.2 &    \ 13.5 &    \ 18.3 &    \ 24.4 \\ 
       LiF &    145.2 &    127.3 &    125.4 &    138.9 \\ 
       LiH &    \ 53.2 &    \ 50.5 &    \ 54.1 &    \ 58.0 \\ 
        N$_2$ &    238.0 &    196.3 &    219.9 &    228.5 \\ 
      N$_2$H$_4$ &    436.4 &    393.8 &    421.5 &    438.6 \\ 
        NH &    \ 79.2 &    \ 74.1 &    \ 81.3 &    \ 83.6 \\ 
       NH$_2$ &    178.0 &    164.6 &    177.3 &    182.0 \\ 
       NH$_3$ &    295.3 &    271.9 &    288.0 &    298.0 \\ 
        NO &    155.2 &    120.2 &    145.5 &    152.8 \\ 
       Na$_2$ &    \ 13.1 &     \ \ 8.7 &    \ 14.2 &    \ 17.0 \\ 
      NaCl &    101.3 &    \ 91.5 &    \ 88.7 &    \ 97.8 \\ 
        O2 &    129.8 &    \ 95.4 &    110.8 &    120.3 \\ 
        OH &    106.6 &    \ 96.5 &    102.0 &    106.7 \\ 
        P$_2$ &    118.1 &   \ 92.2 &    112.1 &    117.2 \\ 
        \br
\end{tabular}
\end{indented}
\end{table}
\begin{table}[tp]
  \begin{indented}
  \item[]
  \begin{tabular}{@{}lcccc}
    \multicolumn{5}{@{}l}%
    {\textbf{Table~\ref{Tab:G2_AE}.} (\textit{continued})} \\
    \br
     Molecule & MP2 & RPA@HF & RPA@PBE & Exp. \\
    \mr
       PH$_2$ &    145.7 &    140.7 &    148.2 &    153.1 \\ 
       PH$_3$ &    231.6 &    221.7 &    231.9 &    243.6 \\ 
        S$_2$ &    100.6 &    \ 79.1 &   \ 89.9 &    101.7 \\ 
        SO &    132.7 &    104.6 &    110.9 &    125.0 \\ 
       SO$_2$ &    273.8 &    206.6 &    232.9 &    258.4 \\ 
       Si$_2$ &   \ 73.7 &    \ 64.6 &    \ 67.7 &    \ 74.7 \\ 
     Si2H6 &    518.4 &    500.9 &    506.6 &    530.6 \\ 
    SiH2($^1A_1$) &    146.4 &    140.8 &    146.5 &    151.7 \\ 
    SiH2($^3B_1$) &    127.0 &    124.0 &    124.7 &    130.9 \\ 
      SiH$_3$ &    219.2 &    213.4 &    217.1 &    227.0 \\ 
      SiH$_4$ &    314.5 &    306.2 &    310.4 &    322.0 \\ 
      SiO &    204.3 &    170.0 &    179.0 &    191.7 \\ 
    \mr
      ME    &    1.0  & -21.5& -13.3&  \\
      MAE   &    5.9  & 21.7 & 13.3 & \\
      MAPE  &     4.3  \% & 13.4\% & 7.6\% & \\
     \br
\end{tabular}
\end{indented}
\end{table}
\subsection{Benchmark MP2 and RPA binding energies for the S22 molecular set}
 \begin{table}[tp]
  \caption{\label{Tab:S22_set} Binding energies (in meV) of the S22 
   molecular set \cite{Jurecka/etal:2006} calculated 
   with MP2, RPA@HF, and CCSD(T). MP2 and RPA results have been obtained
   with FHI-aims and ``\textit{tier} 4 + a5Z-d" basis sets. The reference CCSD(T) results
    are from \cite{Takatani/etal:2010}. } 
 \begin{indented}
 \item[]
  \begin{tabular}{@{}llcccc}
   \br
    &  Molecules & MP2 & RPA@HF & RPA@PBE & CCSD(T) \\
 1 & NH$_3$ dimer &   136 & 116  & 112 & 137 \\
 2 & H$_2$O dimer & 214 & 201 & 182 & 218 \\
 3 & Formic acid dimer & 799& 790 &744 & 816 \\
 4 & Formamide dimer & 681 & 662 & 645 & 700 \\
 5 & Uracil dimer & 881 & 846 & 816 & 898 \\
 6 & 2-pyridoxine.2-pyridoxine & 749 & 666 & 678 & 738 \\
 7 & Adenine thymine & 714 & 639 & 658 & 727   \\
 8 & CH$_4$ dimer & 21 & 11 & 17 & 23 \\
 9 & C$_2$H$_4$ dimer & 67 & 37 & 49 & 65 \\
10 & Benzene.CH$_4$ &  78 & 35 & 49 & 63 \\
11 & Benzene dimer & 211 & 40 & 82 & 114 \\
   & (slip parallel) & & & & \\
12 & Pyrazine dimer & 296  & 110 & 144 &  182 \\
13 & Uracil dimer &  482   &  310 & 379  &  423 \\
14 & Indole.benzene & 345 & 86 & 148 & 199 \\
15 & Adenine.thymine (stack) & 641 & 388 & 422 & 506 \\
16 & Ethene.ethine & 72 & 57 & 54  & 65 \\
17 & Benzene.H$_2$O & 153 & 119 & 123 & 143 \\
18 & Benzene.NH$_3$ & 114 & 74 & 85 & 101 \\
19 & Benzene.HCN & 223 & 178 & 170 & 197 \\
20 & Benzene dimer (T-shape)& 156 & 75 & 96 &118 \\
21 & Indole.benzene (T-shape) & 200 & 182 & 215 & 244 \\
22 & Phenol dimer & 334 & 250 & 266 & 308 \\
\mr
 & ME & 26 & 51 & 39 & \\
 & MAE & 37 & 51 & 39 & \\
 & MAPE & 19\% & 25\% & 16\% & \\
\br
\end{tabular}
\end{indented}
\end{table}
The S22 molecular set proposed by Jure\v{c}ka \textit{et
al.} \cite{Jurecka/etal:2006} has become a standard benchmark database for testing the accuracy of existing and newly developed
methods for the description of weak interactions. S22 represents an
``unbiased" set in the sense that  
it contains molecules of different bonding nature (7 hydrogen bonded,
8 dispersion bonded, and 7 with mixed bonding character) and of different
size (ranging from small ones like the water dimer to relatively large ones
like the Adenine-thymine dimer containing 30 atoms). The MP2 and CCSD(T) 
interaction energies for this set of molecules were already computed by
Jure\v{c}ka \textit{et al.} and extrapolated to the Gaussian 
complete basis set (CBS) limit.  A more consistent and accurate extrapolation for the 
CCSD(T) values was recently carried out by Takatani \textit{et
al.} \cite{Takatani/etal:2010}, which we therefore adopt as reference
here. While MP2 calculations for S22 are common, RPA benchmark
calculations for the whole set have not been performed, yet. 
RPA-type calculations for S22 have recently been reported 
\cite{Zhu/etal:2010,Ren/etal:2011,Eshuis/Furche:2011}. The agreement between
the data from different groups is not perfect, and the basis incompleteness
could be an issue.  Of our own RPA numbers only 
the MAEs and MAPEs have been reported previously \cite{Ren/etal:2011}. 
Now in \Tref{Tab:S22_set} the actual MP2, RPA@HF, and RPA@PBE binding
energies for these molecules as computed using FHI-aims and the
composite ``\textit{tier} 4 + a5Z-d" basis set.  
With our basis setup, the MP2 binding energies are underestimated by $\sim$4.5 meV ($2\%$ on a relative scale) compared to the MP2/CBS results reported in \cite{Jurecka/etal:2006}. We expect a similar convergence 
of the RPA numbers based on our basis set convergence tests shown in previous sections. 
Compared to the CCSD(T) reference data, RPA@PBE, which nowadays dominates
RPA-type production calculations, systematically underestimates all of the three weak bonding categories and gives a MAE of 39 meV and a MAPE of 16\%.  If instead HF is used as a starting point (i.e., as for MP2), the description
of hydrogen bonding improves, while the
description of dispersion bonding worsens. The overall MAE for
RPA@HF is 51 meV and the MAPE 25\%. We are thus faced with the conundrum that RPA can
describe the weak interactions that are beyond the
reach of LDA, GGA, and hybrid functionals, but the accuracy of the two standard RPA schemes is not spectacular. We have analyzed the origin of this underbinding behaviour in \cite{Ren/etal:2011} and 
proposed a simple solution to overcome this problem.

\section{\label{sec:conc}Conclusions and outlook}        
To summarize, we have presented a resolution of identity framework for
the two-electron Coulomb operator that allows efficient, accurate
electronic structure computations based on HF, hybrid functionals, 
MP2, RPA, and $GW$ based on the flexible basis function form of
NAOs. We have shown that NAO basis sets as implemented in FHI-aims are a
competitive choice for approaches involving exact-exchange and/or 
non-local correlation terms, with rather compact basis sets sufficing
for essentially converged results. Our simple ``on-the-fly" scheme to
construct the auxiliary basis functions using Gram-Schmidt
orthonormalization of the ``on-site" products of single-electron
atomic orbitals gives a natural, accurate representation of the
two-electron Coulomb operator for practical calculations. Taken
together, our framework paves the way for an extended usage 
of NAOs in more advanced computational approaches that go beyond LDA
and GGAs. Specifically, we have applied the $G^\text{0}W^\text{0}$ and
MP2 quasiparticle approaches to compute the vertical IEs of a
set of small molecules, and the RPA method to compute the G2-I
atomization energies and the interaction energies for the S22
molecular set. We believe that the well converged numbers reported in 
this work may serve as benchmarks for future studies. Beyond the
specific examples given here, our RI framework as a whole has
already proven to be stable and mature in a number of scientific
applications \cite{Ren/Rinke/Scheffler:2009,Ren/etal:2011,Paier/etal:2011, 
Rossi:PhDthesis,Jenness/etal:2011,Marom/etal:2011a,Marom/etal:2011b}. 

Based on the results above, NAOs emerge as a promising route towards
more compact basis sets for correlated methods. Among our ongoing
developments is the attempt to design more optimized NAO basis sets 
for MP2, RPA, and $GW$ calculations. Another active line of
development is to improve the scaling behaviour of the computational
cost with system size, exploiting the locality of the NAO basis
functions \cite{Wieferink/etal:inpreparation}. Last but not least, we
are working on the extension of the present scheme to periodic
systems. All this combined, we expect NAO-based implementations of
methodologies that go beyond LDA and GGAs to become competitive
alternatives to the traditional implementations that are based on GTO 
or plane-wave basis sets, in particular towards the limit of very
large systems: the benefit of compact basis set size at
a \emph{given} accuracy level should help tackle problem sizes that,
otherwise, could not be done.

\addcontentsline{toc}{section}{Acknowledgments}
\section*{Acknowledgments}
        We acknowledge fruitful discussions with and encouragement by an extensive group of
        friends, coworkers and colleagues over the many years that our
        approach has been in productive use. This work was in part
        funded by the  EU's Sixth Framework Programme through the
        NANOQUANTA (NMP4-CT-2004-500198) network of excellence and
        the EU's Seventh Framework Programme through the European
        Theoretical Spectroscopy Facility e-Infrastructure   grant
        (no. 211956).  
  \begin{appendix}
   \addcontentsline{toc}{section}{Appendix A. Matrix elements for numeric atom-centered orbitals}
   \section*{\label{sec:appA}Appendix A. Matrix elements for numeric atom-centered orbitals}
   \addcontentsline{toc}{subsection}{Appendix A.1. Coulomb potential of a numerical radial function}
\subsection*{Appendix A.1. Coulomb potential of a numerical radial function}
\label{sec:Coulomb}

To reduce the (formally) six-dimensional Coulomb integrations to three
dimensional ones, we first solve Poisson's equation for  
each $P_\mu(\bfrp)$ (classical electrostatics). We define
$Q_{\mu}(\bfr)$ as 
 \begin{equation}
  Q_{\mu}(\bfr) = \int d{\bfrp} v(\bfr-\bfrp)P_\mu(\bfrp)\, .
  \label{Eq:coulomb_auxil_conv}
 \end{equation}
Using the Laplace expansion of the Coulomb potential $v(\bfr-\bfrp)=1/|{\bf r-r^\prime}|$, 
\begin{equation}
v(\bfr-\bfrp)=\sum_{lm}\frac{4\pi}{2l+1}Y_{lm}(\hat{\bf r})Y^\ast_{lm}(\hat{\bf r^\prime})
\times \frac{r_{<}^l}{r_{<}^{l+1}},
\label{Eq:coulomb_expansion}
\end{equation}
where $r_{<}=\text{min}(r,r')$ and $r_{>}=\text{max}(r,r')$, the
integration of the angular part in (\ref{Eq:coulomb_auxil_conv})
can be done analytically, yielding
 \begin{equation}
  Q_{\mu}(\bfr) = \frac{\alpha_{al\kappa}(r)}{r}Y_{lm}(\hat{\bfr}_a).
 \end{equation}
The radial part $\alpha_{al\kappa}(r)$ is given by a simple one-dimensional 
(numerical) integration \cite{Arfken/Weber:2001,Delley:1990,Blum/etal:2009} 
  \begin{equation}
     \alpha_{al\kappa}(r) = \frac{4\pi}{2l+1} 
    \left[ \left(\int _0^{r} d r_<
\xi_{al\kappa}(r_<) \frac{r_<^{l+1}}{r^{l}}\right) + 
\left(\int _r^{\infty}  d r_>
\xi_{al\kappa}(r_>) \frac{r^{l+1}}{r_>^{l}}\right) \right].
  \end{equation}
Thus, the three-center and two-center Coulomb integrals 
(\ref{Eq:coulomb_matrix})  and (\ref{Eq:V_coefficient}) reduce
to the three-dimensional integrals in (\ref{Eq:3center_integral})
and  (\ref{Eq:2center_integral}).

\subsection*{Appendix A.2. Grid-based three-center and two-center integrals}
\addcontentsline{toc}{subsection}{Appendix A.2. Grid-based three-center and two-center integrals}
\label{sec:three-center}

The three-center and two-center integrals in the present work are
computed by grid-based integrations using overlapping, atom-centered
spherical grids and the same technology that is used in many
quantum-chemical applications for the exchange-correlation matrix in
DFT \cite{Becke:1988,Delley:1990}. The integration grid points
$\bfr=\bfr (a,s,t)$ are uniquely specified by 
the atomic center $a$, the radial shell number $s$, and 
angular point $t$. $w(s,t)$ is the corresponding integration
weight. Details of our own implementation (FHI-aims) are given in
\cite{Blum/etal:2009,Havu/etal:2009}. 
Since these are true three-center integrals, we restrict 
the integration domain for a particular integral element to the grids 
associated with the atoms on which the basis functions in  
question are centered.  For instance, denoting the respective atoms by
$a_1$, $a_2$, $a_3$, the three-center integrals  in
(\ref{Eq:3center_integral}) can then be discretized as 
\begin{equation}
   (ij|\mu) = \sum_{a={a_1,a_2,a_3}} \sum_{s,t}p_3(a,\bfr) w(s,t) \phi_i(\bfr)\phi_j(\bfr)Q_\mu(\bfr), 
\end{equation}
where $p_3(a,\bfr)$ is a three-center partition function that satisfies 
\begin{equation}
  \sum_{a=a_1,a_2,a_3}p_3(a,\bfr)=1  \nonumber
\end{equation}
everywhere in the overlapping region of the three functions, and is
zero otherwise. The underlying numerical grids can in principle be
increased up to arbitrary accuracy if needed. The two-center integrals
(\ref{Eq:2center_integral}) can be performed in a similar fashion
using overlapping grids, or with the spherical Bessel transform
techniques explained below.   

   \addcontentsline{toc}{subsection}{Appendix A.3. Two-center integration in Fourier space}
\subsection*{Appendix A.3. Two-center integration in Fourier space}
\label{sec:two-cent-integr}

As mentioned in section~\ref{sec:numerical_integral},
two-center integrals of numeric atom-centered basis functions like
$V_{\mu\nu}$ in (\ref{Eq:coulomb_matrix}) and $S_{\mu\nu}$ in (\ref{Eq:ABF_overlap})
can be efficiently calculated in Fourier space as described by Talman \cite{Talman84-4center,Talman03-MCI}. 
This and the following (\ref{sec:logsbt}) subsections give the details of our
implementation.  We first describe the general procedure here, and the central
ingredient of our implementation, the logarithmic spherical Bessel transform 
(logarithmic SBT, logSBT), will be presented in the next subsection.

As is well-known, the overlap of two functions $f(\bm r)$ and $g(\bm r)$ can
be expressed in Fourier space as
\begin{equation}
  \label{eq:Cb-rec}
  \int  f(\bm r)  g(\bm r - \bm R) \opd\bm r
  = \int \tilde f(\bm k) \tilde g(-\bm k)
  \ep{\im\bm k\cdot\bm R} \opd\bm k
  .
\end{equation}
The Fourier transform $\tilde f(\bm k)$ in~\eqref{eq:Cb-rec} of an atomic
function $f(\bm r) = f(r) Y_{lm}(\hat{\bm r})$ has the same angular momentum
in real and Fourier space for symmetry reasons
\begin{equation}
  \label{eq:FT}
  \tilde f(\bm k)
  = (2\pi)^{-\frac{3}{2}} \int 
  f(\bm r) \ep{-\im\bm k\cdot\bm r} \,\opd \bm r
  = \im^{-l} \tilde f(k) Y_{lm}(\hat{\bm k})
  .
\end{equation}
The radial part $\tilde f(k)$ is given by the SBT of $f(r)$
\begin{equation}
  \label{eq:SBT}
  \tilde f(k)
  = \sqrt{\frac{2}{\pi}} \int_0^\infty j_{l}(kr) f(r)\,r^2 \opd r
  .
\end{equation}
If $f(r)$ is tabulated on a logarithmic grid, its SBT can be calculated
efficiently on an equivalent logarithmic grid using the fast logSBT
algorithm \cite{Talman78-SBT, Hamilton00-fftlog, Talman09-numSBT} as described
in the next section.

The three-dimensional integral~\eqref{eq:Cb-rec} can be separated by expanding
the plane wave $\ep{\im\bm k\cdot\bm R}$ in spherical harmonics and spherical
Bessel functions
\begin{equation}
  \label{eq:relations}
  \ep{\im \bm k\cdot\bm R}
  = (2\pi)^{\frac{3}{2}} \sum_{L=0}^\infty
  \im^L \sqrt{\frac{2}{\pi}} j_L(kR)
  \sum_M Y_{LM}(\hat{\bm k}) Y_{LM}^*(\hat{\bm R})
  .
\end{equation}
The separation yields
\begin{equation}
  \label{eq:Talman-V}
    \int f(\bm r)  g(\bm r - \bm R)\, \opd\bm r
    = (2\pi)^{\frac{3}{2}} \sum_L \im^{-l+l'+L} I_L(R) A_L(\hat{\bm R})
\end{equation}
with a radial integral
\begin{equation}
  \label{eq:Talman-IL}
  I_L(R)
  = \sqrt{\frac{2}{\pi}} \int_0^\infty
  j_L(kR) \tilde f(k) \tilde g(k)\, k^2 \opd k
\end{equation}
and an angular integral
\begin{equation}
  \label{eq:Talman-A}
  A_L(\hat{\bm R})
  = \sum_M Y_{LM}^*(\hat{\bm R})\,
  C\bigl(lm; l'm'; LM\bigr)
  .
\end{equation}
The triple-$Y$ integrals
\begin{equation}
  \label{eq:triple-Y}
  C\bigl(lm; l'm'; LM\bigr)
  := \int_{\Omega} 
  Y_{lm}(\hat{\bm k}) Y_{l'm'}(\hat{\bm k}) Y_{LM}(\hat{\bm k})
  \, \opd \hat{\bm k}
  .
\end{equation}
in~\eqref{eq:Talman-A} can be calculated efficiently using recursion
formulae \cite{Talman03-MCI}. They are nonzero only for $L = \abs{l-l'},
\abs{l-l'}+2, \ldots, (l+l')$.  If two atom-centered functions do not overlap,
the overlap integrals $I_L(R)$ of course vanish.

For any given distance $R$, the radial integrals $I_L(R)$
in~\eqref{eq:Talman-IL} can be calculated directly using the trapezoidal rule on
the logarithmic grid when $j_L(kR)$ is evaluated in logarithmic Fourier space as described in the next section.  
If integrals of the same atom-centered functions
for many differing distances are needed, one can compute these more efficiently
by interpreting~\eqref{eq:Talman-IL} as an SBT of $\tilde P(k) = \tilde f(k) \tilde g(k)$. 
By applying the logSBT, and interpolate for all needed distances $R$, one can obtain all  
the integrals at tight-binding cost, meaning that efficient recursion formula (together
with spline evaluations) can be used intead of evaluating each integral numerically.


Coulomb interactions of atomic functions can be calculated with comparable
ease where the integrand in~\eqref{eq:Talman-IL} is multiplied with 
the Coulomb kernel $4\pi/k^2$
\begin{equation}
  \label{eq:Talman-VL}
  V_L(R)
  = 4\pi \sqrt{\frac{2}{\pi}} \int_0^\infty
  j_L(kR) \frac{\tilde f(k) \tilde g(k)}{k^2}\,k^2\opd k.
\end{equation}
The Coulomb interaction generally does not vanish even if the two charge
densities do not overlap.  However, it has a simple multipolar behaviour and explicit integration of~\eqref{eq:Talman-VL} can thus be avoided.
The function $V_L(R)$ can formally be interpreted as the far field of a charge
distribution of angular momentum $L$ whose radial part $P(r)$ is given by its
SBT $\tilde P(k) = \tilde f(k)\tilde g(k)$.  Therefore, it only depends on the
multipole moment of $P$, which is encoded in its limiting behaviour
for small $k$.  From this, it can be shown that $V_L(R)$ vanishes for all
$L<l+l'$.  For $L=l+l'$ we get
\begin{equation}
  \label{eq:Talman-VL-far}
  V_{l+l'}(R)
  = \frac{4\pi}{R^{l+l'+1}} \sqrt{\frac{2}{\pi}}
  \frac{(2l+2l'-1)!!}{(2l-1)!!\, (2l'-1)!!}\,
  p_f p_g
\end{equation}
with $(2n-1)!! = 1\cdot 3 \cdots (2n-1)$ and with multipole moments
\begin{equation}
  \label{eq:multipole-moment-p_f}
  p_f = \frac{1}{2l+1} \int_0^\infty r^{2+l} f(r)\,\opd r.
\end{equation}
Therefore, the Coulomb interaction of non-overlapping functions depends only
on the product of their multipole moments and can also be calculated at
tight-binding cost.

As a side remark we emphasize that two ABFs do not interact via the Coulomb interaction if they do not overlap and
at least one of the two multipole moments is zero.  As shown by Betzinger in
\cite{Betzinger10-FLAPW}, all but one of the ABFs for a given atom
and a given angular momentum $lm$ can then be chosen to be multipole free by means
of a suitable unitary transformation.

\addcontentsline{toc}{subsection}{Appendix A.4. Logarithmic spherical Bessel transform}
\subsection*{Appendix A.4. Logarithmic spherical Bessel transform}
\label{sec:logsbt}

This section describes our implementation of the logSBT
algorithm.  For a more extensive description we refer the interested reader to
Talman \cite{Talman09-numSBT} and Hamilton \cite{Hamilton00-fftlog}.

The SBT as defined in~\eqref{eq:SBT} can be written as the integral over a
kernel $\mathcal J(kr)$ and a right-hand-side $\mathcal F(r)$:
\begin{equation}
  \label{eq:SBT-kernel}
  \tilde f(k)
  = k^{-\alpha} \int_0^\infty \frac{\opd r}{r}
  \underbrace{\sqrt{\frac{2}{\pi}} j_l(kr) (kr)^\alpha}_{=: \mathcal J(kr)}
  \, \underbrace{r^{3-\alpha} f(r)}_{=: \mathcal F(r)}.
\end{equation}
The choice of the power bias parameter $0 \le \alpha\le 3$ is crucial for the
numerical accuracy and stability of this method and will be discussed further
below.  The basic idea of the logSBT is that in logarithmic coordinates
($\rho=\log r$ and $\kappa = \log k$) the kernel reads $\mathcal J(\kappa +
\rho)$ and~\eqref{eq:SBT-kernel} turns into a convolution, which can be
efficiently calculated using fast Fourier transorms (FFTs).  Please note that
$dr/r=d\rho$ in~\eqref{eq:SBT-kernel}.

As pointed out by Hamilton \cite{Hamilton00-fftlog}, this procedure is exact if
both $\mathcal F(\rho) = \ep{(3-\alpha)\rho} f(\ep{\rho})$ and the corresponding
SBT term $\tilde{\mathcal F}(\kappa) = \ep{\alpha\kappa} \tilde f(\ep{\kappa})$
are periodic in logarithmic space and analytic expressions for the logarithmic
Fourier transform of the kernel are used.  Periodicity can be achieved,
e.\,g., by choosing $\alpha$ near 1.5 and using a sufficiently wide
logarithmic grid.  Under these circumstances, both $\mathcal F(\rho)$ and
$\tilde{\mathcal F}(\kappa)$ smoothely drop to zero on both ends, which
therefore can safely be connected.

Unfortunately, the scaling factor $k^{-\alpha}$ turns out to be quite problematic.  By
design of the algorithm, the absolute error of $\tilde F(\kappa) = k^{\alpha}
f(k)$ before the final scaling is typically of the order of machine
precision, i.\,e., about $10^{-15}$.  This is true even if the magnitude of
the exact value is much smaller than that.  After scaling, however, the absolute error
can get arbitrarily large because $k$ can be very small on a wide logarithmic
grid.

Talman \cite{Talman09-numSBT} circumvents this problem by using two separate
$\alpha$ for small and large $k$ and joining the two results where they differ
least.  The small-$\alpha$ ($\alpha=0$) calculation cannot be done assuming
periodicity for $l=0$ because $\mathcal F(\rho)$ does not decay to zero for
$\rho\rightarrow-\infty$.  Therefore, a trapezoidal rule is used for the
integration, which works well for small $k$ where the Bessel function is
smooth on a logarithmic scale.  This does not break with the spirit of logSBT, because the trapezoidal rule can be
formulated using FFTs, too.

In order to avoid the second transform, we take a different approach.  In
practice, one needs the SBT only for one kind of integral and it is sufficient
to calculate $k^\alpha \tilde f(k)$ to high \emph{absolute} accuracy for a
single given $\alpha$, which can be used as power bias for the transform.

This works well for all cases but $\alpha=0$ and $l=0$.  Here, we cannot
simply resort to the trapezoidal rule because it is invalid for high $k$ where
the Bessel function oscillates rapidly.  Instead, we separate $j_0(\ep{\tau})$
into a smooth part proportional to $\erfc(\tau/\Delta\tau_0)$ and a properly
decaying rest.  We use the first part for a trapezoidal rule and the second
part for the log-periodic algorithm.  Fortunatly, these two schemes differ
only in the way the kernel is constructed so that the two kernels can simply
be added up to a ``hybrid'' kernel.  The actual transform is not affected and
numerically not more expensive than an ordinary logSBT.

Just like Talman \cite{Talman09-numSBT} we double the domain during the
transforms for $l=0$ and $l=1$ in order to avoid the need for large domains
for a proper decay behaviour.


   \addcontentsline{toc}{section}{Appendix B. Ionization energies of the a set of atoms and molecules}
   \section*{\label{sec:appB}Appendix B. Ionization energies of the a set of atoms and molecules}
   In Table~\ref{Tab:G2_IEs} the individual numbers for the \textit{vertical} ionization potentials for
 50 atoms and molecules as computed with
6 different computational approaches are presented. The calculations are performed with FHI-aims and 
``\textit{tier}+a5Z-d" basis set.
\begin{table}\footnotesize
\noindent
\caption{\label{Tab:G2_IEs}The \textit{vertical} ionization potentials (in eV) for 50 atoms and molecules 
 (taken from G2 ion test set \cite{Curtiss/etal:1998}) calculated with HF, MP2, MP2-QP, 
 $G^\text{0}W^\text{0}$@HF, $G^\text{0}W^\text{0}$@PBE0, and $G^\text{0}W^\text{0}$@PBE in comparison to the experimental values, taken from the the NIST database \cite{nist_database}.  The mean absolute errors (MAE) for the three approaches are also shown.}
\begin{tabular}{@{}l@{}c@{}c@{}c@{}c@{}c@{}c@{}c}
\br\\[-1.5ex]
Molecule & \multicolumn{1}{c}{~~~~Exp.~~~~ }& 
\multicolumn{1}{c}{~~HF~~} &
 \multicolumn{1}{c}{~~MP2~~} & 
 \multicolumn{1}{c}{~MP2-QP~} & 
\multicolumn{1}{c}{$G^\text{0}W^\text{0}$@HF} & 
\multicolumn{1}{c}{$G^\text{0}W^\text{0}$@PBE0} &
\multicolumn{1}{c}{$G^\text{0}W^\text{0}$@PBE} \\[0.2ex] 
\mr \\[-1.5ex]
Al         &      ~5.98 &     ~5.95 & ~5.85 &  ~6.08 &    ~6.24 &   ~5.94 &   ~5.64 \\ 
Ar         &     15.76 &    16.08 &  15.87 &  15.61 &    16.08 &    15.51 &    15.21 \\ 
B          &     ~8.30 &    ~8.68 &  ~8.33&  ~8.79 &    ~8.73 &    ~8.11 &    ~7.65 \\ 
BCl$_3$       &     11.64 &    12.48 & 12.58&   11.65 &    12.37 &    11.63 &    11.25 \\ 
BF$_3$        &     15.96 &    18.04 & 16.19&   15.02 &    16.77 &    15.79 &    15.21 \\ 
Be         &     ~9.32 &    ~8.48 & ~8.87&   ~8.98 &    ~9.16 &    ~9.26 &    ~9.03 \\ 
C          &     11.26 &    11.95 & 11.33&   11.79 &    11.68 &    10.93 &    10.47 \\ 
C$_2$H$_2$       &     11.49 &  11.19 & 11.75&   11.43 &    11.76 &    11.29 &    11.01 \\ 
C$_2$H$_4$       &     10.68 &    10.23 &  10.77&   10.37 &    10.83 &    10.47 &    10.22 \\ 
C$_2$H$_4$S      &     ~9.05 &    ~9.43 &  ~9.27&   ~8.74 &    ~9.44 &    ~8.94 &    ~8.71 \\ 
C$_2$H$_5$OH     &     10.64 &    12.05 &  11.30&  10.08 &    11.47 &    10.63 &    10.20 \\ 
C$_6$H$_6$       &     ~9.25 &    ~9.15 &  ~9.88&  ~9.08 &    ~9.63 &    ~9.20 &    ~9.00 \\ 
CH$_2$CCH$_2$    &     10.20 &    10.31 &  10.55&    9.94 &    10.70 &    10.12 &    ~9.85 \\ 
CH$_2$S       &     ~9.38 &    ~8.24 & ~9.76 &   ~8.04 &    ~8.22 &    ~9.27 &    ~9.01 \\ 
CH$_3$        &     ~9.84 &    10.47 & ~9.78 &   10.61 &    10.28 &    ~9.59 &    ~9.24 \\ 
CH$_3$Cl      &     11.29 &    11.87 & 11.63&   11.20 &    11.88 &    11.31 &    11.03 \\ 
CH$_3$F       &     13.04 &    14.46 & 14.18&    12.98 &    14.15 &    13.28 &    12.77 \\ 
CH$_3$SH      &     ~9.44 &    ~9.67 & ~9.56&   ~9.25 &    ~9.81 &    ~9.31 &    ~9.06 \\ 
CH$_4$        &     13.60 &    14.85 & 14.46 &   14.18 &    14.89 &    14.27 &    13.98 \\ 
CHO        &     ~9.31 &    11.13 & ~9.36&   ~9.92 &    10.64 &    ~9.61 &    ~9.14 \\ 
CO         &     14.01 &    15.10 &  14.55 &  14.06 &    14.85 &    13.78 &    13.30 \\ 
CO$_2$        &     13.78 &    14.83 & 14.86&   13.29 &    14.48 &    13.68 &    13.21 \\ 
CS$_2$        &     10.09 &    10.14 & 10.86 &   10.03 &    10.55 &    10.02 &    ~9.72 \\ 
Cl         &     12.97 &    13.09 &  12.90&   13.21 &    13.32 &    12.83 &    12.51 \\ 
Cl$_2$        &     11.49 &    12.08 &  11.66&  11.38 &    12.13 &    11.49 &    11.04 \\ 
ClF        &     12.77 &    11.63 & 11.34&   11.08 &    11.72 &    11.15 &    10.72 \\ 
F          &     17.42 &    18.50 & 17.42&   17.30 &    17.73 &    17.07 &    16.71 \\ 
FH         &     16.12 &    17.70 & 16.47&   14.93 &    16.39 &    15.83 &    15.39 \\ 
H          &     13.61 &    13.61 & 13.61&   13.61 &    13.61 &    13.04 &    12.52 \\ 
He         &     24.59 &    24.98 & 24.41&   24.58 &    24.68 &    24.01 &    23.59 \\ 
Li         &     ~5.39 &    ~5.34 & ~5.38 &  ~5.38 &    ~5.68 &    ~5.84 &    ~5.67 \\ 
Mg         &     ~7.65 &    ~6.88 & ~7.40&   ~7.44 &    ~7.56 &    ~7.64 &    ~7.71 \\ 
N          &     14.54 &    15.54 & 14.66&   14.98 &    14.85 &    14.06 &    13.51 \\ 
N$_2$         &     15.58 &    16.71 & 15.48&   17.22 &    17.27 &    15.45 &    14.86 \\ 
NH$_3$        &     10.82 &    11.70 & 11.06&   10.29 &    11.36 &    10.70 &    10.32 \\ 
Na         &     ~5.14 &    ~4.97 & ~5.11&   ~5.09 &    ~5.37 &    ~5.51 &    ~5.51 \\ 
NaCl       &     ~9.80 &    ~9.68 & ~9.42&   ~9.98 &    ~9.59 &    ~9.09 &    ~8.79 \\ 
Ne         &     21.56 &    23.14 & 21.63&   20.22 &    21.76 &    21.10 &    20.54 \\ 
O          &     13.61 &    14.20 & 13.48&   14.64 &    13.90 &    13.37 &    13.04 \\ 
O$_2$         &     12.30 &    15.22 & 11.85&   12.57 &    13.71 &    12.33 &    11.68 \\ 
OCS        &     11.19 &    11.47 & 11.95&   11.16 &    11.70 &    11.16 &    10.88 \\ 
OH         &     13.02 &    13.98 & 13.14&   13.21 &    13.38 &    12.79 &    12.41 \\ 
P          &     10.49 &    10.67 &  10.56&  10.78 &    10.72 &    10.24 &     9.94 \\ 
P$_2$         &     10.62 &    10.10 & 10.86&   10.59 &    10.70 &    10.35 &    10.13 \\ 
\br
\end{tabular}
\end{table}
\begin{table}\footnotesize
\begin{tabular}{@{}l@{}c@{}c@{}c@{}c@{}c@{}c@{}c}
\multicolumn{8}{@{}l}{\textbf{\Tref{Tab:G2_IEs}.} \textit{(continued)}} \\
\br\\[-1.5ex]
Molecule & \multicolumn{1}{c}{~~~~Exp.~~~~ }& 
\multicolumn{1}{c}{~~HF~~} &
 \multicolumn{1}{c}{~~MP2~~} & 
 \multicolumn{1}{c}{~MP2-QP~} & 
\multicolumn{1}{c}{$G^\text{0}W^\text{0}$@HF} & 
\multicolumn{1}{c}{$G^\text{0}W^\text{0}$@PBE0} &
\multicolumn{1}{c}{$G^\text{0}W^\text{0}$@PBE} \\[0.2ex] 
\mr \\[-1.5ex]
PH$_3$        &     10.59 &    10.58 & 10.58&   10.47 &    10.90 &    10.44 &    10.22 \\ 
S          &     10.36 &    10.33 &  10.15&  11.04 &    10.69 &    10.31 &    10.12 \\ 
S$_2$         &     ~9.55 &    10.38 & ~9.38&   ~9.74 &    10.21 &    ~9.39 &    ~9.05 \\ 
SH$_2$        &     10.50 &    10.49 & 10.53&   10.34 &    10.74 &    10.27 &    10.06 \\ 
Si         &     ~8.15 &    ~8.20 & ~8.10 &   ~8.33 &    ~8.38 &    ~8.01 &    ~7.76 \\ 
SiH$_4$       &     12.30 &    13.24 & 12.81&   12.90 &    13.33 &    12.68 &    12.29 \\[0.2ex] 
\br
\end{tabular}
\end{table}

   \addcontentsline{toc}{section}{Appendix C. Modified Gauss-Legendre grid}
   \section*{\label{sec:appC}Appendix C.  Modified Gauss-Legendre grid}
   
For the integrals over the imaginary frequency axis (e.g., for the RPA correlation energy, equation (\ref{Eq:E_c_rpa_ri}) ),
we use a modified Gauss-Legendre quadrature.
The Gauss-Legendre quadrature provides a way to numerically evaluate an integral on the interval $[-1:1]$
\begin{equation}
\int_{-1}^{1} f(x) dx \approx \sum_{i=1}^n w_i f(x_i),
\end{equation}
where $x_i$ and $w_i$ are the integration points and the corresponding weights, respectively. 
For our purposes a transformation procedure is applied to map the integration range from $[-1:1]$ to $[0:\infty]$ 
whereby the $x_i$ and $w_i$ have to be changed accordingly. Specifically, we use the modification proposed for the evaluation of the Casimir-Polder integral \cite{modCasimir-Polder}: 
\begin{equation}
\tilde{x}_i = x_0^{(1+x_i)/(1-x_i)}, 
\end{equation}
with $x_0$ set to 0.5.  The weights for the tranformed grid are then given by 
\begin{equation}
\tilde{w}_i = 2 w_i x_0/(1-w_i)^2.
\end{equation}

This modified Gauss-Legendre scheme allows a quick convergence of the frequency integration with a 
relatively small number of frequency points. 
In our implementation, a 40-point grid gives micro-Hartree total energy accuracy  
for the systems investigated in this work.

  \end{appendix}
\addcontentsline{toc}{section}{References}
\section*{References}
\bibliography{./CommonBib}

\begin{thebibliography}{100}
\bibitem{basis_strategy}
Here and in the following, we refrain from employing any basis extrapolation
  strategies as often advocated in the Quantum Chemistry community, and rather
  compare to directly computed values from the largest available standard GTO
  basis sets. It is well known that basis set extrapolation can yield good
  results especially for mid-sized GTO basis sets. However, at the meV-level
  convergence with very large basis sets that we are aiming for here, it is not
  clear what degree of residual noise should be expected from a basis set
  extrapolation.
\bibitem{MP2-QP_problem}
For heavy elements, on the other hand, we found that the discrepancy between
  MP2-QP and MP2 becomes large. The discussion of this issue is beyond the
  scope of this paper and will be reported elsewhere.
\end{thebibliography}
\end{document}